\documentclass[preprint]{aastex631}

\usepackage{threeparttable}
\usepackage{soul}
\usepackage{hyperref}
\usepackage{longtable}
\usepackage{amsmath}
\shortauthors{Yu et al.}
\usepackage{multirow}
\usepackage[utf8]{inputenc}
\usepackage{CJK}

\received{September 6, 2022}
\accepted{March 14, 2023}
\submitjournal{ApJS}

\begin{document}
\title{Material Properties of Organic Liquids, Ices, and Hazes on Titan}
\correspondingauthor{Xinting Yu}
\email{xinting.yu@utsa.edu}

\author[0000-0002-7479-1437]{Xinting Yu\begin{CJK*}{UTF8}{gbsn}
(余馨婷)\end{CJK*}}
\affiliation{Department of Physics and Astronomy, University of Texas at San Antonio\\
1 UTSA Circle\\
San Antonio, TX 78249, USA}
\affiliation{Department of Earth and Planetary Sciences, University of California Santa Cruz\\
1156 High St\\
Santa Cruz, CA 95064, USA}

\author{Yue Yu\begin{CJK*}{UTF8}{gbsn}
(于越)\end{CJK*}}
\affiliation{Department of Earth and Planetary Sciences, University of California Santa Cruz\\
1156 High St\\
Santa Cruz, CA 95064, USA}

\author{Julia Garver}
\affiliation{Department of Physics, University of California Santa Cruz\\
1156 High St\\
Santa Cruz, CA 95064, USA}

\author[0000-0002-8110-7226]{Jialin Li\begin{CJK*}{UTF8}{gbsn}
(李嘉霖)\end{CJK*}}
\affiliation{Department of Physics, University of California Santa Cruz\\
1156 High St\\
Santa Cruz, CA 95064, USA}

\author{Abigale Hawthorn}
\affiliation{Department of Physics, University of California Santa Cruz\\
1156 High St\\
Santa Cruz, CA 95064, USA}

\author[0000-0002-1883-552X]{Ella Sciamma-O'Brien}
\affiliation{NASA Ames Research Center, Space Science \& Astrobiology Division, Astrophysics Branch\\
Moffett Field, CA 94035, USA}

\author[0000-0002-8706-6963]{Xi Zhang}
\affiliation{Department of Earth and Planetary Sciences, University of California Santa Cruz\\
1156 High St\\
Santa Cruz, CA 95064, USA}

\author[0000-0002-5034-1300]{Erika Barth}
\affiliation{Southwest Research Institute\\
1050 Walnut St Suite 300, Boulder, CO 80302, USA}

\begin{abstract}
Titan has a diverse range of materials in its atmosphere and on its surface: the simple organics that reside in various phases (gas, liquid, ice) and the solid complex refractory organics that form Titan's haze layers. These materials all actively participate in various physical processes on Titan, and many material properties are found to be important in shaping these processes. Future in-situ exploration on Titan would likely encounter a range of materials, and a comprehensive database to archive the material properties of all possible material candidates will be needed. Here we summarize several important material properties of the organic liquids, ices, and the refractory hazes on Titan that are available in the literature and/or that we have computed. These properties include thermodynamic properties (phase change points, sublimation and vaporization saturation vapor pressure, and latent heat), physical property (density), and surface properties (liquid surface tensions and solid surface energies). We have developed a new database to provide a repository for these data and make them available to the science community. These data can be used as inputs for various theoretical models to interpret current and future remote sensing and in-situ atmospheric and surface measurements on Titan. The material properties of the simple organics may also be applicable to giant planets and icy bodies in the outer solar system, interstellar medium, protoplanetary disks, and exoplanets.
\end{abstract}

\section{Introduction}
\begin{figure}
    \centering
    \includegraphics[width=8cm]{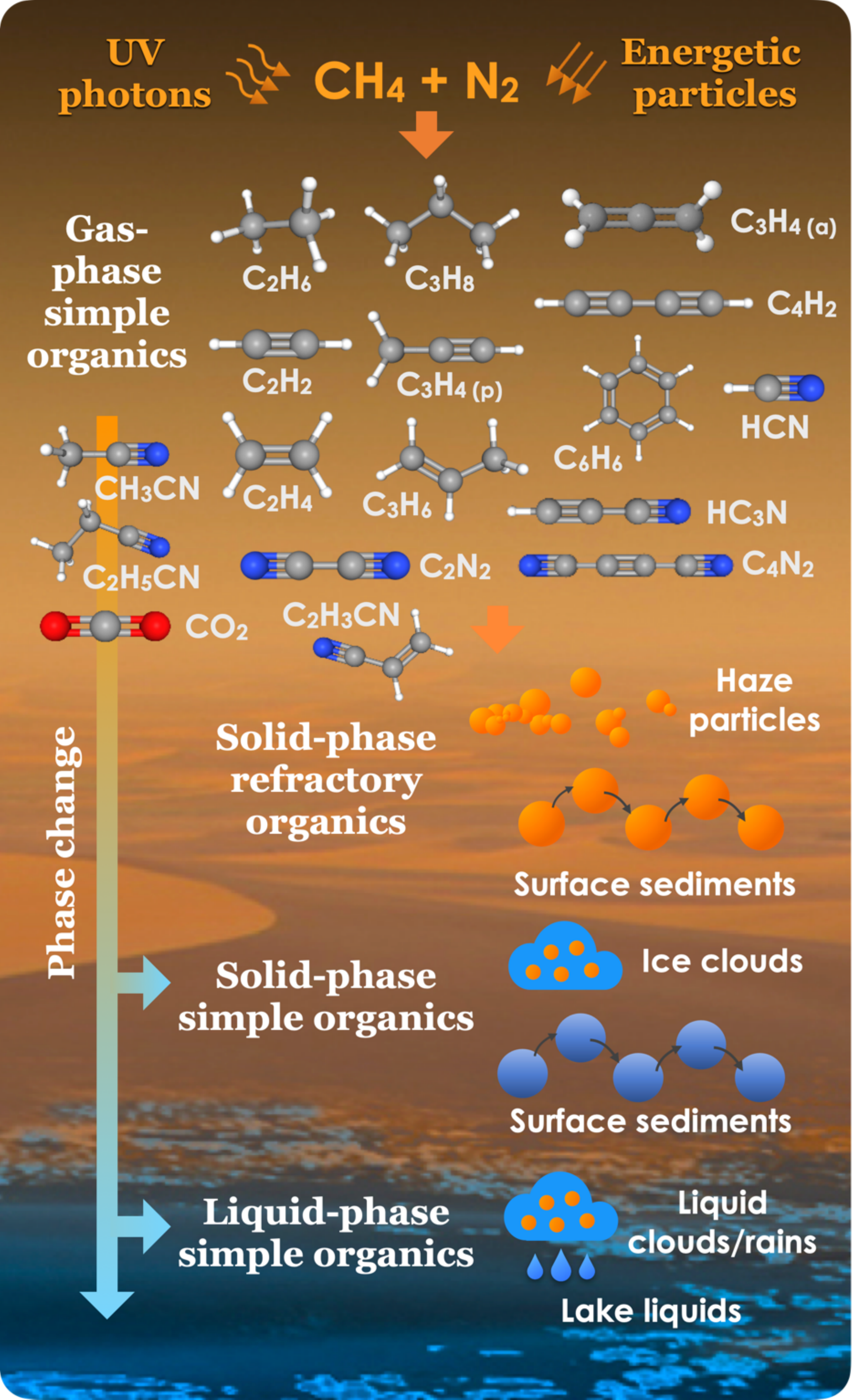}
    \label{fig:photochem}
    \caption{Photochemistry in Titan's upper atmosphere leads to the creation of numerous organic molecules, from simple organics in the gas phase to complex refractory organic solids that form Titan's thick haze layers. The simple organics could undergo phase changes, along with the complex organics, participate in various atmospheric and surface processes on Titan.}
\end{figure}

The methane (CH$_4$) and nitrogen (N$_2$) in Titan's upper atmosphere enable rich photochemistry, creating numerous organic molecules in Titan's atmosphere. At least 18 organic species have been detected with space- and ground-based observations (for a list of the observed gaseous molecules, see Table~\ref{table:abundance} and Figure~\ref{fig:photochem}). The simple organic molecules that have been detected typically have six heavy atoms or less (C or N) and include a range of hydrocarbons and nitriles, e.g., ethane (C$_2$H$_6$), acetylene (C$_2$H$_2$), benzene (C$_6$H$_6$), hydrogen cyanide (HCN), cyanoacetylene (C$_2$N$_2$). Most of the simple organic molecules are in the gas phase when produced in the upper atmosphere. During their descent through the atmosphere, Titan's unique temperature profile allows these gaseous organic molecules to condense into liquids or solids in its stratosphere, forming liquid or ice clouds \citep{1984Icar...59..133S,2005JPCA..109.1382C, 2018SSRv..214..125A}. Various types of organic clouds have been detected in Titan's atmosphere through both ground-based and spacecraft observations by Voyager 1 and Cassini \citep[and references therein]{2018SSRv..214..125A, 2018Icar..310...89V}. The simple organic molecules in Titan's atmosphere can further coagulate and react to form the complex refractory organic particles that constitute Titan's thick haze layers \citep{2005Natur.438..765T}.

It is expected that the organics in Titan's atmosphere will eventually fall towards the surface of Titan. If they land on dry surfaces, species that are able to remain in the solid form will become surface sediments (see Figure~\ref{fig:photochem}). For example, Titan's equatorial regions are largely covered by dunes \citep{2006Sci...312..724L}, and the sand particles are likely derived from complex organic hazes falling from the atmosphere \citep{2007P&SS...55.2025S}. Simple organics in the solid form have been detected on Titan's surface, such as acetylene ice \citep{2016ApJ...828...55S}. At the same time, atmospheric species that remain in liquid forms, when they land on Titan's dry surface, can modify and wet the surface by rainfalls and storms \citep[e.g.,][]{2011Sci...331.1414T}. For organics that land on the surfaces of lakes and seas, the ones in liquid form would become part of the lake liquids, while the ones in solid form could either dissolve in the lake liquids, float on the lake surfaces, or sink and form lakebed sediments \citep{2019NatGe..12..315C, 2020ApJ...905...88Y}. After the lake dries, the dissolved organics could precipitate as solids and form evaporites on Titan's surface \citep{2011Icar..216..136B, 2014Icar..243..191M}.

Because of the diverse range of temperature and pressure regimes that can exist on Titan, from its atmosphere to the base of its lakes and seas, it is anticipated that many organic species are present in multiple phases (gas, liquid, and solid phases). Lots of these species actively participate in various processes such as cloud formation, lake interaction, and sediment transport. The unique material properties of each species are found to play an important role in shaping these different processes \citep[e.g.,][]{2003Icar..162...94B, 2006Icar..182..230B, 2017P&SS..137...20B, 2011Icar..215..732L, 2018SSRv..214..125A, 2018JGRE..123.2310Y, 2019NatGe..12..315C, 2020ApJ...905...88Y, 2020E&PSL.53015996Y, 2022GeoRL..4997913C, 2022GeoRL..4997605L, 2022NatAstro..57.2053F, 2022PSJ.....3....2L} and therefore represent critical input parameters to perform the required calculations to model these processes and prepare for the future in-situ explorations of Titan. For example, the NASA New Frontiers mission Dragonfly, due to arrive at Titan in 2034, will be capable of investigating some of the material properties of the substances it will encounter through in-situ measurements \citep{2021PSJ.....2..130B}. Various material properties are also found to be crucial for the design of the rotorcraft to ensure its success in operation \citep{2019P&SS..17904721B, 2019IEEE.53015996Y}. Thus, a comprehensive database to archive the material properties for all possible material candidates is timely and necessary to better model these processes and support the mission.

So far, there has been no consolidated database that is dedicated to summarizing the material properties of organics on Titan. For the simple organics, many of their material properties have been measured in the laboratory and can be found in published studies. However, these datasets have been scattered in the literature for over a century. Due to the spareness of the data and the expansiveness of the literature involved, modern research studies often assume or estimate values for the material properties like, e.g., the density of simple organics \citep[e.g.,][]{2008ApJ...686.1480B, 2021NatAs...5..289L}, even though there are density measurements for many of the organic ices. Moreover, many properties measured and published in the early 1900s do not have digitized copies, which makes the efforts to utilize these data even more difficult. We have developed a database that will serve as a repository for these published material properties of simple organics, and enable the use of more representative material properties in models. This database will also help identify the gaps between existing material property measurements and those needed for simple organics at Titan-relevant conditions, which will be crucial to guide future laboratory work.

For the complex organics, such as the atmospheric haze particles and the parent material for the organic sands on Titan's surface, no sample returned from Titan is available. Therefore, laboratory-made organic aerosol analogs (``tholins") have been used as the model material. The issue with the material properties of these tholins is different from that of the simple organics. Many laboratory facilities have been synthesizing tholins since the 1970s and have characterized them to different extents (see reviews by \cite{2012chem.....4.1375C} and \cite{2017JGRE..122..432H}, and references therein; \cite{2017Icar..289..214S, 2017ApJ...841L..31H, 2018ApJ...865..133S}). Even though most of the tholins are usually produced from the dominant gas constituents in Titan's atmosphere, N$_2$ and CH$_4$, a broad range of experimental conditions are used that can influence the material properties, including different energy sources, pressures, temperatures, gas exposure times, and gas flow rates. However, modern research studies often adopt the properties of tholins from one particular laboratory \citep[e.g.,][]{2015P&SS..105...65W, 2018Icar..305..314S}. A recent comparative study has demonstrated that tholin samples indeed have varying properties based on their experimental production conditions \citep{2022PSJ.....3....2L}. In view of this recent comparative study, there remain several open questions about 1) whether or which tholins are representative of the actual complex organics on Titan; and 2) whether we can use the properties of tholins to constrain processes involving complex organics. Given that there will be no direct sample from Titan in the near future and that the Dragonfly mission will arrive and take measurements on Titan's surface in the 2030s, we need to provide some material properties as baseline comparisons to the atmospheric hazes and the surface sediments on Titan. We think the best way to enable future science and prepare for Dragonfly is to archive all previously measured material properties of tholins along with their experimental production conditions. This will help us identify whether or not and which laboratory-produced tholins mimic the actual organic hazes/sediments on Titan. Before we find the best analog of Titan aerosols/sediments, it will be important for future modeling studies to include the uncertainty of material properties of tholins produced under different experimental conditions.

In this work, we summarize several important material properties of the simple and complex organics relevant to Titan's exploration that we have aggregated from the literature or newly computed. For the simple organics, we focus on molecules that have already been detected in Titan's stratosphere in the gas phase and may experience phase change on Titan. We also choose to include dicyanoacetylene (C$_4$N$_2$), which has been detected in the solid phase \citep{1987AcSpA..43..421K, 2016GeoRL..43.3088A}, but only an upper limit was established for its gas phase \citep{1997P&SS...45..941S, 2015Icar..248..340J}. Some organic species are predicted by photochemical models to form in Titan's atmosphere, but they have either not been detected (e.g., 1,3-butadiene, \cite{2019Icar..324..120V, 2016ApJ...829...79W}) or only upper limits of their abundances are established (e.g., butane, \cite{2022PSJ.....3...59S}, pyridine and pyrimidine, \cite{2020AJ....160..205N}). In this work, we choose not to include these molecules. Table~\ref{table:abundance} lists the molecules included in this work, their observed abundance ranges, and the associated references. 

For the complex organics, we include existing material property measurements for tholins, their experimental production conditions, and measuring techniques. From here on, to differentiate between the simple organics in various phases and the complex refractory materials, analogs of solid haze particles, we will call the simple organics in the liquid and solid phases, ``organic liquids" and ``organic ices," respectively, and the complex refractory materials as ``organic solids." The material properties investigated in this study include 1) thermodynamic properties (phase change pressure and temperature, saturation vapor pressure for organic liquids and ices, latent heat for vaporization for organic liquids, and latent heat for sublimation for organic ices); 2) physical property (density of organic liquids, ices, and solids), and 3) surface properties (liquid surface tensions for organic liquids and solid surface energies for organic ices and solids). These datasets can be used by the scientific community for modeling, data analysis, comparison to and interpretation of spacecraft observations and in-situ measurements, and preparation/development of future instrumentation.

\begin{table}[h]
\centering
\caption{The 18 detected simple organic compounds included in this work and their stratospheric mixing ratios. The mixing ratios of gaseous organic compounds fluctuate seasonally in Titan's atmosphere \citep[e.g.,][]{2019GeoRL..46.3079T}. Here we adopt the largest possible range of mixing ratios encompassing past Cassini and ground-based observations. References: [1] \citet{2014Icar..231..323L}; [2] \citet{2010JGRE..11512006N}; [3] \citet{2019GeoRL..46.3079T}; [4] \citet{2019ApJ...881L..33L}; [5] \citet{2013ApJ...776L..14N}; [6] \citet{2016Icar..270..409C}; [7] \citet{2020Icar..34413413C}; [8] \citet{2019Icar..319..417T}; [9] \citet{2020AJ....160..178S}; [10] \citet{2017SciA....3E0022P}; [11] \citet{2015ApJ...800L..14C}; [12] \cite{2015Icar..248..340J}.}
    \begin{tabular}{c|c|c|c}
    \toprule
    Compound name & CAS number & Mixing ratio range & Reference  \\
    \hline 
    Methane (CH$_4$) & 74-82-8 & $1\times10^{-2}$ -- $6 \times10^{-2}$ & [1,2] \\
    \hline 
    Ethane (C$_2$H$_6$) & 74-84-0 & 7 $\times$ 10$^{-6}$ -- 3 $\times$ 10$^{-5}$ & [3] \\
    \hline 
    Ethylene (C$_2$H$_4$) & 74-85-1 & 7 $\times$ 10$^{-8}$ -- 6 $\times$ 10$^{-6}$ & [3]\\
    \hline 
    Acetylene (C$_2$H$_2$) & 74-86-2 & $2\times10^{-6}$ --  $1\times10^{-5}$ & [3]  \\
    \hline 
    Diacetylene (C$_4$H$_2$) & 460-12-8 & 9 $\times$ 10$^{-10}$ -- 1 $\times$ 10$^{-7}$ & [3] \\
    \hline 
    Benzene (C$_6$H$_6$) & 71-43-2 & 6 $\times$ 10$^{-11}$ -- 2 $\times$ 10$^{-7}$ & [6,7]  \\
    \hline 
    Propane (C$_3$H$_8$) & 74-98-6 & 2 $\times$ 10$^{-7}$ -- 3 $\times$ 10$^{-6}$ & [6,7] \\
    \hline 
    Propene (C$_3$H$_6$) & 115-07-1 & 1 $\times$ 10$^{-9}$ --  6 $\times$ 10$^{-9}$ & [5] \\
    \hline 
    Allene (C$_3$H$_4$-a) & 463-49-0 & 6.9 $\pm$ 0.8 $\times$ 10$^{-10}$ & [4] \\
    \hline 
    Propyne (C$_3$H$_4$-p) & 74-99-7 & 5 $\times$ 10$^{-9}$ -- 1 $\times$ 10$^{-7}$ & [3]\\
    \hline 
    Hydrogen cyanide (HCN) & 74-90-8 & 4 $\times$ 10$^{-8}$ -- 3 $\times$ 10$^{-4}$ & [6] \\
    \hline 
    Cyanoacetylene (HC$_3$N) & 68010-34-4 & 6 $\times$ 10$^{-11}$ -- 9 $\times$ 10$^{-7}$ & [3,7] \\
    \hline 
    Carbon dioxide (CO$_2$) & 124-38-9 & 1 $\times$ 10$^{-8}$ -- 3 $\times$ 10$^{-8}$ & [3] \\
    \hline 
    Acetonitrile (CH$_3$CN) & 75-05-8 & 4 $\times$ 10$^{-10}$ -- 1 $\times$ 10$^{-8}$ & [8] \\
    \hline 
    Propionitrile (C$_2$H$_5$CN) & 107-12-0 & 8 $\times$ 10$^{-10}$ -- 3 $\times$ 10$^{-9}$ & [11] \\
    \hline 
    Acrylonitrile (C$_2$H$_3$CN) & 107-13-1 & 2 $\times$ 10$^{-10}$ -- 8 $\times$ 10$^{-10}$ & [10] \\
    \hline 
    Cyanogen (C$_2$N$_2$) & 460-19-5 & 5 $\times$ 10$^{-11}$ -- 7 $\times$ 10$^{-9}$ & [9] \\
    \hline  
    Dicyanoacetylene (C$_4$N$_2$) & 1071-98-3 & $<$ 5 $\times$ 10$^{-10}$ & [12]   \\
    \toprule
    \end{tabular}
    \label{table:abundance}
\end{table}

Recent experimental works have shown that a new class of minerals, the co-crystals, could form by co-condensing two or more molecules in Titan's atmosphere or by co-precipitation on Titan's surface \citep[see a review of][]{cable2021titan}. While the co-crystals could also be important components of Titan, the associated laboratory studies are still in their infancy. Thus, in this work, we only include the material properties of pure substances. In the following sections, we describe the efforts associated with archiving and calculating each material property.

\section{Thermodynamic properties}
In the subsequent sections, we review and compute several important thermodynamic properties of organic liquids and ices for the 18 species detected on Titan. In Section~\ref{Phase_change}, we summarize several phase change points of these substances. These phase change points allow us to determine what the physical state of these substances would be at a given pressure and temperature on Titan. In Section~\ref{SVP}, we summarize and compute the sublimation saturation vapor pressure (s-g SVP) of the organic solids and the vaporization saturation vapor pressure (l-g SVP) of the organic liquids as a function of temperature, for each of the 18 organic species detected on Titan. The SVP can help determine the transformation of substances, e.g., when do a vapor reach saturation and form clouds (or when do a species transform from the gaseous state to the liquid or the solid state)? What is the humidity of a certain substance given the temperature and pressure of the location? Lastly, in Section~\ref{Latent_heat}, we compute the sublimation latent heats of the organic ices and the vaporization latent heats of the organic liquids for each of the 18 species at their triple point temperatures. The latent heats of these substances can not only help us understand the energy release/uptake during the phase transformation, but they are also useful in predicting other material properties of these substances (e.g., the surface energies of organic ices in Section~\ref{ice_SE}).

\subsection{Phase change points of organics liquids and ices for species detected on Titan}\label{Phase_change}

In Table~\ref{table:temps}, we summarize several phase change temperatures and pressures for the 18 simple organics listed in Table~\ref{table:abundance}: the triple point temperature and pressure ($T_{tri}$ and $P_{tri}$), the critical point temperature and pressure ($T_{cri}$ and $P_{cri}$), and the crystalline ice phase transition temperature ($T_{tran}$) for applicable species. The triple point of a substance is defined as the temperature ($T_{tri}$) and pressure ($P_{tri}$) where the solid, liquid, and gas (vapor) phases can coexist, and it is also the intersection of the sublimation, fusion, and vaporization curves. The critical point of a substance is the temperature ($T_{cri}$) and pressure ($P_{cri}$) where the boundary between two phases vanishes. All the critical points in Table~\ref{table:temps} represent the liquid-gas critical points, and beyond which a substance can be both in the liquid and the gas phase, or the so-called supercritical liquid. Note that on Titan, none of the 18 detected substances can reach their critical point. For organic ices, we only consider phase changes in the crystalline phase, when there is a sudden change in a crystal lattice structure. These phase transition temperatures are also summarized in Table~\ref{table:temps} for selected crystal transitions. We do not take into consideration the amorphous-crystalline phase transitions in this table. For the known amorphous-crystalline transition information of the ices, see Table~\ref{table:ice_density_data_2} in the Appendix section. Note that in Table~\ref{table:temps}, except for the few data points that are direct experimentally determined \citep{1963JChPh..38...69D,1965JChPh..42..749P,1962JChPh..36..829W,1972chm..thermo..359F}, most of the phase transition temperatures are directly pulled from the published NIST/TRC Web Thermo Tables (WTT) database \citep{2022nist.292..686K}, which are ``determined using an average of a series of accepted experimental and predicted values, weighted by uncertainties." 

\begin{table}[h]
\centering
\addtolength{\leftskip} {-2cm}
\caption{Triple point temperature ($T_{tri}$) and pressure ($P_{tri}$), critical temperature ($T_{cri}$) and pressure ($P_{cri}$), and crystalline phase transition temperature ($T_{tran}$) for the 18 species detected on Titan include in this work (Table~\ref{table:abundance}). Most of the triple point temperature, all of the critical point temperature and pressure, and most of the crystalline phase transition temperature data are from \cite{2022nist.292..686K}, which are averages of a series of accepted experimental and predicted values. The triple point temperature of HC$_3$N is estimated using experimental data from \cite{1963JChPh..38...69D}. The triple point pressures are calculated using saturation vapor pressure functions in Table~\ref{table:SVP}. The crystalline phase transition temperatures of CH$_3$CN, C$_2$H$_5$CN, and C$_2$H$_3$CN are experimental data from \cite{1965JChPh..42..749P}, \cite{1962JChPh..36..829W}, and \cite{1972chm..thermo..359F}.}
\resizebox{7.9in}{!}{
    \begin{tabular}{c|c|c|c|c|c|c}
    \toprule
    Species & $T_{tri}$ (K) & $P_{tri}$ (bar) & $T_{cri}$ (K) & $P_{cri}$ (bar) & $T_{tran}$ (K) & Phase transition note \\
    \hline 
    CH$_4$ & 90.686$\pm$0.010 & 0.117$\pm$0.001 & 190.564$\pm$0.010 & 45.992$\pm$0.073 & 20.509$\pm$0.047 & crystal II-crystal I\\
    \hline 
    \multirow{2}{*}{C$_2$H$_6$} & \multirow{2}{*}{90.3560 $\pm$ 0.0050} & \multirow{2}{*}{$(1.1\pm0.1)\times10^{-5}$} & \multirow{2}{*}{305.322$\pm$0.035} & \multirow{2}{*}{48.72$\pm$0.23} & 89.816$\pm$0.027 & crystal II-crystal I\\
     &  &  &  &  & 89.726$\pm$0.050 & crystal III-crystal II\\
    \hline 
    C$_2$H$_4$ & 103.966$\pm$0.016 & $(1.2\pm0.1)\times10^{-3}$ & 282.350$\pm$0.025 & 50.417$\pm$0.068 & N/A & N/A\\
    \hline 
    C$_2$H$_2$ & 191.66$\pm$0.94 & 1.2$\pm$0.1 & 308.39$\pm$0.35 & 62.06$\pm$0.67 & 133.0$\pm$1.0 & crystal II-crystal I\\
    \hline 
    C$_4$H$_2$ & 237.7$\pm$2.5 & $0.13\pm0.01$ & 410$\pm$16 & 26.0$\pm$10.0 & N/A & N/A \\
    \hline 
    C$_6$H$_6$ & 278.6877$\pm$0.0023 & $(4.7\pm0.1)\times10^{-2}$ & 562.02$\pm$0.13 & 49.06$\pm$0.20 & N/A & N/A\\
    \hline 
    C$_3$H$_8$ & 85.400$\pm$0.054 & $(1.8\pm0.2)\times10^{-9}$ & 369.89$\pm$0.14 & 42.51$\pm$0.21 &80.40$\pm$0.10 & metastable crystal-crystal I  \\
    \hline 
    C$_3$H$_6$ & 87.850$\pm$0.020 & $(8\pm1)\times10^{-9}$ & 364.21$\pm$0.26 & 45.55$\pm$0.76 & N/A & N/A \\
    \hline 
    C$_3$H$_4$-a & 136.70$\pm$0.53 & $(1.8\pm0.1)\times10^{-4}$ & 393.9$\pm$4.0 & 62.80$\pm$4.40 & N/A & N/A\\
    \hline 
    C$_3$H$_4$-p & 169.87$\pm$0.77 & $\sim3.8\times10^{-3}$ & 402.7$\pm$1.5 & 56.58$\pm$0.81 & N/A & N/A \\
    \hline 
    HCN & 259.871$\pm$0.091 & $0.1865\pm0.0015$ & 456.65$\pm$0.50 & 49.70$\pm$3.90 & 170.40$\pm$0.50 & crystal II-crystal I \\
    \hline 
    HC$_3$N & 280$\pm$1 & $\sim0.25$ & 527$\pm$11 & 52.00$\pm$27.00 & N/A & N/A \\
    \hline 
    CO$_2$ & 216.5890$\pm$0.0077 & $5.18\pm0.01$ & 304.128$\pm$0.038 & 73.77$\pm$0.20 & N/A & N/A \\
    \hline 
    CH$_3$CN & 229.349$\pm$0.053 & $\sim1.9\times10^{-3}$ & 545.41$\pm$0.40 & 48.66$\pm$0.27 & 216.9$\pm$0.1 & crystal II-crystal I\\
    \hline 
    C$_2$H$_5$CN & 180.4$\pm$1.2 & $\sim1.7\times10^{-6}$ & 561.26$\pm$0.20 & 42.60$\pm$0.89 & 176.964$\pm$0.005 & crystal II-crystal I   \\
    \hline 
    C$_2$H$_3$CN & 189.642$\pm$0.050 & $(3.5\pm0.5)\times10^{-5}$ & 540.0$\pm$2.0 & 46.50$\pm$2.80 & 162.50$\pm$0.10 & crystal II-crystal I \\
    \hline 
    C$_2$N$_2$ & 245.29$\pm$0.10 & $0.77\pm0.03$ & 397.1$\pm$3.0 & 62.50$\pm$4.00 & N/A & N/A\\
   \hline  
    C$_4$N$_2$ & 293.0$\pm$1.3 & $1.7\times10^{-1}$ & 496$\pm$22 & 66.00$\pm$55.00 & N/A & N/A \\
    \toprule
    \end{tabular}
    \label{table:temps}}
\end{table}

\subsection{Saturation vapor pressures of organic liquids and ices for species detected on Titan}\label{SVP}

In Table~\ref{table:SVP}, we summarize the sublimation and vaporization saturation vapor pressures (SVPs) as functions of temperature for the gas-phase species observed in Titan's atmosphere (Table~\ref{table:abundance}). For consistency with previous work \citep{2009P&SS...57.2053F}, all the saturation vapor pressures ($P_{sat}$) are listed in units of bar, and the temperature in K. Most of the applicable temperature range starts (for vaporization equilibrium, or l-g) or ends (for sublimation equilibrium, or s-g) with the material's triple point (see also Table~\ref{table:temps}). For the vaporization equilibrium (l-g), most of the applicable temperature range ends with the material's critical point (see also Table~\ref{table:temps}). For each saturation vapor pressure function, we define the applicable temperature range as such: temperature values without brackets have existing laboratory measurement data, and temperature values with brackets apply for extrapolated SVPs beyond the laboratory-measured range. The idea is that users should be cautious when using extrapolated SVPs in the bracketed temperature range, as actual SVPs may significantly deviate from the extrapolated values. We use such notation hoping to provide directions for future laboratory efforts to fill in the gaps of the existing SVP measurements. Specifically, there are no reliable sublimation vapor pressure measurements for C$_3$H$_4$-p, C$_3$H$_6$, C$_3$H$_8$, CH$_3$CN, C$_2$H$_5$CN, and C$_2$H$_3$CN ices, and more laboratory work is needed to directly measure the sublimation vapor pressure of these ices at low temperatures. We attempt to establish sublimation vapor pressure functions for these species using thermodynamic and empirical approaches.

In this work, a total of six empirical mathematical functions are used to express SVPs (in bar) as a function of temperature (in K):
\begin{itemize}
    \item The Antoine equation (\citealp{1888ant}, \citealp[used in][]{Yaws2015}), where $A$, $B$, and $C$ are empirically fitted for each compound:
    \begin{equation}
        P_{sat}(T) = 10^{A-B/(T+C)}
        \label{eq:antoine}
    \end{equation}
    \item The extended Antoine equation \citep{1999book...2..121D}, where $A$, $B$, $C$, $n$, $E$, and $F$ are empirically fitted for each compound, $T_0$ is the lower bound of the temperature, and $T_{cri}$ is the critical temperature:
    \begin{equation}
        P_{sat}(T) = 10^{A-B/(T+C)+0.43429[(T-T_0)/T_{cri}]^n+E[(T-T_0)/T_{cri}]^8+F[(T-T_0)/T_{cri}]^{12}}
    \end{equation}
    \item The polynomial equation  \citep[used in][]{1980brownziegler, 2009P&SS...57.2053F}, where $A_i$s ($i=0, 1, ..., n$) are empirically fitted for each compound:
    \begin{equation}
        P_{sat}(T) = e^{A_0+\sum_{i=1}^{n}A_i/T^i}.
    \end{equation}
    \item The Kirchoff-Rankine equation \citep{1858AnP...180..612K, 1866Ran, 2022nist.292..686K}, where $A$, $B$, and $C$ are empirically fitted for each compound:
    \begin{equation}
        P_{sat}(T) = e^{A+B/T+C\mathrm{ln}T}.
    \end{equation}
    \item The Wagner equation (\citealp{1973Cryo...13..470W}, \citealp[used in][]{2022nist.292..686K}), where $A_i$s ($i=1, 2, 3, 4$) are empirically fitted for each compound, and $P_{cri}$ is the critical pressure:
    \begin{equation}
        P_{sat}(T) = P_{cri}e^{T_{cri}/T[A_1(1-T/T_{cri})+A_2(1-T/T_{cri})^{1.5}+A_3(1-T/T_{cri})^{2.5}+A_4(1-T/T_{cri})^5]}.
    \end{equation}
    \item The extended polynomial equation used in this work, where $A_i$s ($i=0, 1, 2, ..., 7$) are empirically fitted for each compound:
    \begin{equation}
        P_{sat}(T) = e^{A_0 + A_1/T+A_2\mathrm{ln}(T)+A_3T+A_4T^{1.5}+A_5T^2+A_6T^3+A_7T^4}.
    \end{equation}
\end{itemize}

Most vapor pressure expressions are directly extracted from literature reviews or existing chemical databases \citep{1980brownziegler,1999book...2..121D, 2009P&SS...57.2053F,2022nist.292..686K, Yaws2015}. Note that we only select to present one set of vapor pressure functions for each phase change of each species. Among existing SVP expressions for a given species, we typically select the one that has the largest applicable temperature range or that best fits the experimental measurements in a broad range of temperatures (see Figures~\ref{fig:s-g_comparsion1}-\ref{fig:l-g_comparsion2} in the Appendix section). For example, for the sublimation SVP of C$_6$H$_6$ ice, we choose to present the expression by \cite{1999book...2..121D}, not the others \citep{2009P&SS...57.2053F, 2022nist.292..686K, 2014Thermo..581.1399R, 2021PSJ.....2..121D, 2022PSJ.....3..120H}, as \cite{1999book...2..121D}'s expression best fits the most recent low-temperature measurements of \cite{2022PSJ.....3..120H}, the adjusted low-temperature measurements of \cite{2021PSJ.....2..121D} ($130<T<160$ K), and previous high-temperature measurements ($T>150$ K). We also rated the approximate order of accuracy of the adopted SVP functions in Table~\ref{table:SVP}. The SVP expressions with ratings of A$^{+}$, A, and B typically have good agreements with laboratory-measured data ($<$10 \%) within the temperature range without brackets. The accuracy of SVP expressions with ratings of C and D are lower ($>$10 \%), due to the larger spread and/or the large uncertainties of the existing laboratory data. More laboratory measurements are needed to help improve the accuracy of these SVPs. 

\begin{table}[h]
    \addtolength{\leftskip} {-1.7cm}
\caption{Adopted saturation vapor pressure expressions ($P_{sat}$) as functions of temperature ($T$) for species in Table~\ref{table:abundance}, s-g represents solid-gas transitions and l-g represents liquid-gas transitions. The ``Temp. (K)" column indicates the temperature range for which the SVP functions can be applied. Note that temperature ranges with brackets correspond to temperatures for which only extrapolated SVPs are available, and temperatures without brackets have experimentally determined SVPs in the temperature range. References: [a] \cite{1999book...2..121D}; [b] \cite{Yaws2015}; [c] \cite{2009P&SS...57.2053F};  [d] \cite{2022nist.292..686K}; [e] \cite{2014Icar..243..494O};  [f] this work; [g] \cite{1963JChPh..38...69D}; [h] \cite{1990chemthermo...2..121D}. The ``Accuracy" column gives the accuracy rating of the fittings within the temperature range where experimental data exist in the form of a rating. The ratings consist of five letters. Each letter corresponds to the rough order of accuracy of the fittings to the experimental data: A$^{+}$ (0.1-1 \%), A (1-5 \%), B (5-10 \%), C (10-50 \%), D ($>$50 \%).} 
\resizebox{8in}{!}{
\begin{threeparttable}
    \begin{tabular}{cccccc}
    \toprule
    Species & Type & Saturation vapor pressure expressions $P_{sat}(T)$ (bar) & Temp. (K) & Ref. & Accuracy\\
    \hline 
    CH$_4$ & s-g & \(10^{4.31972 - 451.64/(T-4.66)}\) &
    [20.509]-48-90.686 ($T_{tri}$) & [a] & 48-70 K, C\\
     &  & & & & 70 K-$T_{tri}$, A\\
    & l-g & \(10^{3.96867-435.453/(T-1.789)}\) & 90.686-190.564 ($T_{cri}$) & [b] & A\\
    \hline 
    C$_2$H$_6$ & s-g & \(10^{6.6388-1009.6/(T-3.15)}\) & [$<$]71-90.3560 ($T_{tri}$) & [a] & C\\
     & l-g & \(10^{4.0768-698.93/(T-12.886)}\) & 90.3560-305.322 ($T_{cri}$) & [b] & B \\
    \hline 
    C$_2$H$_4$ & s-g & \(e^{15.4 - 2206/T -1.216\times10^4/T^2 + 2.843\times10^5/T^3 -2.203\times10^6/T^4}\) & [$<$]77.3-103.966 ($T_{tri}$) & [c] & C\\
    & l-g & \(10^{3.82898 - 581.901/(T-17.787)}\) & 103.966-122 & [a] & C\\
    & l-g & \(10^{3.91382 - 596.526/(T-16.78)}\) & 122-174 & [a] & B\\
    & l-g & \(10^{3.91382 - 596.526/(T-16.78)+0.43429[(T-174)/T_{cri}]^{2.79132}+9.717[(T-174)/T_{cri}]^8+52.77[(T-174)/T_{cri}]^{12}}\) & 174-282.350 ($T_{cri}$) & [a] & A\\
    \hline 
    C$_2$H$_2$ & s-g & \(e^{13.4-2536/T}\) & [$<$]98.55-191.66 ($T_{tri}$) & [c] & B\\
    & l-g & \(62.06e^{T_{cri}/T[-7.5718(1-T/T_{cri})+3.1025(1-T/T_{cri})^{1.5}-3.4594(1-T/T_{cri})^{2.5}+0.32221(1-T/T_{cri})^5]}\) & 191.66-308.39 ($T_{cri}$) & [d] & A\\
    \hline 
    C$_4$H$_2$ & s-g & \(10^{7.3757 - 1958.25/T}\) & [$<$]127-237.7 ($T_{tri}$) & [e] & D \\
    & l-g & \(e^{10.600 - 3007.6/T}\) & 237.7-410 ($T_{cri}$) & [d] & C\\
    \hline 
    C$_6$H$_6$ & s-g & \(10^{6.72391 - 2107.85/(T-16.45)}\) & [$<$]134.8-278.6877 ($T_{tri}$) & [a] & B\\
    & l-g & \(10^{3.98523 - 1184.236/(T-55.623)}\) & 278.6877-376 & [a] & C \\
    & l-g & \(10^{3.98523 - 1184.236/(T-55.623)+0.43429[(T-376)/T_{cri}]^{2.3835}+12.283[(T-376)/T_{cri}]^8+664.01[(T-376)/T_{cri}]^{12}}\) & 376-562.02 ($T_{cri}$) & [a] & B \\
    \hline 
    C$_3$H$_8$ & s-g & \(e^{17.802 - 2985/T - 3.423\times10^4/T^2 + 1.163\times10^6/T^3 - 1.488\times10^7/T^4}\) & [50.0]-[85.400] ($T_{tri}$) & [f] & N/A\\
    & l-g & \(10^{4.1451-889.864/(T-16.066)}\) & 85.400-369.89 ($T_{cri}$) & [b] & B\\
    \hline 
    C$_3$H$_6$ & s-g & \(e^{16.425 - 2691/T - 5.291\times10^4/T^2 + 1.858\times10^6/T^3 - 2.481\times10^7/T^4}\) & [50.0]-[87.850] ($T_{tri}$) & [f] & N/A\\
    & l-g & \(10^{4.1322-859.722/(T-17.255)}\) & 87.850-364.21 ($T_{cri}$) & [b] & B\\
    \hline 
    C$_3$H$_4$-a & s-g & \(10^{6.7534 - 1434.94/T}\) & [$<$]106-136.70 ($T_{tri}$) & [a] & B\\
    & l-g & \(62.80e^{T_{cri}/T[-8.4862(1-T/T_{cri})+5.4394(1-T/T_{cri})^{1.5}-5.3149(1-T/T_{cri})^{2.5}+0.70544(1-T/T_{cri})^5]}\) & 136.70-393.9 ($T_{cri}$) & [d] & C\\
    \hline 
    C$_3$H$_4$-p & s-g & \(e^{16.795-3799.6/T}\) & [$<$169.87] ($T_{tri}$) & [f] & N/A\\
    & l-g & \(10^{4.2817-923.311/(T-32.071)}\) & 169.87-402.7 & [b] & B\\
    \hline 
    HCN & s-g & \(e^{54.889-5969.5/T-6.0411\mathrm{ln}T}\) & [$<$]194.565-259.871 ($T_{tri}$) & [d] & A\\
    & l-g & \(10^{4.8607 - 1432.7/(T-3.7358)}\) & 259.871-456.65 ($T_{cri}$) & [f] & A\\
    \hline 
    HC$_3$N & s-g & \(10^{9.0744-3046.5/(T+36.176)}\) & [$<$]164.9-280 ($T_{tri}$) & [f] & B\\
     & l-g & \(10^{4.6649 - 1470/T}\) & 280-315-(527) ($T_{cri}$) & [g] & C\\
    \hline 
    CO$_2$ & s-g & \(e^{20.0456-3274.96/T-0.6102\mathrm{ln}(T)}\) & [$<$]69.697-216.5890 ($T_{tri}$) & [d] & B\\
     & l-g & \(73.77e^{T_{cri}/T[-7.0206(1-T/T_{cri})+1.5072(1-T/T_{cri})^{1.5}-2.1863(1-T/T_{cri})^{2.5}-3.0645(1-T/T_{cri})^5]}\) & 216.5890-304.128 ($T_{cri}$) & [h] & A$^{+}$ \\
    \hline 
    CH$_3$CN & s-g & \(e^{21.542-5853.5/T-6.9785\times10^{-3}T-1.9407\times10^{-4}T^{1.5}}\) & [20]-[216.9] & [f] & N/A\\
     & s-g & \(e^{23.405-5655.0/T-0.72971\mathrm{ln}(T)-1.6902\times10^{-3}T-1.9407\times10^{-4}T^{1.5}}\) & [216.9]-[229.349] ($T_{tri}$) & [f] & N/A\\
     & l-g & \(10^{4.6691-1583.4/(T-15.263)}\) & 229.349-545.41 ($T_{cri}$) & [b] & B\\
    \hline 
    C$_2$H$_5$CN & s-g & \(e^{25.869-6428.6/T-2.1410\times10^{-2}T-2.0625\times10^{-4}T^{1.5}+2.7238\times10^{-5}T^2}\) & [15]-[176.964] & [f] & N/A\\
    & s-g & \(e^{-366.46+436.73/T+75.442\mathrm{ln}(T)-0.22541T-2.0625\times10^{-4}T^1.5}\) & [176.964]-[180.4] $T_{tri}$) & [f] & N/A\\
     & l-g & \(42.60e^{T_{cri}/T[-8.1477(1-T/T_{cri})+2.4460(1-T/T_{cri})^{1.5}-2.5323(1-T/T_{cri})^{2.5}-2.4144(1-T/T_{cri})^5]}\) & 180.4-561.26 ($T_{cri}$) & [d]  & B\\
    \hline 
    C$_2$H$_3$CN & s-g & \(e^{23.485-5879.0/T-1.1719\times10^{-2}T-1.0523\times10^{-5}T^2+5.0429\times10^{-9}T^3-1.1612\times10^{-12}T^4}\) & [12.5]-[162.5] & [f] & N/A\\
    & s-g & \(e^{20.897-5583.8/T-4.1416\times10^{-2}\mathrm{ln}(T)-5.6792\times10^{-3}T-1.0523\times10^{-5}T^2+5.0429\times10^{-9}T^3-1.1612\times10^{-12}T^4}\) & [162.5]-[189.642] ($T_{tri}$) & [f] & N/A\\
    & l-g & \(10^{4.5332-1477.29/(T-24.71)}\) & 189.642-540.0 ($T_{cri}$) & [b] & B\\
    \hline 
    C$_2$N$_2$ & s-g & \(e^{16.53 - 4109/T}\) & [$<$]156.3-245.29 ($T_{tri}$) & [c] & B\\
     & l-g & \(62.50e^{T_{cri}/T[-7.0700(1-T/T_{cri})+0.019087(1-T/T_{cri})^{1.5}-0.13088(1-T/T_{cri})^{2.5}-3.7467(1-T/T_{cri})^5]}\) & 245.29-397.1 ($T_{cri}$) & [d] & B\\
     \hline  
    C$_4$N$_2$ & s-g & \(e^{19.09-6036/T}\) & [125.0]-147.0-273.2-[293.0] ($T_{tri}$) & [c] & D\\
     & l-g & \(10^{3.6326 - 1156.9/(T - 30.398)}\) & [293.0]-296.0-349.7-[496] ($T_{cri}$) & [f] & A$^{+}$ \\
    \toprule
    \end{tabular}
    \label{table:SVP}
    \end{threeparttable}}
\end{table}

\clearpage

Note that extra caution is always needed when using any SVP expressions beyond the temperature range for which measurements are available, as those are extrapolations to the existing data to estimate the behavior of the material at lower temperatures. By comparing the existing SVP functions in Table~\ref{table:SVP} and in the Appendix Table~\ref{table:SVP_app}, we found that the different SVP functions typically agree well with each other within the temperature range with laboratory measurements and start to deviate (up to orders of magnitude) towards lower temperatures, where no or few SVP measurements have been made. \citet{2019EPSC...13.1276B,Barth2021} also summarizes a comprehensive sublimation SVP and latent heat database that compares adopted vapor pressure functions from various literature sources, focusing on a range of general hydrocarbons and nitriles in the outer solar system; while the work presented here focuses on all the already detected gas-phase species in Titan's atmosphere.

In this work, we also introduce new SVP functions for C$_3$H$_8$ (s-g), C$_3$H$_6$ (s-g), C$_3$H$_4$-p (s-g), HCN (l-g), HC$_3$N (s-g), CH$_3$CN (s-g), C$_2$H$_5$CN (s-g), C$_2$H$_3$CN (s-g), and C$_4$N$_2$ (l-g). Some of these new SVP functions are fitted using more comprehensive experimental data (HCN l-g, HC$_3$N s-g, and C$_4$N$_2$ l-g) using the Antoine equation~\ref{eq:antoine}, and some are theoretically calculated using thermodynamic and empirical approaches as no direct SVP measurements at cryogenic temperatures are available in the literature (C$_3$H$_8$ s-g, C$_3$H$_6$ s-g, C$_3$H$_4$-p s-g, CH$_3$CN s-g, C$_2$H$_3$CN s-g, and C$_2$H$_3$CN s-g). 

\begin{itemize}
\item For HCN, we least-square fit its l-g SVP expression using data from \cite{1925breig...31..449, 1926xxx..48..299c, 1926jap..1.59c}.

\item For HC$_3$N, we least-square fit its s-g SVP expression using data from \cite{1963JChPh..38...69D,1994JGR....9917069B,2020chcp.book.....L}.

\item For C$_4$N$_2$, we least-square fit its l-g SVP expression using data from \cite{1957OrChem...59..133S}.

\item For C$_3$H$_6$ and C$_3$H$_8$, \cite{1978advance..xxx} theoretically computed the sublimation vapor pressure and latent heat from 50 K up to the triple points of the two species using the thermodynamic approach. This allows us to derive the s-g SVP functions for C$_3$H$_6$ and C$_3$H$_8$, by least-square fitting the computed data using fourth-degree polynomials.

\item  For CH$_3$CN, C$_2$H$_5$CN, and C$_2$H$_3$CN, we are able to derive their sublimation vapor pressure as a function of temperature using the thermodynamic approach presented in \cite{2009P&SS...57.2053F}. The phase equilibrium for a pure substance is given by the Clausius-Clapeyron equation:

\begin{equation}
    \frac{dP_{sub}(T)}{dT} = \frac{L_{sub}(T)}{T(V_v-V_s)},
\label{eq:cc0}
\end{equation}
 where $P_{sub}(T)$ is the sublimation saturation vapor pressure of a species, $L_{sub}(T)$ is the molar sublimation latent heat or enthalpy as a function of temperature (with a unit of J$\cdot$mol$^{-1}$). The latent heat is the heat absorbed when a pure substance transfers from the solid to the vapor phase at constant pressure. $V_v$ and $V_s$ are the molar volumes of the vapor and the solid phase. The volume of the condensed phase is typically much less significant compared to the volume of the vapor phase, thus, we may assume $V_v\gg V_s$. If we also assume the gas follows perfect gas behavior, the volume of the vapor phase can be expressed using the ideal gas law ($V_v=RT/P$), where $R$ is the universal gas constant (8.314 J$\cdot$K$^{-1}$ mol$^{-1}$). The Clausius-Clapeyron equation can be approximated as:
\begin{equation}
    \frac{dP_{sub}(T)}{P_{sub}} =  \frac{L_{sub}(T)dT}{RT^2},
\label{eq:cc}
\end{equation}
This allows us to integrate Equation~\ref{eq:cc} to obtain the sublimation vapor pressure of a species:
\begin{equation}
    \mathrm{ln}(P_{sub}(T)) = \int_{T_0}^{T}\frac{L_{sub}(T')}{RT'^2}dT' + \mathrm{ln}(P_{sub}(T_0)).
\label{eq:cc1}
\end{equation}
If we know the sublimation vapor pressure at a certain temperature $T_0$ and the sublimation latent heat as a function of temperature $L_{sub}(T)$, we can compute the sublimation pressure of a species as a function of temperature. For CH$_3$CN, C$_2$H$_5$CN, and C$_2$H$_3$CN, we choose the triple point temperature and pressure calculated using the vaporization vapor pressure function, then we need to determine their latent heat of sublimation as a function of temperature, which can be written as:
\begin{equation}
    L_{sub}(T)=H_v(T)-H_s(T),
\label{eq:latent_heat1}
\end{equation}
where $H_v(T)$ and $H_s(T)$ are the enthalpies of the species in the gas and solid phase. For an ideal gas, $H_v(T)=\int_{T_0}^{T}C_{p,v}(T')dT'+H_v(T_0)$, and for the solid $H_s(T)=\int_{T_0}^{T}C_{p,s}(T')dT'+H_s(T_0)$, where $C_{p,v}(T)$ and $C_{p,s}(T)$ are the molar heat capacities of the gas and the solid (with a unit of J$\cdot$K$^{-1}$ mol$^{-1}$). Thus we can rewrite the latent heat as:
\begin{equation}
    L_{sub}(T)=\int_{T_0}^{T}(C_{p,v}(T')-C_{p,s}(T'))dT'+L_{sub}(T_0).
\label{eq:latent_heat2}
\end{equation}
For CH$_3$CN, C$_2$H$_5$CN, and C$_2$H$_3$CN, we are able to find their solid molar heat capacity data ($C_p,s(T)$) measured at cryogenic temperatures below the triple point and the computed gas molar heat capacity data ($C_p,v(T)$). These data are least-square fitted to facilitate integration in Equation~\ref{eq:latent_heat2}. We also need the latent heat of sublimation at a certain temperature in Equation~\ref{eq:latent_heat2}. In this case, we choose the triple point of CH$_3$CN, C$_2$H$_5$CN, and C$_2$H$_3$CN. Because the latent heat of sublimation is the summation of the latent heat of fusion and vaporization, $L_{sub}=L_{fus}+L_{vap}$, we can use the latent heat of fusion measured at the triple point $L_{fus}(T_{tri})$ and the latent heat of vaporization $L_{vap}(T_{tri})$ computed using the vaporization vapor pressure function to obtain the latent heat of sublimation at the triple point $L_{sub}(T_{tri})$. The fitted molar heat capacity functions and latent heat data needed to perform the s-g SVP integration for CH$_3$CN, C$_2$H$_5$CN, and C$_2$H$_3$CN are summarized in Table~\ref{table:selected_property}.

\item For C$_3$H$_4$-p, the only existing sublimation vapor pressure measurement for its solid phase is likely measured upon supercooled C$_3$H$_4$ liquid \citep[see discussion in][]{2009P&SS...57.2053F}. We are unable to locate any heat capacity data for solid C$_3$H$_4$-p. Thus, we cannot use the above thermodynamic approach to calculate its sublimation saturation vapor pressure. \cite{1989SCIPPs..38...69D} estimated the latent heat of fusion of C$_3$H$_4$-p using the method of \cite{1991JOrg.....8.7431C}. This allows us to empirically derive the sublimation saturation vapor pressure function of C$_3$H$_4$-p. We start with Equation~\ref{eq:cc1}, where we choose the triple point temperature and pressure calculated using the vaporization vapor pressure function. Then by assuming a constant latent heat of sublimation independent of temperature $L_{sub}(T)=L_{sub}(T_{tri})$, we can directly integrate Equation~\ref{eq:cc1} using the sum of the latent heat of fusion ($L_{fus}$) provided by \cite{1989SCIPPs..38...69D} and the latent heat of vaporization ($L_{vap}$) calculated by using the vaporization SVP function of \cite{Yaws2015} ($L_{sub}=L_{fus}+L_{vap}$).

\end{itemize}

\begin{table}[]
\addtolength{\leftskip} {-1.7cm}
\caption{Selected property values for CH$_3$CN, C$_2$H$_5$CN, and C$_2$H$_3$CN for s-g SVP calculations through the thermodynamic approach and for C$_3$H$_4$-p through the empirical approach. The notations in the ``Acc" (accuracy) column are the same as Table~\ref{table:SVP}.}
\resizebox{7.9in}{!}{
    \begin{tabular}{lccc}
    \toprule
    CH$_3$CN selected property value & Unit & Reference & Acc.\\
    \hline
        $C_{p,s1}(T)=0.30962T+6.0668,\ 216.9<T<229.349$ ($T_{tri}$) & J$\cdot$K$^{-1}$mol$^{-1}$ & \cite{1965JChPh..42..749P} & A$^+$\\
        $C_{p,s2}(T)=0.39755T,\ 20<T<216.9$ & J$\cdot$K$^{-1}$mol$^{-1}$ & \cite{1965JChPh..42..749P} & C\\
        $C_{p,v}(T)=-6.0507\times10^{-3}T^{1.5}+0.28151T,\ 0<T<900$ & J$\cdot$K$^{-1}$mol$^{-1}$ & \cite{1961JMoSp...5..212K} & B\\
        $L_{vap}(T_{tri})=34.79$ & kJ$\cdot$mol$^{-1}$ & \cite{Yaws2015}  \\
        $L_{fus}(T_{tri})=8.17$ & kJ$\cdot$mol$^{-1}$ & \cite{1965JChPh..42..749P}\\
        $T_{tran}=216.9$ & K & \cite{1965JChPh..42..749P}\\
        $L_{trans}(T_{transition})=0.898$ & kJ$\cdot$mol$^{-1}$ & \cite{1965JChPh..42..749P}\\
    \hline
    C$_2$H$_5$CN selected property value & Unit & Reference\\
    \hline
        $C_{p,s1}(T)=4.1003T-627.22,\ 176.964<T<180.4$ ($T_{tri}$) & J$\cdot$K$^{-1}$mol$^{-1}$ & \cite{1962JChPh..36..829W} & A$^+$\\
        $C_{p,s2}(T)=-1.3588\times10^{-3}T^2+0.70822T,\ 15<T<176.964$ & J$\cdot$K$^{-1}$mol$^{-1}$ & \cite{1962JChPh..36..829W} & C \\
        $C_{p,v}(T)=-6.4302\times10^{-3}T^{1.5}+0.35221T,\ 0<T<600$ & J$\cdot$K$^{-1}$mol$^{-1}$ & \cite{1955JChPh..23..434D} & A$^+$\\
        $L_{vap}(T_{tri})=42.36$ & kJ$\cdot$mol$^{-1}$ & \cite{2022nist.292..686K}  \\
        $L_{fus}(T_{tri})=5.05$ & kJ$\cdot$mol$^{-1}$ & \cite{1962JChPh..36..829W}\\
        $T_{tran}=176.964$ & K & \cite{1962JChPh..36..829W}\\
        $L_{trans}(T_{transition})=1.71$ & kJ$\cdot$mol$^{-1}$ & \cite{1962JChPh..36..829W}\\
    \hline
    C$_2$H$_3$CN selected property value & Unit & Reference\\
    \hline
        $C_{p,s1}(T)=0.42969T+0.34433,\ 162.50<T<189.642$ ($T_{tri}$) & J$\cdot$K$^{-1}$mol$^{-1}$ & \cite{1972chm..thermo..359F} & A\\
        $C_{p,s2}(T)=0.53012T,\ 12.5<T<162.50$ & J$\cdot$K$^{-1}$mol$^{-1}$ & \cite{1972chm..thermo..359F} & C\\
        $C_{p,v}(T)=-1.9309\time10^{-10}T^4+5.0312\times10^{-7}T^3-5.2491\times10^{-4}T^2+0.33526T,\ 0<T<1000$ & J$\cdot$K$^{-1}$mol$^{-1}$ & \cite{1972chm..thermo..359F} & A\\
        $L_{vap}(T_{tri})=37.39$ & kJ$\cdot$mol$^{-1}$ & \cite{Yaws2015}  \\
        $L_{fus}(T_{tri})=6.23$ & kJ$\cdot$mol$^{-1}$ & \cite{1972chm..thermo..359F}\\
        $T_{tran}=162.50$ & K & \cite{1972chm..thermo..359F}\\
        $L_{tran}(T_{transition})=1.19$ & kJ$\cdot$mol$^{-1}$ & \cite{1972chm..thermo..359F}\\
    \toprule
    C$_3$H$_4$-p selected property value & Unit & Reference\\
    \hline
        $L_{vap}(T_{tri})=26.24$ & kJ$\cdot$mol$^{-1}$ & \cite{Yaws2015}  \\
        $L_{fus}(T_{tri})=5.35$ & kJ$\cdot$mol$^{-1}$ & \cite{1989SCIPPs..38...69D}\\
    \toprule
    \end{tabular}}
    \label{table:selected_property}
\end{table}

\begin{figure}
    \centering
    \makebox[\textwidth][c]{\includegraphics[width=1.15\textwidth]{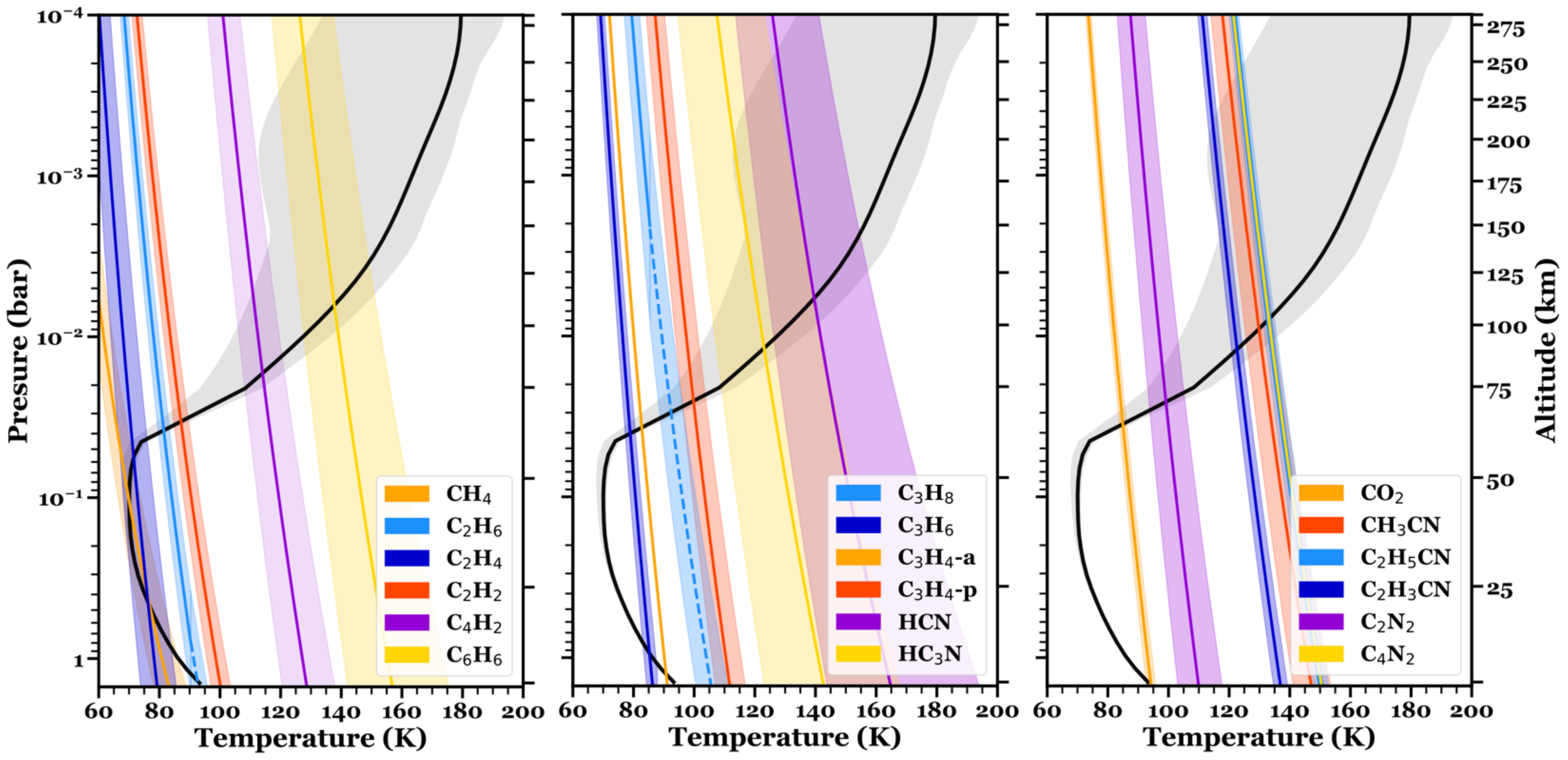}}
    \caption{Condensation curves for detected gaseous organic species in Titan's atmosphere. The black line in each panel represents Titan's mean temperature profile, and the shaded grey area represents the range of retrieved temperature profiles using Cassini/CIRS data throughout the Cassini mission \citep{2019GeoRL..46.3079T}. Note that the altitude axis on the right represents the altitude for the mean temperature-pressure profile. The solid condensation curves are plotted with the average mixing ratio, and the shaded regions are plotted using the observed upper and lower bond mixing ratios listed in Table~\ref{table:abundance}. Solid condensation curves represent solid-gas (s-g) transitions and dash lines represent liquid-gas transitions (l-g). (left) Condensation curves for CH$_4$, C$_2$H$_6$, C$_2$H$_4$, C$_2$H$_2$, C$_4$H$_2$, and C$_6$H$_6$. (center) Condensation curves for C$_3$H$_6$, C$_3$H$_8$, C$_3$H$_4$-a (allene), C$_3$H$_4$-p (propyne), HCN, and HC$_3$N. (right) Condensation curves for CO$_2$, CH$_3$CN, C$_2$H$_5$CN, C$_2$H$_3$CN, C$_2$N$_2$, and C$_4$N$_2$.}  
    \label{fig:condense_curves}
\end{figure}

Using the saturation vapor pressure functions in Table~\ref{table:SVP} and species abundances in Table~\ref{table:abundance}, we can now construct condensation curves for all the detected gas-phase species in Titan's atmosphere. We plot the condensation curves using the ratio between the saturation vapor pressure expressions ($P_{sat}(T)$) and the mixing ratios of each species ($X$). In Figure~\ref{fig:condense_curves}, we show the condensation curves for all 18 detected gas-phase organic species (Table~\ref{table:abundance}). Condensation occurs when the condensation curve of one species intersects the pressure-temperature (P-T) profile. To account for the seasonal and latitudinal changes on Titan, in addition to the mean P-T profile (shown as the solid black curve in Figure~\ref{fig:condense_curves}), we also include the possible range of P-T profile using the retrieved Cassini/Composite Infrared Spectrometer (CIRS) data \citep[see the supporting information of][]{2019GeoRL..46.3079T}, which is marked as the shaded grey region. Note that the P-T profile measured by the Huygens-Atmospheric Structure Instrument (HASI) \citep{2005Natur.438..785F} is closer to the lower limit of the P-T profiles of \cite{2019GeoRL..46.3079T}. We choose to present the data of \cite{2019GeoRL..46.3079T} because Cassini/CIRS measurements are most sensitive to the stratospheric temperature (10$^{-2}$ to 10$^{-4}$ bar), and the data covers the entire Cassini mission, taking into account temporal and latitudinal variations of the P-T profiles. Solid condensation curves represent solid-gas (s-g) transitions, and dashed lines represent liquid-gas (l-g) transitions. We also include the possible seasonal/latitudinal variations of the abundances of each species by using the largest possible range of stratospheric mixing ratios of each species, as provided in Table~\ref{table:abundance}, to compute the condensation curves. The variation of the condensation curves is represented by the shaded regions with the same color as the condensation curve for each compound. Note that we used vertically constant mixing ratios to plot these condensation curves. However, since most of the species condense in and below the stratosphere, the stratospheric mixing ratios ($\sim$10$^{-2}$-10$^{-3}$) we used should closely reflect the mixing ratio of each species before condensation. Based on Figure~\ref{fig:condense_curves}, all the 18 observed gas-phase species would condense in the stratosphere of Titan's atmosphere. While most species would condense as ices (with the solid black condensation curves intersecting the P-T profile), C$_3$H$_8$ from the upper atmosphere and evaporated C$_2$H$_6$ from Titan's surface would condense as liquids (with their dashed condensation curve intersecting the P-T profile). 

In Table~\ref{table:condensing_pt}, we summarize the mean condensation temperature ($T_{cond}$) and pressure ($P_{cond}$) for each species, i.e., the intersecting temperature and pressure of the mean P-T profile and condensation curve constructed with the mean mixing ratio of a species. We also summarize the maximum and minimum condensation temperature, pressure, and altitude given the varying P-T profile and condensation curves due to the variation of species abundances. Note that in this work, we only consider the condensation curves of pure species. It has been shown that nitrogen dissolved in methane/ethane serves as an anti-freezer \citep{2006P&SS...54.1177A,2010ApJ...712L..40W}. Because of this effect, it is expected that methane condenses as a liquid near the surface of Titan and that methane ice clouds only form higher in the atmosphere ($\sim$20-30 km) where the temperature falls below the freezing point of the CH$_4$-N$_2$ mixture and N$_2$ exsolves from the mixture \citep{1992Icar...97..187T, 2006Natur.442..432T,2010Icar..206..787W}. This highlights the importance of determining the saturation vapor pressure for mixed species in the laboratory, especially ternary mixtures, see theoretical calculations in \citep{2017NatAs...1E.102C, 2018tankargel} and some preliminary experimental works \citep{2015LPI....46.2763C, 2016DPS....4850203H}. 

\begin{table}
\addtolength{\leftskip} {-2.5cm}
\caption{Condensation temperatures, pressures, and altitudes for each species identified on Titan (Table~\ref{table:abundance}). We extract the mean-, min-, and max condensing temperatures ($T_{cond}$), pressures ($P_{cond}$), and altitudes ($h_{cond}$) at the intersections of the condensation curves and Titan's mean-, min-, and max P-T profile in Figure~\ref{fig:condense_curves}, respectively.}
\begin{tabular}{crcccrcccrccc}
\toprule
Species && \multicolumn{3}{c}{$T_{cond}$ (K)}  && \multicolumn{3}{c}{$P_{cond}$ (mbar)} && \multicolumn{3}{c}{$h_{cond}$ (km)}\\
 &&  Mean & Min & Max && Mean & Min & Max && Mean & Min & Max\\
\hline
\multirow{2}{*}{CH$_4$} && 70.0 & 69.1 & 71.9 && 1.1$\times$10$^2$& 0.45$\times$10$^2$ & 1.8$\times$10$^2$ && 44.8 & 36.1 & 57.3 \\
&& 76.9 & 71.9 & 86.4 && 0.48$\times$10$^3$ & 0.18$\times$10$^3$ & 1.0$\times$10$^3$ && 20.3 & 6.14 & 36.1\\
\hline
C$_2$H$_6$ (ice) && 81.5 & 79.2 & 83.4 && 38 & 30 & 41 && 61.5 & 60.7 & 64.4\\
C$_2$H$_6$ (liquid) && 92.5 & 89.3 & 93.0 && 1.4$\times$10$^3$ & 1.2$\times$10$^3$ & 1.4$\times$10$^3$ && 0.998 & 0.000 & 3.99\\
\hline
C$_2$H$_4$ && 71.6 & 67.7 & 76.1 && 0.57$\times$10$^2$ & 0.38$\times$10$^2$ & 1.5$\times$10$^2$ && 54.6 & 39.2 & 60.2 \\
\hline
C$_2$H$_2$ && 87.1 & 84.3 & 89.5 && 34 & 25 & 37 && 63.8 & 62.5 & 68.4\\
\hline
C$_4$H$_2$ && 114 & 106 & 121 && 17 & 2.8 & 23 && 80.2& 72.9 & 122\\
\hline
C$_6$H$_6$ && 138 & 119 & 151 && 6.4 & 0.077 & 13 && 109 & 89.0 & 234\\
\hline
C$_3$H$_8$ (liquid) && 91.6 & 87.6 & 95.6 && 30 & 20 & 35 && 66.0 & 63.9 & 73.0\\
\hline
C$_3$H$_6$ && 78.9 & 77.1 & 80.6 && 40 & 33 & 43 && 60.5 & 59.8 & 62.8 \\
\hline
C$_3$H$_4$-a && 82.7 & 82.1 & 83.0 && 37 & 30 & 39 && 62.0 & 61.7 & 64.3\\
\hline
C$_3$H$_4$-p && 99.7 & 95.2 & 104 && 25 & 12 & 30 && 70.0 & 67.3 & 84.5\\
\hline
HCN && 140 & 117 & 157 && 5.8 & 0.058 & 14 && 112 & 87.8 & 244 \\
\hline
HC$_3$N && 123 & 106 & 141 && 12 & 0.19 & 23 && 89.8 & 73.3 & 204\\
\hline
CO$_2$ && 85.0 & 83.3 & 86.3 && 35 & 27 & 38 && 62.9 & 62.2 & 66.3\\
\hline
CH$_3$CN && 130 & 117 & 135 && 9.0 & 0.21 & 14 && 98.1 & 87.1 & 200\\
\hline
C$_2$H$_5$CN && 133 & 122 & 136 && 7.9 & 0.20 & 11 && 102 & 93.1 & 202 \\
\hline
C$_2$H$_3$CN && 123 & 113 & 125 && 12 & 0.41 & 16 && 89.4 & 82.5 & 180\\
\hline
C$_2$N$_2$ && 99.2 & 93.1 & 105 && 26 & 11 & 31 && 69.7 & 66.2 & 87.0\\
\hline
C$_4$N$_2$ && 133 & 123 & 134 && 7.9 & 0.22 & 11 && 102 & 94.9 & 199 \\
\toprule
\end{tabular}
\label{table:condensing_pt}
\end{table}

\clearpage

\subsection{Latent heats of organic liquids and ices for species detected on Titan}\label{Latent_heat}

The latent heat for each condensate ($L$) can be found using the Clausius-Clapeyron equation (or Equation~\ref{eq:cc0}):
\begin{equation}
    L_{vap/sub} = T(V_v-V_c)\frac{dP_{sat}(T)}{dT},
\label{eq:latent_heat}
\end{equation}
where $V_c$ is the volume of the condensed phase (solid or liquid). The sublimation or vaporization latent heat ($L_{sub}$ or $L_{vap}$) has a unit of J$\cdot$mol$^{-1}$, though it is usually expressed in unit of kJ$\cdot$mol$^{-1}$. When the gas follows ideal gas behavior ($V_v=RT/P$) and the volume of the condensed phase is insignificant compared to the vapor phase ($V_v\gg V_c$), Equation~\ref{eq:latent_heat} can be reduced to:
\begin{equation}
    L_{vap/sub} = \frac{RT^2}{P_{sat}(T)}\frac{dP_{sat}(T)}{dT}.
\label{eq:latent_heat_var}
\end{equation}

When the vapor phase no longer behaves like an ideal gas law, typically towards low temperature and high pressure where intermolecular forces and the volume of the gas molecules are not negligible, we can use the van der Waals equation of state to calculate the volume of the vapor phase $V_v$ in Equation~\ref{eq:latent_heat}:
\begin{equation}
    \left(P_{sat}(T)+a/V_v^2\right)\left(V_v-b\right) = RT,
\label{eq:vanderwaals}
\end{equation}
where $a$ and $b$ are empirical van der Waals constants for each substance. The van der Waals constants of the 18 Titan-relevant species considered in this study are listed in Table~\ref{table:vDw_constants}. Note that the appropriate universal gas constant to be used here, which corresponds with the unit of the van der Waals constants in Table~\ref{table:vDw_constants}, is 0.08314 L$\cdot$bar$\cdot$K$^{-1}$mol$^{-1}$. For C$_4$H$_2$, HC$_3$N, and C$_4$N$_2$, their van der Waals constants are calculated using their critical temperatures ($T_{cri}$) and pressures ($P_{cri}$) listed in Table~\ref{table:temps}:
\begin{equation}
    a = \frac{27R^2T_{cri}^2}{64P_{cri}}
   \quad\mathrm{and}\quad 
    b = \frac{RT_{cri}}{8P_{cri}}.
\label{eq:vdWconstants}
\end{equation}

As an example, the ideal gas assumption does not hold for the air on the surface of Titan compared to Earth due to the lower temperature (94 K versus 300 K) and higher pressure (1.5 bar versus 1 bar). The volume of gas near the surface of Titan (assuming it is a mixture of 95 \% N$_2$ and 5 \% CH$_4$ at 94 K, 1.5 bar) calculated using the ideal gas law is 2.87 \% higher than the actual gas volume calculated using the van der Waals equation. While for air near the surface of Earth (assuming 80 \% N$_2$ and 20 \% O$_2$ at 300 K, 1 bar), the volume of air assuming ideal gas is only 0.07 \% higher than the actual gas volume.

\begin{table}[!htbp]
\centering
\caption{Van der Waals constants for each Titan-relevant species listed in Table~\ref{table:abundance}. References: [a] \cite{2014chcp.book.....H}, [b] \cite{1972hcp.book.....H}, [c] this work.}
\begin{tabular}{cccc}
    \toprule
    Species & $a$  & $b$ & Ref. \\
    & (L$^2\cdot$bar$\cdot$mol$^{-2}$) & (L$\cdot$mol$^{-1}$) & \\
    \hline 
    CH$_4$ & 2.303 & 0.0431 & [a]\\
    \hline 
    C$_2$H$_6$ & 5.580 & 0.0651 & [a]\\ 
    \hline 
    C$_2$H$_4$ & 4.612 & 0.0582 & [a]\\
    \hline 
    C$_2$H$_2$ & 4.516 & 0.0522 & [a]\\
    \hline 
    C$_4$H$_2$ & 18.856 & 0.1639 & [c]\\
    \hline 
    C$_6$H$_6$ & 18.82 & 0.1193 & [a]\\
    \hline 
    C$_3$H$_8$ & 9.39 & 0.0905 & [a]\\
    \hline 
    C$_3$H$_6$ & 8.442 & 0.0824 & [a]\\ 
    \hline 
    C$_3$H$_4$-a & 8.235 & 0.07467 & [b]\\
    \hline 
    C$_3$H$_4$-p & 8.40 & 0.0744 & [b] \\
    \hline 
    HCN & 11.29 & 0.0881 & [a]\\
    \hline 
    HC$_3$N & 15.576 & 0.1053 & [c]\\
    \hline 
    CO$_2$ & 3.658 & 0.0429 & [a]\\
    \hline 
    CH$_3$CN & 17.89 & 0.1169 & [b] \\ 
    \hline 
    C$_2$H$_5$CN & 21.57 & 0.1369 & [b]\\ 
    \hline 
    C$_2$H$_3$CN & 18.37 & 0.1222 & [b] \\
    \hline 
    C$_2$N$_2$ & 7.803 & 0.6952 & [b]\\
    \hline 
    C$_4$N$_2$ & 10.871 & 0.0781 & [c] \\
    \toprule
    \end{tabular}
    \label{table:vDw_constants}
\end{table}

In this work, we calculated latent heats of sublimation ($L_{sub}(T_{tri})$) and vaporization ($L_{vap}(T_{tri})$) at the triple point of each substance. Considering the volume of the condensed phase negligible compared to the vapor phase, we use the following equation and Equation~\ref{eq:vanderwaals} to calculate the latent heats:
\begin{equation}
    L_{vap/sub} = TV_v\frac{dP_{sat}(T)}{dT}.
\label{eq:latent_heatx}
\end{equation}
The saturation vapor pressure functions, $P_{sat}(T)$ are listed in Table~\ref{table:SVP} and the van der Waals constants are listed in Table~\ref{table:vDw_constants} for these calculations. The calculated latent heats values can be found in Table~\ref{table:latent_heat}. The latent heat values at $T_{tri}$ are used for subsequent surface energy calculations of organic ices in Section~\ref{ice_SE}.

\begin{table}[!htbp]
\centering
\caption{Calculated latent heat of sublimation ($L_{sub}$) and vaporization ($L_{vap}$) of the condensates at the triple point of each of the 18 species detected on Titan and listed in Table~\ref{table:abundance}.}
\begin{tabular}{ccc}
    \toprule
    Species & $L_{sub}(T_{tri})$  & $L_{vap}(T_{tri})$ \\
    & (kJ$\cdot$mol$^{-1}$) & (kJ$\cdot$mol$^{-1}$) \\
    \hline 
    CH$_4$ & 9.569 & 8.640 \\ 
    \hline 
    C$_2$H$_6$ & 20.75 & 18.20\\ 
    \hline 
    C$_2$H$_4$ & 19.69 & 16.21\\
    \hline 
    C$_2$H$_2$ & 20.72 & 16.63\\ 
    \hline 
    C$_4$H$_2$ & 37.28 & 24.88\\ 
    \hline 
    C$_6$H$_6$ & 45.51 & 35.34\\ 
    \hline 
    C$_3$H$_8$ & 28.30 & 25.85\\
    \hline 
    C$_3$H$_6$ & 27.60 & 25.49\\ 
    \hline 
    C$_3$H$_4$-a & 27.47 & 23.86\\
    \hline 
    C$_3$H$_4$-p & 31.59 & 26.86 \\
    \hline 
    HCN & 36.44 & 28.13\\ 
    \hline 
    HC$_3$N & 45.43 & 27.96\\ 
    \hline 
    CO$_2$ & 24.87 & 16.13\\ 
    \hline 
    CH$_3$CN & 42.96 & 34.79 \\ 
    \hline 
    C$_2$H$_5$CN & 47.41 & 42.36\\ 
    \hline 
    C$_2$H$_3$CN & 43.62 & 37.39 \\
    \hline 
    C$_2$N$_2$ & 34.60 & 24.69\\ 
    \hline 
    C$_4$N$_2$ & 50.02 & 27.50 \\ 
    \toprule
    \end{tabular}
    \label{table:latent_heat}
\end{table}

\section{Physical properties - density}
Density is an important material property that can shape the physical evolution of substances on Titan, such as settling and buoyancy, etc. Density is also often required to predict other properties of a substance. In the subsequent sections, we summarize the densities of Titan-relevant organic liquids, along the saturation curve, of Titan-relevant organic ices, and of Titan haze analogs.

\subsection{Densities of organic liquids for the species detected on Titan}

In Table~\ref{table:density_liquid}, we summarize the saturation liquid densities as a function of temperature for the 18 Titan-relevant organic species listed in Table~\ref{table:abundance}. For some liquids, we adopt existing the liquid density expressions provided in \cite{1989SCIPPs..38...69D, 2020chcp.book.....L}. For the other liquids, we least-square fit the liquid density ($\rho_l$) as a function of temperature using the following expression:
\begin{equation}
\rho_l(T) = \rho_l(T_{cri}) + A_0(1-T/T_{cri})^{0.35} + \displaystyle\sum_{n=1}^i{A_i(1-T/T_
{cri})^i}.
\end{equation}
For most of the organic liquids (CH$_4$, C$_2$H$_6$, C$_2$H$_4$, C$_3$H$_8$, C$_3$H$_6$, CO$_2$), we fit the saturation liquid density data from \cite{2020chcp.book.....L}. For C$_4$H$_2$, we fit the saturation liquid density data from \cite{1925Hebd...5..289L, 1926JCh..36..829W, JPhchem...57.2053F}. For C$_6$H$_6$, we fit the saturation liquid density data from \cite{2022nist.292..686K}. For C$_3$H$_4$-p, we fit the saturation liquid density data from \cite{1921JAmC...32..116M, 1931CJRes...5..306M, 1986Chem...351.8G, 2022nist.292..686K,2020chcp.book.....L}. The liquid density data that are used for the fitting and the deviation between the fitting and each experimental data point are summarized in Table~\ref{table:liquid_density_data} in the Appendix section.

\begin{table}[!htbp]
\caption{Density expressions of Titan-relevant organic liquids along the saturation line. The ``Temp. (K)" column has the same meaning as Table~\ref{table:SVP}. [a] This work; [b] \cite{1989SCIPPs..38...69D}; [c] \cite{2022nist.292..686K}; [d] \cite{1963JChPh..38...69D}; [e] \cite{2020chcp.book.....L}.}
\addtolength{\leftskip} {-1.7cm}
\resizebox{7.9in}{!}{
\begin{tabular}{crccc}
\toprule
Species &&Liquid density, $\rho_l(T)$ (kg$\cdot$m$^{-3}$)  & Temp (K) & Ref \\
\hline
CH$_4$ && $162.70+291.23(1-T/T_{cri})^{0.35}+142.42(1-T/T_{cri})-121.96(1-T/T_{cri})^2+107.38(1-T/T_{cri})^3$ & $T_{tri}$-$T_{cri}$ & [a]\\
C$_2$H$_6$ && $206.20+390.80(1-T/T_{cri})^{0.35}+161.80(1-T/T_{cri})-90.702(1-T/T_{cri})^2+88.144(1-T/T_{cri})^3$ & $T_{tri}$-$T_{cri}$ & [a]\\
C$_2$H$_4$ && $214.20+402.42(1-T/T_{cri})^{0.35}+177.02(1-T/T_{cri})-99.181(1-T/T_{cri})^2+101.65(1-T/T_{cri})^3$ & $T_{tri}$-$T_{cri}$ & [a]\\
C$_2$H$_2$ && $63.816/0.27448^{1+(1-T/T_{cri})^{0.28752}}$ & $T_{tri}$-$T_{cri}$ & [b] \\
C$_4$H$_2$ && $2165.6-5.2315T$ & $T_{tri}$--278--[$T_{cri}$] & [a] \\
C$_6$H$_6$ && $304.70+596.41(1-T/T_{cri})^{0.35}+292.04(1-T/T_{cri})-296.31(1-T/T_{cri})^2+378.24(1-T/T_{cri})^3$ & $T_{tri}$-$T_{cri}$ & [a] \\
C$_3$H$_8$ && $220.50+429.60(1-T/T_{cri})^{0.35}+167.26(1-T/T_{cri})-87.547(1-T/T_{cri})^2+96.621(1-T/T_{cri})^3
$ & $T_{tri}$-$T_{cri}$ & [a]\\
C$_3$H$_6$ && $229.60+447.91(1-T/T_{cri})^{0.35}+181.53(1-T/T_{cri})-102.10(1-T/T_{cri})^2+121.01(1-T/T_{cri})^3$ & $T_{tri}$-$T_{cri}$ & [a] \\
C$_3$H$_4$-a && $243.20+508.18(1-T/T_{cri})^{0.35}+66.874(1-T/T_{cri})+232.07(1-T/T_{cri})^2-159.04(1-T/T_{cri})^3$ & $T_{tri}$-$T_{cri}$ & [c] \\
C$_3$H$_4$-p && $244.90+469.73(1-T/T_{cri})^{0.35}+409.34(1-T/T_{cri})-658.50(1-T/T_{cri})^2+600.71(1-T/T_{cri})^3$ & $T_{tri}$-$T_{cri}$ & [a] \\
HCN && $195.00+451.55(1-T/T_{cri})^{0.35}+1249.0(1-T/T_{cri})-3452.1(1-T/T_{cri})^2+3809.9(1-T/T_{cri})^3$ & $T_{tri}$-$T_{cri}$ & [c] \\
HC$_3$N && $1189.6-1.2850T$ & $T_{tri}$--315.2--[$T_{cri}$] & [d] \\
CO$_2$ && $467.60+925.62(1-T/T_{cri})^{0.35}+465.51(1-T/T_{cri})-412.21(1-T/T_{cri})^2+509.61(1-T/T_{cri})^3$ & $T_{tri}$-$T_{cri}$ & [a] \\
CH$_3$CN && $247.37+556.17(1-T/T_{cri})^{0.35}+214.71(1-T/T_{cri})+34.189(1-T/T_{cri})^2+34.820(1-T/T_{cri})^3$ & $T_{tri}$-$T_{cri}$ & [c] \\
C$_2$H$_5$CN && $244.85+447.66(1-T/T_{cri})^{0.35}+601.25(1-T/T_{cri})-658.21(1-T/T_{cri})^2+497.05(1-T/T_{cri})^3$ & $T_{tri}$-$T_{cri}$ & [c]\\
C$_2$H$_3$CN && $252.09+757.67(1-T/T_{cri})^{0.35}-448.60(1-T/T_{cri})+1307.2(1-T/T_{cri})^2-944.27(1-T/T_{cri})^3$ & $T_{tri}$-$T_{cri}$ & [c]\\
C$_2$N$_2$ && $352.99+734.98(1-T/T_{cri})^{0.35}+191.04(1-T/T_{cri})+216.84(1-T/T_{cri})^2-248.71(1-T/T_{cri})^3$ & $T_{tri}$-$T_{cri}$ & [c] \\
C$_4$N$_2$ && 970.8 & 298.15 & [e] \\
\toprule
\end{tabular}}
\label{table:density_liquid}
\end{table}

\subsection{Densities of organic ices for the species detected on Titan}

\begin{table}
\caption{Adopted intrinsic density expressions as a function of temperature for Titan-relevant organic ices. The ``Temp. (K)" column has the same meaning as Table~\ref{table:SVP}. [a] refitted data of \cite{2020IUCrj292..683M}; [b] \cite{1978Chry..36..829W}; [c] fitted, this work;  [d] estimated, this work; [e] \cite{1958JChPh..28..425S}; [f] \cite{1962JChPh..36.2654B}; [g] \cite{2022LPICo2678.2321M}; [h] \cite{1958ActaCry..28..425S}; [i] \cite{1998xxx..48..299c}; [j] \cite{1953ActCry...6..350H}. }
\addtolength{\leftskip} {-1.8cm}
\resizebox{7.9in}{!}{
\begin{tabular}{ccccc}
\toprule
Species & Crystal type & Ice intrinsic density, $\rho_s(T)$ (kg$\cdot$m$^{-3}$) & Temp (K) & Ref \\
\hline
\multirow{2}{*}{CH$_4$}  & I & $1.0624\times10^5/(201.74-1.2024\times10^{-2}T+1.8596\times10^{-3}T^2)$ & 20.509--$T_{tri}$ & [a]\\
 & II & $8.4992\times10^5/(1603.6-1.1690\times10^{-3}T^3+9.0131\times10^{-5}T^4)$ & [$<$]8--20.509 & [a]\\
\hline
\multirow{2}{*}{C$_2$H$_6$}  & I, II & 669 & 89.726--$T_{tri}$ & [b]\\ 
& III & $737.91+0.21576T-6.1118\times10^{-3}T^2$ & ($<$)5--89.726 & [c]\\
\hline
C$_2$H$_4$ & I & $795.15-0.69734T$ & [$<$]20.4--$T_{tri}$ & [c]\\
\hline
C$_2$H$_2$ & I & $853.63-0.64829T$ & [$<$]77--188.15--[$T_{tri}$] & [c]\\
\hline
C$_4$H$_2$ & I & $1194.3-0.63366T$ & [$\leq$$T_{tri}$] & [d] \\
\hline
C$_6$H$_6$ & I & $1127.8-8.1121\times10^{-2}T-1.1497\times10^{-3}T^2$ & [$<$]20.4--$T_{tri}$ & [c]\\
\hline
C$_3$H$_8$ & I & $832.98-0.96372T$ & [$<$]20.4--$T_{tri}$ & [c]\\
\hline
C$_3$H$_6$ & I & 806 & 77 & [e]\\
\hline
C$_3$H$_4$-a & I & 860 & 95 & [f]\\
\hline
C$_3$H$_4$-p& I & 882 & 90 & [g] \\
\hline
HCN & I & $1234.6-1.3477T$ & [$<$]153.15--233.15--[$T_{tri}$] & [c]\\
\hline
HC$_3$N & I & 1075.3 & 248.15 & [h]\\
\hline
CO$_2$ & I & $1711.5-8.7556\times10^{-2}T-3.3884\times10^{-3}T^2$ & [$<$]20.4--194.15--[$T_{tri}$] & [c]\\
\hline
\multirow{2}{*}{CH$_3$CN} & I (or $\alpha$) & $816.57+2.4008T-6.7142\times10^{-3}T^2$ & 201--$T_{tri}$ & [c] \\
 & II (or $\beta$) & $1127.5-4.3029\times10^{-2}T-1.5440\times10^{-3}T^2$  & [$<$]5--216.9 & [c] \\
\hline
C$_2$H$_5$CN & I & $1060.7-0.46187T$ & [$<$]100--150--[$T_{tri}$] & [c]\\
\hline
C$_2$H$_3$CN & I & 1027.4 & 153 & [i]\\
\hline
C$_2$N$_2$ & I & $1285.9-0.20032T$ & [$<$]15--178.15--[$T_{tri}$] & [c]\\
\hline
C$_4$N$_2$ & I & 1230 & 278 & [j] \\
\toprule

\end{tabular}}
\label{table:density_solid}
\end{table}

The existing density data for Titan-relevant organic ices are usually more scarce, and the measured values sometimes may also seem inconsistent with each other. The apparent inconsistencies in the measured ice density are mainly caused by the different experimental diagnostic techniques used and how each technique defines the density (see \cite{2017Icar..297...97Y} for all the different density definitions).

There are two main groups of ice density measurement techniques. The first group of techniques measures the intrinsic density or the material density of the ice. This is the highest density that can be measured for a material, i.e., density with no pore space. Techniques that constrain the crystal structure, such as neutron or X-ray diffraction, give the intrinsic density, with the evaluated volume being the crystal of a unit cell with no pore space. The second group of techniques measures the average density. The average density is usually lower than the intrinsic density, as pore space may be included in addition to the volume of the ice itself. Techniques that measure average density include the quartz crystal microbalance (QCM)/double laser interferometry combination technique, estimation from the refractive index of the ice, and pycnometry technique. With the QCM/double laser interferometry technique, the mass of the ice sample is monitored and measured by the QCM, and the volume is determined by monitoring the thickness of the ice layers using laser interferometry. Therefore, if pore space develops during the ice growth process, the volume calculated using the thickness of the ice inevitably includes pore space. The average density of the ice can also be estimated from the measured refractive index of the ice (n$_{vis}$) using the Lorentz-Lorenz equation (see Equation~\ref{eq:lorentz}). Because the measured refractive index correlates with the porosity of the ice, the derived ice density is the average density. The pycnometry technique directly measures the volume of the condensed ice. As a consequence, the measured volume may include pore space, if any develops during the ice growth process. The porosity of the ice can be strongly affected by the growth processes of the ice, the formation temperatures used in the ice production experiments, and the phase of the ice (crystalline or amorphous) \cite[e.g,][]{2016ApJ...827...98L, 2018ASSL..451...51S}. Note that diffraction techniques cannot measure the density of low-temperature amorphous ice as there is no unit cell structure. Thus, techniques such as QCM/interferometry can serve as diagnostic tools to assess the density and porosity of amorphous ices.

To establish relationships that can be easily used by the community (i.e., the ice density variation with temperature), in this section, we use the intrinsic density data with no or minimal pore space that are measured at different temperatures. Here we choose to use the intrinsic density measured by the diffraction techniques, as this technique measures the unit cell structure of ice crystals without pore space. Because visual inspection of pore space (or bubbles) of the ice sample is possible using the pycnometry technique and procedures have been developed to sufficiently reduce porosity \citep[e.g.,][]{1930PhChem.147..466H}, many of the pycnometry-determined densities are also relatively close to the intrinsic density, e.g., \cite{1958JChPh..28..425S} for C$_2$H$_6$ and CH$_4$ ices, \cite{1971JChem..17904721B, 1964Sov...32..116M} for CH$_4$ ices. Thus, we choose to use density data that are measured by diffraction techniques and pycnometry to create fits for ice intrinsic density as a function of temperature. We summarize all the density measurements we used to fit for the intrinsic density-temperature relationships in Table~\ref{table:ice_density_data_1}. 
Since ice under different growth conditions may have different porosity, the measured average density may span a range of values, even at the same temperature. For example, the measured average density of crystalline C$_2$H$_4$ ice at 60 K is almost 30 \% higher in \cite{2022Icar..37314799Y} (796 kg$\cdot$m$^3$) than in \cite{2017Icar..296..179S} (629 kg$\cdot$m$^3$). Thus, in this work, we choose to only present all the existing average density measurements, as shown in Table~\ref{table:ice_density_data_2}. 

In Table~\ref{table:density_solid}, we summarize the fitted intrinsic density expressions as a function of temperature for all Titan-relevant ices (cf. compounds listed in Table~\ref{table:abundance}). For most ices, we have generated fits using all the published diffraction and pycnometry data that we could find. For methane ice, however, even though there have been many studies to determine its density using diffraction techniques, the results are not always in agreement -- possibly due to differences in the purity of the ice. We choose to only present the fit using the recent measurements by \cite{2020IUCrj292..683M} from 8-88~K because they cover the largest temperature range among all the CH$_4$ ice densities we found. For CH$_3$CN ice, it is important to note that even though the transition temperature is at 216.9 K (see Table~\ref{table:temps}), the crystal I or the $\beta$ form of CH$_3$CN ice is found to grow below the transition temperature \citep{ActaCry...57.2053F}, thus there is an overlap of the temperature range for the expressions in Table~\ref{table:density_solid}. For a full list of intrinsic density data that are used to create fittings in Table~\ref{table:density_solid} (i.e., measured by diffraction and pycnometry), see Table~\ref{table:ice_density_data_1} in the Appendix section. All the ice densities measured in Table~\ref{table:ice_density_data_1} are for crystalline ices. The rest of the average density measurements that are not used to create the fittings can be found in Table~\ref{table:ice_density_data_2} (i.e., measured by QCM or estimated using refractive index of the ice). Note that for amorphous ices, the average density values tend to be much lower compared to the adopted intrinsic density fit, indicating high porosity. Specifically, the porosity can be as high as 40-50 \% for amorphous C$_2$H$_6$, C$_2$H$_4$, CO$_2$, and C$_2$H$_5$CN ices, 30-35 \% for amorphous HCN and CH$_3$CN ices, and 20 \% for amorphous C$_3$H$_8$ ice. While crystalline ices tend to have much lower porosity, from just a few percent to 20 \%, depending on the growth process of the ices \citep{2016ApJ...827...98L, 2022Icar..37314799Y}.

C$_4$H$_2$ is the only organic ice for which we could not find any published density measurement. Future laboratory work is required to directly determine the intrinsic density of C$_4$H$_2$ ice. In this work, we estimate the density of C$_4$H$_2$ ice ($\rho_s(T)$) using a simple solid-liquid density relationship. We use the liquid density expression of C$_4$H$_2$ in Table~\ref{table:density_liquid} to first calculate its liquid density value at the triple point ($\rho_l(T_{tri})$). Then we can use the relationship determined by \cite{2004Jchem..351.8866G}, which is found to be accurate within 6 \% for most organics:

\begin{equation}
    \rho_s(T)=\left(1.28-0.16\frac{T}{T_{tri}}\right)\rho_l(T_{tri}),
\end{equation}
where both $\rho_l(T_{tri})$ and $\rho_s(T)$ have units of kg$\cdot$m$^{-3}$.

As shown in Table~\ref{table:density_solid}, many organic ices only have ice intrinsic density measurements at one temperature (C$_3$H$_6$, C$_3$H$_4$-a, C$_3$H$_4$-p, HC$_3$N, C$_2$H$_3$CN). More laboratory density measurements performed at different temperatures are needed to create fits for ice density variation as a function of temperature.

\subsection{Densities of Titan haze analogs}
Laboratory-made tholins have been serving as the model material for the refractory haze particles in Titan's atmosphere. Out of the many laboratories that have synthesized tholins, the four groups that have measured the densities of their tholin samples (see a summary in Table~\ref{table:density_tholin}) are: the University of Colorado Boulder group (UC Boulder, setup name: ``Tolbert setup" -- named after the PI), NASA Ames Research Center group 1 (NASA ARC 1, setup name: tholin-Radio frequency (RF) system), Johns Hopkins University group (JHU, setup name: Planetary Haze Research (PHAZER)), and Laboratoire Atmosph\`eres, Milieux, Observations Spatiales (LATMOS, setup name: Production d'A\'erosols en Microgravit\'e par Plasma REactif (PAMPRE)). The densities obtained by these groups are summarized in Table~\ref{table:density_tholin}, along with some experimental parameters specific of each sample (setup name, energy source, gas mixture, pressure, type of density measurement). For a list of the general experimental parameters (pressure, temperature, reaction time, gas irradiation time, gas flow rate) that are used for each setup, please refer to Table~\ref{table:additional_parameters}.

\begin{table}
\caption{Measured densities of Titan haze analogs, including the type of density measured, the setup in which the sample were produced, and some experimental parameters (energy source and gas mixture composition). $\mathsection$: The UV lamp emits continuously from 115-400 nm. References: [a] \cite{2013ApJ...770L..10H}; [b] \cite{2006PNAS..10318035T}; [c] \cite{2008Icar..194..186S}; [d] \cite{2012Icar..218..247I}; [e] \cite{2017ApJ...841L..31H}; [f] \cite{2018JGRE..123..807L}; [g] \cite{2016MNRAS.462S..89B}.}
\addtolength{\leftskip} {-1.8cm}
\resizebox{7.9in}{!}{
\begin{tabular}{crcrccccc}
\toprule
Laboratory location && Energy source && Gas mixture & Pressure & Density, $\rho_s$ & Measured density & Ref.\\
(Setup) &&  &&  & (mbar) &  (kg$\cdot$m$^{-3}$) &  definition & \\
\hline
&& \multirow{7}{*}{UV lamp$^\mathsection$} && N$_2$:CH$_4$ (99.99:0.01) & $830-850$ & 480$\pm$70 & \multirow{7}{*}{Average density} & [a]\\
 &&&& N$_2$:CH$_4$ (99.9:0.1) & $830-850$& 650$\pm$110 && [a] \\
&&  && N$_2$:CH$_4$ (99.9:0.1) & $800$ & 790$\pm$90 & & [b]\\
 &&&& N$_2$:CH$_4$ (99:1)& $830-850$& 950$\pm$150 &  & [a]\\
 &&&& N$_2$:CH$_4$ (98:2)& $830-850$& 760$\pm$150 & (AMS \& SMPS) & [a]\\
UC Boulder &&&& N$_2$:CH$_4$ (95:5)& $830-850$& 740$\pm$130 && [a]\\
(Tolbert setup) &&&& N$_2$:CH$_4$ (90:10) & $830-850$ & 500$\pm$100 && [a]\\
\cline{3-9}
 && \multirow{5}{*}{Spark discharge} && N$_2$:CH$_4$ (99.9:0.1)  & \multirow{5}{*}{$830-850$} &	820$\pm$230 & \multirow{5}{*}{Average density} & \multirow{5}{*}{[a]}\\
&&&& N$_2$:CH$_4$ (99:1) & & 980$\pm$220 && \\
&&&& N$_2$:CH$_4$ (98:2) & & 1130$\pm$100 &&\\
&&&& N$_2$:CH$_4$ (95:5) & & 660$\pm$110 &  (AMS \& SMPS) & \\
 &&&& N$_2$:CH$_4$ (90:10)& & 400$\pm$240 && \\
\hline
 && && \multirow{4}{*}{N$_2$:CH$_4$ (90:10)} & 0.4 & 1380 & & [c] \\
NASA ARC 1 && RF cold plasma &&& 1.5 & 1270  & Average density & [c] \\
(tholin-RF system) && irradiation && & 0.26 & 1386.6  & (Pycnometry) & [d]\\
 &&  && & 1.6 &	1311.3 &  & [d]\\
\hline
 && \multirow{7}{*}{AC plasma discharge} && N$_2$:CH$_4$ (95:5) & \multirow{7}{*}{2.67} & 1343$\pm$2 & & \multirow{7}{*}{[e]}\\
&&&& N$_2$:CH$_4$:CO (94.95:5:0.05) && 1348$\pm$2 & & \\
JHU &&&& N$_2$:CH$_4$:CO (94.8:5:0.2) && 1364$\pm$3 & Average density &\\
(PHAZER)&&&& N$_2$:CH$_4$:CO (94.5:5:0.5) && 1379$\pm$3 & (Pycnometry)&\\
&&&& N$_2$:CH$_4$:CO (94:5:1) && 1395$\pm$3 &&\\
&&&& N$_2$:CH$_4$:CO (92.5:5:2.5) && 1416$\pm$3 &&\\
&&&& N$_2$:CH$_4$:CO (90:5:5) && 1424$\pm$3 &&\\
\hline
 && \multirow{4}{*}{DC plasma discharge} && N$_2$:CH$_4$ (98:2) & \multirow{4}{*}{0.2--10} & 1450$\pm$20 &  & [f]\\ 
LATMOS &&&& \multirow{2}{*}{N$_2$:CH$_4$ (95:5)} && 1440$\pm$30 & Average density & [f]\\
(PAMPRE) &&&&  && 1440$\pm$10 &  (Pycnometry) & [g]\\
&&&& N$_2$:CH$_4$ (92:8) && 1340$\pm$30 && [f]\\
\hline
\multicolumn{5}{c}{\textbf{Recommended density range}} & \multicolumn{3}{c}{\textbf{400-1450 kg$\cdot$m$^{-3}$}}\\
\toprule
\end{tabular}}
\label{table:density_tholin}
\end{table}

Note that only two methods have been used to characterize the densities of tholins produced in various laboratories. The UC Boulder group used a combination system including an aerosol mass spectrometer (AMS) and a scanning mobility particle sizer (SMPS) that measure the average density of the produced tholin particles in-situ. With this method, the volume in the density determination could include volume larger than the material itself if the particles are not spherical and have internal pores. The measured average density of these tholin samples ranges from 400-1130~kg$\cdot$m$^{-3}$. The other three groups all used helium pycnometry to measure the densities of their tholin sample collected from substrates/walls of the setup. Since the helium gas used in pycnometry can fill pores as small as 3 \AA \citep{2017Icar..297...97Y}, the volume in the measured density only includes the internal pores that are not connected to the outside. The pycnometry-measured average densities of these tholin samples range from 1270-1450~kg$\cdot$m$^{-3}$.

It is interesting that there is a clear gap between the average density measured by the UC Boulder group and the other three lab groups. However, it is hard to evaluate whether their density values are lower than the other groups because of the additional volume taken into account in their average density measurements or because their tholin samples are intrinsically different than the ones from the other groups. Previous works proposed that both the deposition technique and the pressure may lead to density differences of the tholin samples. Specifically, depositing particles on substrates/walls may reduce internal voids/porosity to increase density \citep{2013ApJ...770L..10H, 2017ApJ...841L..31H} and lower pressure seems to create tholin samples with higher density \citep{2008Icar..194..186S, 2012Icar..218..247I,2019NatGe..12..315C}. Overall, we recommend using the widest possible density range, from 400--1450~kg$\cdot$m$^{-3}$, for Titan's haze particles to account for the uncertainty of the properties of tholins made in different laboratories. For certain processes such as aerosol sedimentation or comparative planetology, this density range is likely sufficient to describe the haze behavior, as the density of hazes is much larger than Titan's atmosphere and much smaller than most refractory materials on other planets (silicates, metals). For some other processes, we need more laboratory work to help constrain the density values of the actual haze particles on Titan. For example, we still cannot distinguish whether tholin particles would float or sink into the methane-ethane-nitrogen lakes as the density of the hazes (400-1500~kg$\cdot$m$^{-3}$) overlaps the density range of the lake liquids ($\sim$450-700~kg$\cdot$m$^{-3}$) \citep{2019NatGe..12..315C, 2020ApJ...905...88Y}.

\section{Surface properties}
The surface properties govern the strength of cohesion and adhesion, rising from the intermolecular structures of substances. The surface free energy is the intrinsic material property that characterizes the surface properties of a substance. The surface free energy is defined as the energy needed to create a unit area of the surface or half of the energy needed to separate two surfaces of a unit area. On the one hand, the surface free energies can affect interactions between the same substances through cohesion. Thus, it is important for atmospheric processes including haze formation through particle coagulation, and also surface processes including sediment transport, as wind shear stress has to overcome sediment cohesion to initiate sediment transport \citep{2017Icar..297...97Y, 2020ApJ...905...88Y}. On the other hand, the surface free energies can affect interactions between two different substances through adhesion. In order to assess processes such as cloud formation through condensate-haze interactions and lake-haze interactions on Titan \citep{2020ApJ...905...88Y}, the surface free energies are therefore an important parameter to consider in order to understand how these substances interact with each other.

Note that the surface free energies are often referred to differently for a liquid and a solid. For a liquid, its surface free energy is often referred to as the surface tension ($\gamma_l$), with a unit of mN$\cdot$m$^{-1}$, while for a solid, its surface free energy is often referred to as the surface energy ($\gamma_s$), with a unit of mJ$\cdot$m$^{-2}$ (equivalent to mN$\cdot$m$^{-1}$). In the following sections, we summarize and compute the surface free energies of the Titan-relevant organic liquids, ices, and solids.

\subsection{Surface tensions of Titan-relevant organic liquids}

For liquid condensates, we list their liquid surface tension expressions as a function of temperature ($\gamma_{liq}(T)$) in Table~\ref{table:surface_tension}. Most surface tension expressions provided in this table are directly taken from \cite{Yaws2014}, except for HC$_3$N, C$_4$N$_2$, and C$_4$H$_2$, which have no surface tension measurements. Note that \cite{Yaws2014} does provides an estimation of the surface tension of C$_4$H$_2$. In this work, we predicted the surface tensions of these three compounds using three methods recommended in \cite{Poling2001}: 1) the MacLeod-Sugden method \citep{1923Tranf..32..116M, 1924Tranf..32..116M}; 2) the Brock-Bird method \citep{1955brockbird, 1963miller}; and 3) the Pitzer method \citep{1958Curl,1995Pitzer}.

The MacLeod-Sugden method estimates the surface tension of a material by using its liquid density ($\rho_l$) (Table~\ref{table:density_liquid}) and the calculated parachor based on its chemical structure:
\begin{equation}
\gamma_l=\left(\frac{P\rho_l}{M}\right)^4,
\label{eq:parachor}
\end{equation}
where the parachor $P$ can be calculated using Table 25 in \citet{1953Tranf..32..116M} based on the structure of the chemical. \textit{P} is 148.2 for C$_4$H$_2$ , 141.2 for HC$_3$N, and 192.8 for C$_4$N$_2$. The conventional unit of the parachor is (mN$\cdot$m$^{-1}$)$^{1/4}\times$(cm$^3\cdot$mol$^{-1}$). \textit{M} is the molar mass of the chemical (with a unit of g$\cdot$mol$^{-1}$ in Equation~\ref{eq:parachor}), and $\rho_l$ is its liquid density (with a unit of g$\cdot$cm$^{-3}$ in Equation~\ref{eq:parachor}). 

The Brock-Bird method predicts the surface tensions of a material using the critical pressure ($P_{cri}$) and temperature ($T_{cri}$) (Table~\ref{table:temps}) and the boiling point temperature ($T_{b}$): 
\begin{equation}
\begin{split}
&\gamma_l=P_{cri}^{2/3}T_{cri}^{1/3}Q(1-T/T_{cri})^{11/9},\\
&Q=0.1196\left[1+\frac{T_{b}/T_{cri}\mathrm{ln}(P_{cri}/1.01325)}{1-T_{b}/T_{cri}}\right]-0.279,
\label{eq:bbmethod}
\end{split}
\end{equation}
where $T_{b}$ can be calculated using the SVP equations in Table~\ref{table:SVP} by equating the pressure to 1 bar. 

The Pitzer method predicts the surface tensions of a material using $P_{cri}$ and $T_{cri}$ (Table~\ref{table:temps}), and the boiling point temperature ($T_{b}$): 
\begin{equation}
\begin{split}
&\gamma_l=P_{cri}^{2/3}T_{cri}^{1/3}\frac{1.86+1.18\omega}{19.05}\left[\frac{3.75+0.91\omega}{0.291-0.08\omega}\right]^{2/3}(1-T/T_{cri})^{11/9},\\
&\omega=-log_{10}[P(T=0.7T_{cri})/P_{cri}]-1.0,
\label{eq:pitzer}
\end{split}
\end{equation}
where $\omega$ is the acentric factor, which can be calculated using the vapor pressure at $0.7\ T_{cri}$ using Table~\ref{table:SVP}.

In this work, to accommodate for the uncertainty of the different estimation methods, we report an average of the predicted surface tension values. For C$_4$H$_2$, we averaged the predicted surface tension values by all three methods at its triple point, and the averaged value is within 0.4 \% of the provided values by \cite{Yaws2014}. For HC$_3$N, we report an average of the surface tensions predicted by all three methods (Equations~\ref{eq:parachor}-\ref{eq:pitzer}). For C$_4$N$_2$, because we do not have data or extrapolations we can use for its liquid density at the triple point, we report an average of the surface tensions predicted by only the latter two methods (Equations~\ref{eq:bbmethod}-\ref{eq:pitzer}).

We also calculated the surface tensions using all three methods for the rest of the 15 compounds and compared the results to the surface tension values provided by \cite{Yaws2014}, see Table~\ref{table:calc_SE_liq} in the Appendix section. For some compounds (e.g., C$_2$H$_6$, C$_2$H$_4$, and C$_6$H$_6$), the calculated surface tensions are very close to what is provided by \cite{Yaws2014} at the triple point (within 10 \%). But for some other compounds, especially triple-bonded substances such as C$_3$H$_4$-a, C$_3$H$_4$-p, and C$_2$N$_2$, the calculated surface tensions are up to 70 \% different compared to \cite{Yaws2014} at the triple point. Thus, the estimated surface tension values of C$_4$H$_2$, HC$_3$N, and C$_4$N$_2$ should be considered with caution. More laboratory experiments are needed to accurately determine the surface tensions of these three compounds. 

\begin{table}
\caption{Surface tension expressions ($\gamma_{l}(T)$) as a function of temperature of Titan-relevant organic liquids. The ``Temp. (K)" column has the same meaning as Table~\ref{table:SVP}. [a] \cite{Yaws2014}; [b] this work.}
\addtolength{\leftskip} {-1.5cm}
    \begin{tabular}{cccccccccc}
    \toprule
    Species & Surface tension, $\gamma_{l}(T)$ & Temp &  $\gamma_{l}(T_{tri})$  & Ref. & $\gamma_{l}^d(T_{tri})$ & $\gamma_{l}^p(T_{tri})$ & Ref.\\
     & (mN$\cdot$m$^{-1}$) & (K) & (mN$\cdot$m$^{-1}$) & & (mN$\cdot$m$^{-1}$) & (mN$\cdot$m$^{-1}$) & \\
    \hline 
    CH$_4$ & \(37.432(1-T/190.564)^{1.09200}\) & $T_{tri}$-$T_{cri}$ & 18.487 & [a] & 18.487 & 0 & [a]\\
    \hline 
    C$_2$H$_6$ & \(49.630(1-T/305.322)^{1.20650}\)& $T_{tri}$-$T_{cri}$ & 32.500 & [a] & 32.500 & 0 & [a]\\
    \hline 
    C$_2$H$_4$ & \(53.198(1-T/282.350)^{1.27840}\)& $T_{tri}$-$T_{cri}$ & 29.576 & [a] & 29.576 & 0 & [a]\\
    \hline 
    C$_2$H$_2$ & \(56.510(1-T/308.39)^{1.13460}\)& $T_{tri}$-$T_{cri}$ & 18.768 & [a] & 18.768 & 0 & [a]\\
    \hline 
    C$_4$H$_2$ & \(68.107(1-T/478.02)^{1.22222}\)& [$T_{tri}$]-[$T_{cri}$] & 29.388 & [a] & 29.388 & 0 & [a]\\
    \hline 
    C$_6$H$_6$ & \(71.970(1-T/562.02)^{1.23890}\)& $T_{tri}$-$T_{cri}$ & 30.806 & [a] & 30.806 & 0 & [a]\\
    \hline 
    C$_3$H$_8$ & \(49.179(1-T/369.89)^{1.22222}\)& $T_{tri}$-$T_{cri}$ & 35.681 & [a] & 35.681 & 0 & [a]\\
    \hline 
    C$_3$H$_6$ & \(53.547(1-T/364.21)^{1.20580}\)& $T_{tri}$-$T_{cri}$ & 38.387 & [a] & 38.387 & 0 & [a]\\
    \hline 
    C$_3$H$_4$-a & \(49.123(1-T/393.9)^{1.16200}\)& $T_{tri}$-$T_{cri}$ & 29.935 & [a] & 29.935 & 0 & [a]\\
    \hline 
    C$_3$H$_4$-p & \(57.277(1-T/402.7)^{1.18980}\)& $T_{tri}$-$T_{cri}$ & 29.845 & [a] & 29.845 & 0 & [a]\\
    \hline 
    HCN & \(52.310(1-T/456.65)^{1.01980}\)& $T_{tri}$-$T_{cri}$ & 22.169 & [a] & 14.5 & 7.7 & [b]\\
    \hline 
    HC$_3$N & 25.5$\pm$1.9 & $T_{tri}$ & 25.5$\pm$1.9 & [b] & 24.2 & 1.3$\pm$1.9 & [b]\\
    \hline 
    CO$_2$ & \(79.971(1-T/304.128)^{1.2617}\) & $T_{tri}$-$T_{cri}$ & 16.616 & [a] & 16.616 & 0 & [a]\\
    \hline 
    CH$_3$CN & \(68.332(1-T/545.41)^{1.09700}\) & $T_{tri}$-$T_{cri}$ & 37.557 & [a] & 23.9 & 13.7 & [b]\\
    \hline 
    C$_2$H$_5$CN & \(62.590(1-T/561.26)^{1.13640}\) & $T_{tri}$-$T_{cri}$ & 40.284 & [a] & 27.9 & 12.4 & [b]\\
    \hline 
    C$_2$H$_3$CN & \(63.112(1-T/540.0)^{1.07050}\)& $T_{tri}$-$T_{cri}$ & 39.718 & [a] & 27.5 & 12.2 & [b] \\
    \hline 
    C$_2$N$_2$ & \(73.193(1-T/397.1)^{1.20100}\)& $T_{tri}$-$T_{cri}$ & 23.064 & [a] & 23.064 & 0 & [a]\\
    \hline 
    C$_4$N$_2$ & 46.2$\pm$2.5 & $T_{tri}$ & 46.2$\pm$2.5 & [b] & 46.2$\pm$2.5 & 0 & [b]\\
    \toprule
    \end{tabular}
    \label{table:surface_tension}
\end{table}

The total surface tension can be divided into different components to reflect specific independent intermolecular forces \citep{1964I&EChem...56...40}. This is especially important when considering the interactions between two compounds, because only the surface tension components from the same intermolecular forces would interact with each other. The most commonly used model, the Owens, Wendt, Rabel, and Kaelble (OWRK) model \citep{1969JAPSci...11..1741, 1970JAdh...2...66,1971FL...77..997}, divides the surface tension into a dispersive ($\gamma_l^d$) and a polar component ($\gamma_l^p$): 
\begin{equation}
    \gamma_l=\gamma_l^d+\gamma_l^p.
\end{equation}
The dispersive component includes non-polar London dispersive forces between molecules, and the polar component includes polar dipole-dipole and H-bonding interactions. 

Most of the organic liquids on Titan are completely non-polar, thus, their polar components of the surface tensions ($\gamma_l^p$) are zero, and the dispersive components of the surface tensions are equal to their total surface tensions ($\gamma_l^d=\gamma_l^{tot}$). These non-polar substances include CH$_4$, C$_2$H$_6$, C$_2$H$_4$, C$_2$H$_2$, C$_4$H$_2$, C$_6$H$_6$, C$_3$H$_8$, C$_3$H$_6$, C$_3$H$_4$-a, C$_3$H$_4$-p, CO$_2$, C$_2$N$_2$, and C$_4$N$_2$. Some of the nitriles do have polar intermolecular force contributions, including HCN, CH$_3$CN, HC$_3$N, C$_2$H$_5$CN, and C$_2$H$_3$CN. Thus, they have non-zero surface tension polar components. To calculate their surface tension partitioning pattern, we need to know one of the surface tension compounds, and then the other component is the remainder of the total surface tension. Here, we estimate the dispersive surface tension components of these five organic liquids using the Lifshitz theory \citep{1992per...5..289L, 2011Israelachvili}, see also \cite{2021NatAs...5..822Y}:
\begin{equation}
\gamma_l^d=\frac{kT}{32\pi d_0^2}\left(\frac{\varepsilon-1}{\varepsilon+1}\right)^2+\frac{h\nu_{e}}{128\sqrt{2}\pi d_0^2}\frac{(n_{vis}^2-1)^2}{(n_{vis}^2+1)^{3/2}},
\label{eq:lifshitz}
\end{equation}
where $k$ is the Boltzmann constant, $T$ is the temperature of the system (here all the calculations are performed at the triple point temperature of each substance), $d_0$ is the equilibrium separation distance for two surfaces at contact and is found to be around 0.157$\pm$0.009 nm for most materials \citep{2006VanOss}, $\varepsilon$ is the static dielectric constants of the liquid, $h$ is Planck's constant, $\nu_e$ is the main electronic absorption frequency of the molecule, and $n_{vis}$ is the refractive index of the liquid at the visible wavelengths.

Table~\ref{table:eandn} lists all the required values ($\nu_e$, $\varepsilon$, and $n_{vis}$) for this calculation for the five organic liquids. The main electronic absorption frequency of the molecule ($\nu_e$, in Hz) can be calculated using $n_e=$ ionization potential$/h$, where the molecule's ionization potential is in eV and is $4.1357\times10^{-15}$ eV$\cdot$Hz$^{-1}$. For the organic liquids, we assume $n_{vis} = n_D$, where $n_D$ is the refractive index at the sodium D line (589 nm). We use $n_D$ because, for most materials, it is the only wavelength where measurements are available in the literature.

\begin{table}
\caption{The main electronic absorption frequency ($\nu_e$) and the ionization potential, static dielectric constant ($\varepsilon$), and refractive index at the visible wavelength ($n_{vis}$) for the Titan-relevant organic liquids and ices with polar intermolecular force contribution. [a] \cite{2023NIST}; [b] \cite{2020chcp.book.....L}; [c] \cite{1972JPChem...5..212K}; [d] \cite{1936JAChem..28..425S}, [e] this work, estimated, [f] \cite{Wurflinger1980}, [g] \cite{Wohlfarth1996}, [h] \cite{2022nist.292..686K}. $\dagger$: Note that for HCN, \cite{1936JAChem..28..425S} measured its dielectric constants up to its melting point. However, we discard the data above 100 K due to the presence of impurities in the sample, which causes a sharp increase in the measured dielectric constant near the melting point.}
\addtolength{\leftskip} {-1.9cm}
\resizebox{7.9in}{!}{
    \begin{tabular}{crccc}
    \toprule
    Species && Main electronic absorption & Ionization  & Ref.\\
     &&  frequency, $\nu_e$ ($10^{15}$ Hz) & potential (eV) & \\
    \hline 
    HCN && 3.288 & 13.60 & \multirow{5}{*}{[a]}  \\
    HC$_3$N && 2.810 & 11.62 &  \\
    CH$_3$CN && 2.950 & 12.20 &  \\
    C$_2$H$_5$CN && 2.865 & 11.85 &   \\
    C$_2$H$_3$CN && 2.638 & 10.91 &  \\
    \toprule
    Species && Static dielectric constant, $\varepsilon$ & Temp (K) & Ref.\\
    \hline 
    HCN (liquid) && $3733.1-23.18T+3.6963\times10^{-2}T^2$ & $T_{tri}$-299 & [b]  \\
    HCN (solid) && 2.33$^\dagger$ & 87 & [c, d]\\
    \hline 
    HC$_3$N (liquid) && $918.03-4.9149T+6.9104\times10^{-3}T^2$ & $T_{tri}$-314 & [b]\\
    HC$_3$N (solid) && 4 & N/A & [e]\\
    \hline 
    CH$_3$CN (liquid) && $297.24-1.5508T+2.2591\times10^{-3}T^2$ & 280-333 & [b] \\
    CH$_3$CN (solid) && 4 & 217-$T_{tri}$ & [f]\\
    \hline 
    C$_2$H$_5$CN (liquid) && $82.222-0.22937T+1.7424\times10^{-4}T^2$ & 213-473 & [b] \\
    C$_2$H$_5$CN (solid) && 4 & N/A & [e]\\
    \hline 
    C$_2$H$_3$CN (liquid) && $111.09-0.36806T+3.4879\times10^{-4}T^2$ & 233-413 & [b]\\
    C$_2$H$_3$CN (solid) && 4 & N/A & [e] \\
    \toprule
    Species && Refractive index at the visible wavelength, $n_{vis}$ & Temp (K)  & Ref.\\
    \hline
    HCN (liquid) && 1.2614 & 293 & [b]\\
    HCN (solid) && 1.3425 & $T_{tri}$ & [e]\\
    \hline
    HC$_3$N (liquid) && 1.38699 & 290 & [g]\\
    HC$_3$N (solid) && 1.5320 & $T_{tri}$ & [e]\\
    \hline
    CH$_3$CN (liquid) && $1.49138-5.0236\times10^{-4}T$ & 278-328 & [h]\\
    CH$_3$CN (solid) && 1.4612 & $T_{tri}$ & [e]\\
    \hline
    C$_2$H$_5$CN (liquid) && $1.5114-4.9622\times10^{-4}T$ & 292-327 & [h]\\
    C$_2$H$_5$CN (solid)  && 1.4719 & $T_{tri}$ & [e]\\
    \hline
    C$_2$H$_3$CN (liquid) && $1.5241-4.5398\times10^{-4}T$ & 293-308 & [h]\\
    C$_2$H$_3$CN (solid) && 1.5176 & $T_{tri}$ & [e]\\
    \toprule
    \end{tabular}}
    \label{table:eandn}
\end{table}

\subsection{Surface energies of Titan-relevant organic ices}\label{ice_SE}

For the organic ices, because there lack direct laboratory measurements of their solid surface energy ($\gamma_s$), we estimate their total surface energies by using some of their known properties through two empirical approaches \citep{1997P&SS...45..611G}. The first empirical approach uses the surface tension of the compound in the liquid phase ($\gamma_{l}(T_{tri})$), and its latent heats of sublimation ($L_{sub}(T_{tri})$) and vaporization ($L_{vap}(T_{tri})$) at the triple point temperature to compute the solid surface energy \citep{1978xxx..12..315C}: 

\begin{equation}
    \gamma_{s} = \left(\frac{L_{sub}(T_{tri})}{L_{vap}(T_{tri})}\right)^2\gamma_{l}(T_{tri}).
    \label{eqn:method1}
\end{equation}

The second method estimates the surface energy of the ice using the following equation \citep{1953JAtS...10..416M}:

\begin{equation}
    \gamma_{s} = \frac{1}{4}\frac{L_{sub}(T_{tri})}{N_A}\sigma_s - \left(\frac{1}{4}\frac{L_{vap}(T_{tri})}{N_A}\sigma_s - \gamma_{l}\right)\frac{L_{vap}(T_{tri})}{L_{sub}(T_{tri})},
    \label{eqn:method2}
\end{equation}
where $N_A$ is Avogadro's constant. In addition to $\gamma_{l}(T_{tri})$, $L_{sub}(T_{tri})$, and $L_{vap}(T_{tri})$, the second method also uses the surface density of molecules of the ice ($\sigma_s$), which is a function of the density ($\rho_s$) and molar mass ($M$) of the ice:
\begin{equation}
    \sigma_s = \frac{1}{d^2},\ where\ 
    d^3 = \frac{M}{N_A\rho_s}.
\end{equation}

We summarize the calculated total surface energy values of organic ices using both methods through Equation~\ref{eqn:method1} and \ref{eqn:method2} in Table~\ref{table:surface energy}. Similar to the organic liquids, the non-polar organic ices have zero polar surface energy components ($\gamma_s^p=0$) and their non-polar surface energy components are equal to their total surface energy ($\gamma_s^d=\gamma_s^{tot}$). For the organic ices with polar intermolecular contributions, we can also estimate their dispersive components of the surface energies using the Lifshitz theory (Equation~\ref{eq:lifshitz}) and their bulk properties (the static dielectric constants of ices and the refractive index of the ice at the visible wavelengths, see also Table~\ref{table:eandn}). The dielectric constants of HC$_3$N, C$_2$H$_5$CN, and C$_2$H$_3$CN ices have never been measured in the laboratory and future work is needed to determine the properties of these ices. Because the first term of Equation~\ref{eq:lifshitz}, where the dielectric constant $\varepsilon$ is used, contributes only $\sim$1 \% of the total dispersive surface energy, the exact selection of $\varepsilon$ is not so important. Here, we assume the dielectric constants of HC$_3$N, C$_2$H$_5$CN, and C$_2$H$_3$CN to be the same as CH$_3$CN, given their similar $n_{vis}$ and thus similar dielectric constants at high frequency $\varepsilon_\infty=n_{vis}^2$. 

There are some existing measurements on the refractive indices for these organic ices with polar intermolecular contributions \citep{2010ApJS..191...96M, 2019Icar..333..183N, 2020Icar..33813548H, 2022Icar..37314799Y}. However, most of the ices measured in those works include porosity to some extent (see discussion in Section 3.2), which will make the measured refractive indices lower than that for the ices without porosity. Here we choose to estimate the refractive index of all the polar ices using the Lorentz-Lorenz equation with the known density of the liquid phase (Table~\ref{table:density_liquid}), the known refractive index of the liquid phase (Table~\ref{table:eandn}), and the known intrinsic density of the ice phase (Table~\ref{table:density_solid})):
\begin{equation}
    \frac{n_s^2-1}{n_s^2+2}\frac{1}{\rho_s}=\frac{n_l^2-1}{n_l^2+2}\frac{1}{\rho_l},
\label{eq:lorentz}
\end{equation}
where $n_s$ and $n_l$ are the refractive indices of the material in the solid and the liquid phase, and $\rho_s$ and $\rho_l$ are the densities of the solid and the liquid, respectively. Note that the Lorentz-Lorenz equation was originally developed for nonpolar molecules but has been found to be applicable to polar molecules \citep{2021ApJ...906...81D}. 

\begin{table}[h]
\renewcommand{\arraystretch}{0.9}
\caption{Calculated solid surface energy ($\gamma_{s}$) of Titan-relevant organic ices using the two empirical approaches (Equations~\ref{eqn:method1} and \ref{eqn:method2}) and their associated surface energy partitioning components.}
\begin{tabular}{crcccrccc}
\toprule
    \centering
     Species && \multicolumn{3}{c}{Method 1 (Equation~\ref{eqn:method1})} && \multicolumn{3}{c}{Method 2 (Equation~\ref{eqn:method2})}\\
    \cline{3-5} \cline{7-9}
     && $\gamma_{s}$($T_{tri}$)  & $\gamma_s^d$($T_{tri}$) & $\gamma_s^p$($T_{tri}$)  && $\gamma_{s}$($T_{tri}$)  & $\gamma_s^d$($T_{tri}$) & $\gamma_s^p$($T_{tri}$)\\
     && (mJ$\cdot$m$^{-2}$) & (mJ$\cdot$m$^{-2}$) & (mJ$\cdot$m$^{-2}$)  && (mJ$\cdot$m$^{-2}$)  & (mJ$\cdot$m$^{-2}$) & (mJ$\cdot$m$^{-2}$)\\
    \hline 
    CH$_4$ && 22.676 & 22.676 & 0 && 21.820 & 21.820 & 0\\
    \hline 
    C$_2$H$_6$ && 42.245 & 42.245 & 0 && 39.715 & 39.715 & 0\\
    \hline 
    C$_2$H$_4$ && 43.638 & 43.638 & 0 && 40.734 & 40.734 & 0\\
    \hline 
    C$_2$H$_2$ && 29.135 & 29.135 & 0 && 35.194 & 35.194 & 0\\
    \hline 
    C$_4$H$_2$ && 65.981 & 65.981 & 0 && 65.979 & 65.979 & 0\\
    \hline 
    C$_6$H$_6$ && 51.088 & 51.088 & 0 && 53.503 & 53.503 & 0\\
    \hline 
    C$_3$H$_8$ && 42.765 & 42.765 & 0 && 41.776 & 41.776 & 0 \\
    \hline 
    C$_3$H$_6$ && 45.005 & 45.005 & 0 && 44.054 & 44.054 & 0\\
    \hline 
    C$_3$H$_4$-a && 39.679 & 39.679 & 0 && 41.427 & 41.427 & 0\\
    \hline 
    C$_3$H$_4$-p && 41.282 & 41.282 & 0 && 45.729 & 45.729 & 0 \\
    \hline 
    HCN && 37.202 & 21.6 & 15.6 && 61.707 & 21.6 & 40.1 \\
    \hline 
    HC$_3$N && 67.3$\pm$5.0 & 40.0 & 27.3$\pm$5.0 && 79.4$\pm$3.1 & 40.0 & 39.4$\pm$3.1\\
    \hline 
    CO$_2$ && 39.501 & 39.501 & 0 && 56.284 & 56.284 & 0\\
    \hline 
    CH$_3$CN && 57.268 & 32.9 & 24.4 && 67.543 & 32.9 & 34.6 \\
    \hline 
    C$_2$H$_5$CN && 50.462 & 33.2 & 17.3 && 55.252 & 33.2 & 22.1\\
    \hline 
    C$_2$H$_3$CN && 54.056 & 35.7 & 18.4 && 58.744 & 35.7 & 23.0\\
    \hline 
    C$_2$N$_2$ && 45.294 & 45.294 & 0 && 58.020 & 58.020 & 0\\
    \hline 
    C$_4$N$_2$ && 153$\pm$8 & 153$\pm$8 & 0 && 91.5$\pm$4.5 & 91.5$\pm$4.5 & 0\\
\toprule
    \end{tabular}
    \label{table:surface energy}
\end{table}

\subsection{Surface energies of Titan haze analogs}

Three laboratories have characterized the surface energies of their laboratory-made Titan haze analogs, including the JHU group (setup name: PHAZER), another NASA Ames Research Center group (NASA ARC 2, setup name: COSmIC), and the University of Northern Iowa group (UNI, setup name: Photochemical Aerosol Chamber, PAC). Two methods have been used to measure the surface energies of tholins: an indirect method called the sessile drop contact angle method (CA method), which probes the surface energy of the material through liquids with known surface tensions, and a direct method called the surface force apparatus (SFA method), which directly measures the cohesion force between the material to derive its surface energy. In Table~\ref{table:SE_tholin}, we list the measured surface energy values of the tholin samples from each setup, the measurement technique, and some of the experimental parameters associated with the production of each tholin sample (setup name, energy source, and gas mixture). The additional experimental parameters can be found in Table~\ref{table:additional_parameters}. In general, tholins have higher surface energies compared to common polymers \citep[20-50 mJ$\cdot$m$^{-2}$, e.g.,][]{1969JAPSci...11..1741}. In addition to the experimental conditions that make the tholin samples, the measuring environment is found to be important in determining the measured surface energy values of the samples \citep{2022PSJ.....3....2L}. For example, the surface properties of the samples may deviate from their true values if the measurements are performed in ambient air over an inert atmosphere/vacuum, where water vapor and other containment like hydrocarbons/oils adsorb on the surface of samples.  Here, the recommended surface energy values of tholins are taken from \cite{2022PSJ.....3....2L}, compiled using surface energy data of the four samples made with 5 \% CH$_4$/95 \% N$_2$, where the measurements were conducted in a dry nitrogen environment to avoid water adsorption and contamination from the ambient atmosphere.

\begin{table}
\caption{Measured surface energies ($\gamma_s$) of Titan haze analogs and the associated experimental setup with some experimental tholin production parameters (energy source and gas mixture composition) and the measuring environment (``air" stands for ambient air environment, and ``inert" stands for inert atmosphere such as dry nitrogen). $\dagger$: Note that the actual surface energy of tholin sample measured by surface force apparatus may vary between 50-100 mJ$\cdot$m$^{-2}$
\citep{2020ApJ...905...88Y}. References: [a] \cite{2020ApJ...905...88Y}; [b] \cite{2022PSJ.....3....2L}.}
\addtolength{\leftskip} {-2.2cm}
\resizebox{7.9in}{!}{
\begin{tabular}{crcrccccccc}
\toprule
Laboratory location && Energy source && Gas mixture & $\gamma_s^{tot}$ & $\gamma_s^d$ & $\gamma_s^p$ & Measurement  & Measurement & Ref.\\
(Setup) &&  &&  &  (mJ$\cdot$m$^{-2}$) & (mJ$\cdot$m$^{-2}$) & (mJ$\cdot$m$^{-2}$) & technique & environment & \\
\hline
 && \multirow{3}{*}{AC plasma discharge} && \multirow{3}{*}{N$_2$:CH$_4$ (95:5)} & 68.1$\pm$2.5 & 39.6$\pm$1.4 & 28.5$\pm$2.2 & \multirow{2}{*}{Contact angle} & air & [a]\\
JHU &&&&  & 73.3$\pm$1.3 & 37.3$\pm$1.5 & 36.0$\pm$1.5 &  & inert & [b]\\
(PHAZER) &&&& & 66$^\dagger$ & N/A & N/A & SFA & inert & [a]\\
\cline{3-11}
  && \multirow{2}{*}{UV lamp} && \multirow{2}{*}{N$_2$:CH$_4$ (95:5)} & 66.0$\pm$4.7 & 41.1$\pm$0.9 & 24.9$\pm$4.7 & \multirow{2}{*}{Contact angle} & air & [a]\\
&&  && & 47.0$\pm$2.9 & 31.8$\pm$1.8 & 15.2$\pm$2.9 &  & inert & [b]\\
\hline
NASA ARC 2 && \multirow{2}{*}{DC plasma discharge} && N$_2$:CH$_4$ (95:5) & 65.8$\pm$1.9 & 45.7$\pm$2.1 & 20.1$\pm$1.8 & \multirow{2}{*}{Contact angle} & inert & \multirow{2}{*}{[b]}\\
(COSmIC) && 
&& N$_2$:CH$_4$:C$_2$H$_2$ (94.5:5:0.5) & 53.4$\pm$5.2 & 44.4$\pm$1.0 & 9.0$\pm$5.1 & & inert & \\
\hline
UNI (PAC) && UV lamp && N$_2$:CH$_4$ (95:5) & 49.7$\pm$5.8 & 33.9$\pm$2.9 & 15.8$\pm$5.7 & Contact angle & inert & [b]\\
\hline
\multicolumn{5}{c}{\textbf{Recommended average surface energy, $\gamma_{avg}$}} & \textbf{59.0} & \textbf{37.2} & \textbf{21.8} \\
\multicolumn{5}{c}{\textbf{Recommended low-end surface energy, $\gamma_{low}$}} & \textbf{47.0} & \textbf{31.8} & \textbf{15.2} &&&\\
\multicolumn{5}{c}{\textbf{Recommended high-end surface energy, $\gamma_{high}$}} & \textbf{81.7} & \textbf{45.7} & \textbf{36.0} &&&\\
\toprule
\end{tabular}}
\label{table:SE_tholin}
\end{table}

\section{Conclusion}
In this work, we summarize and calculate the following important material properties of Titan-relevant organic liquids, ices, and Titan haze analogs:
\begin{itemize}
    \item The thermodynamic properties of the organic liquids and ices for the species detected on Titan, including triple point and critical point temperatures and pressures, crystalline ice phase transition temperatures, sublimation and vaporization saturation vapor pressures as functions of temperature, and sublimation and vaporization latent heats at the triple point of each species.
    \item Intrinsic densities for the Titan-relevant organic liquids and organic ices as functions of temperature, all previously measured intrinsic and average density data for the Titan-relevant organic liquids and organic ices, and all previously measured density values for Titan haze analogs, ``tholins". 
    \item The surface properties of Titan-relevant organic liquids and ices, and Titan haze analogs, including the surface tensions of organic liquids as functions of temperature, the surface energies of organic ices at the triple point temperature, and the surface energy of the haze analogs.
\end{itemize}
These fundamental material properties could not only be applicable to materials on Titan, but also to icy bodies in the solar system, ices in the interstellar medium, protoplanetary disks, and exoplanets. Other material properties such as the crystal structures of these compounds, the thermal diffusion coefficient of these species in relevant Titan-lake-like solvents, and parameters for clathrates would also be very useful for Titan's context (see for instance \cite{2021A&A...653A..80C}), which we would like to include in future works. Because the field is evolving rapidly with more laboratory measurements being conducted, this database will change. Thus, we have created a publicly available database website to serve as a repository for all the data presented in this study: \href{https://titanmaterials.sites.ucsc.edu/}{https://titanmaterials.sites.ucsc.edu/}. We will keep updating this database for the community, taking into account any new laboratory measurements in the future.

\section{Acknowledgements}
X.Y. and J. G., are supported by the 51 Pegasi b Postdoctoral Fellowship from the Heising-Simons Foundation. Y.Y and J. L. is supported by the summer research support offered by the Other Worlds Laboratory at UC Santa Cruz. X.Y. and J.L. are also supported by the NASA Cassini Data Analysis Program Grant 80NSSC21K0528. X.Y., E.S.O, and A.H. are supported by the NASA Cassini Data Analysis Program Grant 80NSSC22K0633. E.S.O acknowledges the support of the ``NASA Center for Optical Constants" ISFM. X.Z. is supported by NASA Solar System Workings grant 80NSSC19K0791. We thank the librarians at UC Santa Cruz for providing excellent and extensive interlibrary loan services that make the literature search in this work possible. We also thank Allen (Boshu) Qiao at UT San Antonio for performing data check for our public-accessible website.

\section{Data availability}
In addition to the public-accessible website (\href{https://titanmaterials.sites.ucsc.edu/}{https://titanmaterials.sites.ucsc.edu/}) we built to archive the database, we also archived all the data on Zenodo: \href{https://doi.org/10.5281/zenodo.7539342}{https://doi.org/10.5281/zenodo.7539342}. The tables that have only numbers are archived as .csv files. The tables that include functions are archived as Jupyter notebook files (.ipynb).

\section{Appendix}

\begin{table}[h]
    \addtolength{\leftskip} {-1.3cm} 
\caption{Additional saturation vapor pressure expressions that are evaluated in this work. References: [a] \cite{2009P&SS...57.2053F}; [b] \cite{2022nist.292..686K}; [c] \cite{1999book...2..121D}; [d] \cite{Yaws2015}; [e] \cite{2021PSJ.....2..121D}; [f] \cite{2014Thermo..581.1399R}; [g] \cite{2022PSJ.....3..120H}; [h] this work; [i] \cite{1926jap..1.59c}; [j] \cite{1963JChPh..38...69D}; [k] \cite{1989IJT....10..739O}; [l] \cite{1964zh...865..1989}; [m] \cite{1947stull..32..M}. Notations are the same as Table~\ref{table:SVP}.} 
\resizebox{7.5in}{!}{
\begin{threeparttable}
    \begin{tabular}{ccccc}
    \toprule
    Species & Type & Saturation vapor pressure expressions $P_{sat}$ (bar) & Temp. (K) & Ref.\\
    \hline 
    CH$_4$ & s-g & \(e^{10.51-1110/T-4341/T^2+1.035\times10^5/T^3-7.91\times10^5/T^4}\) &
    [20.509]-48.15-90.686 ($T_{tri}$) & [a]\\
     & s-g & \(e^{7.134-1109.5/T+0.65607\mathrm{ln}(T)}\) &
    [20.509]-49.55-90.686 ($T_{tri}$) & [b]\\
    & l-g & \(10^{3.7687 - 395.744/(T-6.469)}\) & [90.686]-92-115 & [c]\\
    & l-g & \(10^{3.7687 - 395.744/(T-6.469)+0.43429[(T-115)/T_{cri}]^{2.51347}-6.0941[(T-115)/T_{cri}]^8+369.43[(T-115)/T_{cri}]^{12}}\) & 115.0-190.564 ($T_{cri}$) & [c]\\       \hline 
    C$_2$H$_6$ & s-g & \(e^{15.11-2207/T-2.411\times10^4/T^2+7.744\times10^5/T^3-1.161\times10^7/T^4+6.763\times10^7/T^5}\) & [$<$]83.35-88.75-[90.3560] ($T_{tri}$) & [a] \\
     & s-g & \(e^{18.7925-2706.8/T}\) & [$<$]80.808-89.726-[90.3560] ($T_{tri}$) & [b] \\
    & l-g & \(10^{4.0567 - 687.3/(T-14.46)}\) & 90.35-133 & [c]\\
    & l-g & \(10^{3.95405 - 663.72/(T-16.469)}\) & 133-190 & [c]\\
    & l-g & \(10^{3.95405 - 663.72/(T-16.469)+0.43429[(T-190)/T_{cri}]^{2.79768}-55.054[(T-190)/T_{cri}]^8+2992.1[(T-190)/T_{cri}]^{12}}\) & 190-305.322 ($T_{cri}$)& [c]\\    \hline 
    C$_2$H$_4$ & s-g & \(e^{14.752-2305.9/T+0.15361\mathrm{ln}(T)}\) & [$<$] 77.5-103.966 ($T_{tri}$) & [b] \\
    & l-g & \(10^{4.0936-649.806/(T-10.42)}\) & 103.966-282.350 ($T_{cri}$) & [d]\\
    \hline 
    C$_2$H$_2$ & s-g & \(10^{6.08413-1148.97/(T-0.31)}\) & [$<$]138.0-191.66 ($T_{tri}$) & [c]\\
     & s-g & \(e^{60.618-3951.8/T-7.574\mathrm{ln}(T)}\) & [$<$]133.0-191.66 ($T_{tri}$) & [b]\\
     & l-g & \(10^{3.67374 - 528.67/(T-44.36)}\) & 191.66-211-[308.39] ($T_{cri}$) & [c]\\
    \hline 
    C$_4$H$_2$ & s-g & \(10^{8.22178-2707.52/(T+59.54)}\) & [$<$]188.0-237.7 ($T_{tri}$) & [c]\\
     & s-g & \(e^{16.074-4308.6/T}\) & [$<$]194.045-237.7 ($T_{tri}$) & [b]\\
    & l-g & \(10^{3.33975-707.254/(T-71.198)}\) & 237.7-285-[410] ($T_{cri}$) & [c]\\
    & l-g & \(10^{3.6154-754.853/(T-74.882)}\) & 237.7-303.0-[410] ($T_{cri}$) & [d]\\
    \hline 
    C$_6$H$_6$ & s-g & \(e^{17.35-5663/T}\) & [$<$]184.3-278.6877 ($T_{tri}$) & [a]\\
    & s-g & \(e^{60.797-7273.7/T-6.7019\mathrm{ln}(T)}\) & [$<$]183.012-278.6877 ($T_{tri}$) & [b]\\
    & s-g & \(10^{6.015-2042/T}\) & 146.0-157.6-[278.6877] ($T_{tri}$) & [e]\\
    & s-g & \(10^{6.345-2150/T}\) & [$<$]134.8-146 & [e]\\
    & s-g & \(0.04785e^{(1-278.674/T)exp(3.064696-1.907097\times10^{-4}T-5.468886\times10^{-7}T^2)}\) & [$<$]150-278.6877 ($T_{tri}$) & [f]\\
    & s-g & \(e^{18.98-5978.2/T}\) & [$<$]134.8-157.6-[278.6877] ($T_{tri}$) & [g]\\
    & l-g & \(10^{3.9392-1090.43/(T-76.004)}\) & 278.6877-376.0-[562.02] & [d]\\
    \hline 
    C$_3$H$_8$ & l-g & \(e^{6.673-574.5/T-2.96\times10^5/T^2+1.795\times10^7/T^3-4.463\times10^8/T^4}\) & 85.400-231.05-[369.89] ($T_{cri}$) & [h]\\
    & l-g & \(10^{4.6956-1030.7/(T-7.79)}\) & 85.400-167 & [c]\\
    & l-g & \(10^{3.92828-803.997/(T-26.108)}\) & 167-237 & [c]\\
    & l-g & \(10^{3.92828-803.997/(T-26.108)+0.43429[(T-237)/T_{cri}]^{2.55753}+50.655[(T-237)/T_{cri}]^8-1408.9[(T-237)/T_{cri}]^{12}}\) & 237-369.89 ($T_{cri}$)& [c]\\ 
    \hline 
    C$_3$H$_6$ & l-g & \(e^{6.527-461.46/T-3.022\times10^5/T^2+1.823\times10^7/T^3-4.593\times10^8/T^4}\) & 87.850-225.4-[364.21] ($T_{cri}$) & [h]\\
    & l-g & \(10^{4.48447-934.227/(T-14)}\) & [87.850]-89-163 & [c]\\
    & l-g & \(10^{3.95606-789.624/(T-25.57)}\) & 163-240 & [c]\\
    & l-g & \(10^{3.95606-789.624/(T-25.57)+0.43429[(T-240)/T_{cri}]^{2.67417}+22.1292[(T-240)/T_{cri}]^8-199.34[(T-240)/T_{cri}]^{12}}\) & 240-364.21 ($T_{cri}$)& [c]\\ 
    \hline 
    C$_3$H$_4$-a & s-g & \(e^{16.026-3393.3/T+0.041846\mathrm{ln}(T)}\) & [$<$]127-136.70 ($T_{tri}$) & [b]\\
     & l-g & \(10^{4.3495 - 971.15/(T - 16.5)}\) & 136.70-174 & [c]\\
     & l-g & \(10^{3.6752 - 734.57/(T - 38.41)}\) & 174-253 & [c]\\
    & l-g & \(10^{3.6752 - 734.57/(T - 38.41)+0.43429[(T-253)/T_{cri}]^{1.136}-265[(T-253)/T_{cri}]^8+1.6325\times10^4[(T-253)/T_{cri}]^{12}}\) & 253-393.9 ($T_{cri}$) & [c]\\    
     & l-g & \(10^{3.6695-674.848/(T-55.808)}\) & [136.70]-177.5-393.9 ($T_{cri}$) & [d]\\
    \hline 
    C$_3$H$_4$-p & l-g & \(10^{4.24555 - 935.09/(T-29.57)}\) & [169.87]-175-276-402.7 ($T_{cri}$) & [c]\\
    \hline 
    HCN & s-g & \(e^{13.93-3624/T-1.325\times10^5/T^2+6.314\times19^6/T^3-1.128\times10^8/T^4}\) & [$<$]235.2-259.871 ($T_{tri}$) & [a]\\
    & l-g & \(10^{4.67417-1340.79/(T-11.592)}\) & 259.871-319.4-[456.65] ($T_{cri}$) & [i]\\
    & l-g & \(49.7e^{T_{cri}/T[-9.2183(1-T/T_{cri})+4.846(1-T/T_{cri})^{1.5}-5.2(1-T/T_{cri})^{2.5}+4.1085(1-T/T_{cri})^5]}\) & 259.871-456.65 ($T_{cri}$) & [b]\\
    & l-g & \(10^{5.1309-1602.88/(T+13.743)}\) & 259.871-456.65 ($T_{cri}$) & [d]\\
    \hline 
    HC$_3$N & s-g & \(e^{13.01 - 4426/T}\) & [50.0]-164.9-201.5-[280] ($T_{tri}$) & [a]\\
    & s-g & \(10^{7.30 - 2210/T}\) & [$<$]247-280 ($T_{tri}$) & [j]\\
     & s-g & \(52e^{T_{cri}/T[-7.3158(1-T/T_{cri})+1.7462(1-T/T_{cri})^{1.5}-2.2814(1-T/T_{cri})^{2.5}-2.7759(1-T/T_{cri})^5]}\) & [$<$]160-280 ($T_{tri}$) & [b]\\
     & l-g & \(10^{3.3419-712.924/(T-102.005)}\) & 280-336.89-527 ($T_{tri}$) & [d]\\
     & l-g & \(52e^{T_{cri}/T[-7.3158(1-T/T_{cri})+1.7462(1-T/T_{cri})^{1.5}-2.2814(1-T/T_{cri})^{2.5}-2.7759(1-T/T_{cri})^5]}\) & 280-527 ($T_{cri}$) & [b] \\
    \hline 
    CO$_2$ & s-g & \(e^{14.76 -2571/T -7.781\times10^4/T^2 + 4.325\times10^6/T^3 - 1.207\times10^8/T^4 + 1.350\times10^9/T^5}\) & [$<$]69.69-194.7-[216.5890] ($T_{tri}$)& [a] \\
     & l-g & \(10^{4.6571 - 836.06/(T-4.927)}\) & 216.5890-304.128 ($T_{cri}$) & [k] \\
    \hline 
     CH$_3$CN & l-g & \(48.66e^{T_{cri}/T[-8.027(1-T/T_{cri})+1.9887(1-T/T_{cri})^{1.5}-1.5119(1-T/T_{cri})^{2.5}-2.5934(1-T/T_{cri})^5]}\) & 229.349-545.41 ($T_{cri}$) & [b]\\
    \hline 
    C$_2$H$_5$CN  & l-g & \(10^{4.1883-1275.05/(T-65.154)}\) & [180.4]-275.44-561.26 ($T_{cri}$)  & [d] \\
    \hline 
    C$_2$H$_3$CN & l-g & \(10^{4.06661-1255.94/(T-41.853)}\) & [189.642]-222-351-[540.0] ($T_{cri}$) & [l]\\
    & l-g & \(46.5e^{T_{cri}/T[-8.2647(1-T/T_{cri})+3.119(1-T/T_{cri})^{1.5}-2.962(1-T/T_{cri})^{2.5}-2.3068(1-T/T_{cri}]^5}\) & [189.642]-222-540.0 ($T_{cri}$) & [b]\\
    \hline 
    C$_2$N$_2$ & s-g & \(e^{26.335-4336.5/T-1.628\mathrm{ln}(T)}\) & [$<$]179.979-245.29 ($T_{tri}$) & [b]\\
    & l-g & \(10^{4.51661-1041.52/(T-21.288)}\) & 245.29-397.1 ($T_{tri}$) & [m]\\
    & l-g & \(10^{4.0691-779.237/(T-60.228)}\) & 245.29-397.1 ($T_{tri}$) & [d]\\
    \toprule
    \end{tabular}
    \label{table:SVP_app}
    \end{threeparttable}}
\end{table}

\begin{figure}
    \centering
    \includegraphics[width=\textwidth]{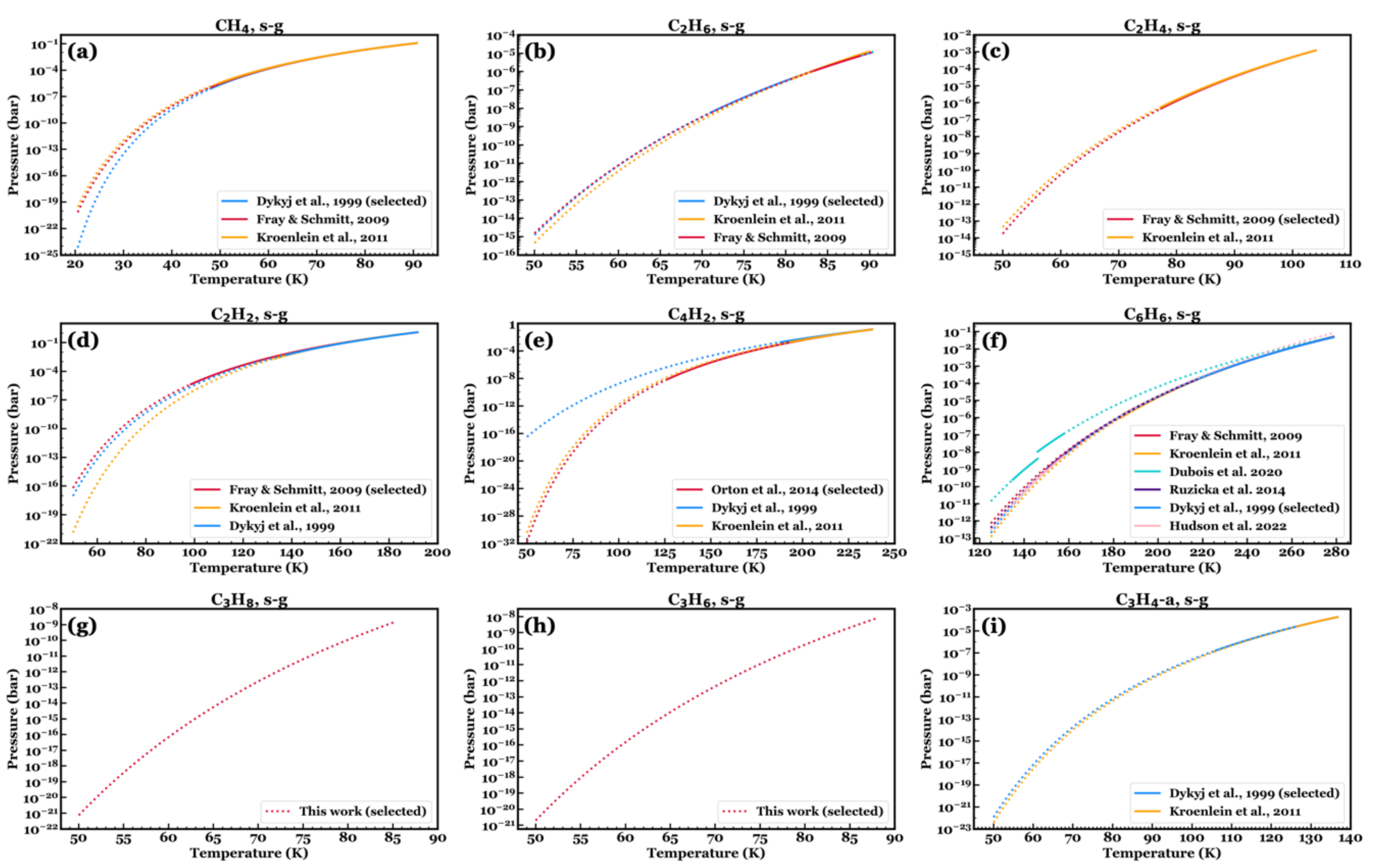}
    \caption{Comparison of the selected solid-gas (s-g) saturation vapor pressure (SVP) functions in Table~\ref{table:SVP} to the other evaluated s-g SVP functions in Table~\ref{table:SVP_app}, for the first half of the species in Table~\ref{table:abundance}. The solid lines represent the applicable temperature range (or the temperature without brackets in Table~\ref{table:SVP} and \ref{table:SVP_app}). The different SVP functions agree relatively well for the applicable temperature range. The dotted lines represent the extrapolated temperature range for each SVP function (or the temperature without brackets in Table~\ref{table:SVP} and \ref{table:SVP_app}). The different SVP functions start to deviate from each other for the extrapolated temperature range. Note that for benzene, we plot the unadjusted low-temperature s-g SVP expressions in  \cite{2021PSJ.....2..121D}, the adjusted s-g SVP expressions of \cite{2021PSJ.....2..121D} in \cite{2022PSJ.....3..120H} are similar to the independently measured SVP expressions in \cite{2022PSJ.....3..120H}.} 
    \label{fig:s-g_comparsion1}
\end{figure}

\begin{figure}
    \centering
    \includegraphics[width=\textwidth]{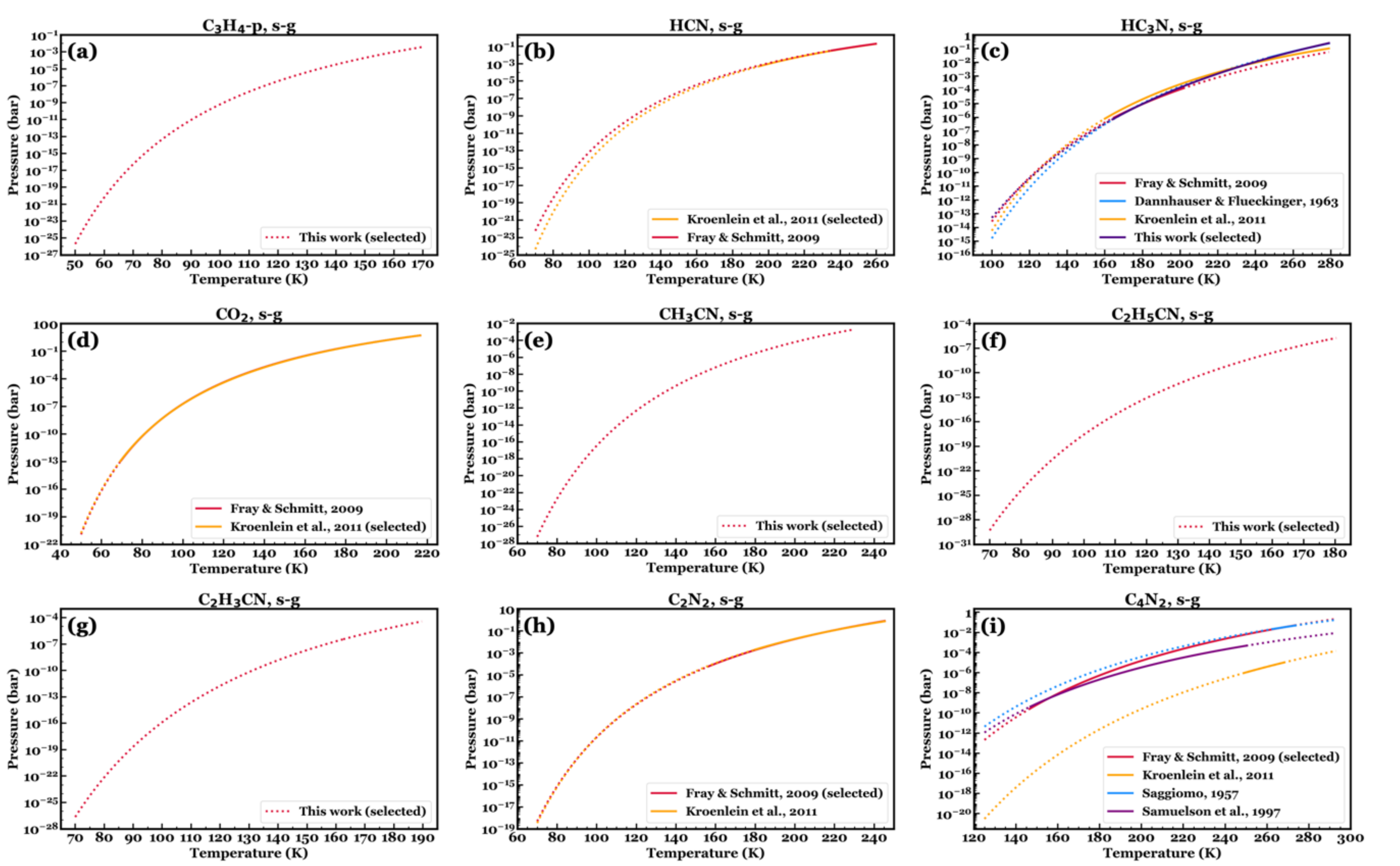}
    \caption{Comparison of the selected solid-gas (s-g) saturation vapor pressure (SVP) functions in Table~\ref{table:SVP} to the other evaluated s-g SVP functions in Table~\ref{table:SVP_app}, for the second half of the species in Table~\ref{table:abundance}. Notations are the same as Figure~\ref{fig:s-g_comparsion1}. The different SVP functions agree relatively well for the applicable temperature range, except for C$_4$N$_2$, which has one outlier s-g SVP. Similar to Figure~\ref{fig:s-g_comparsion1}, the different SVP functions start to deviate from each other for the extrapolated temperature range.} 
    \label{fig:s-g_comparsion2}
\end{figure}

\begin{figure}
    \centering
    \includegraphics[width=\textwidth]{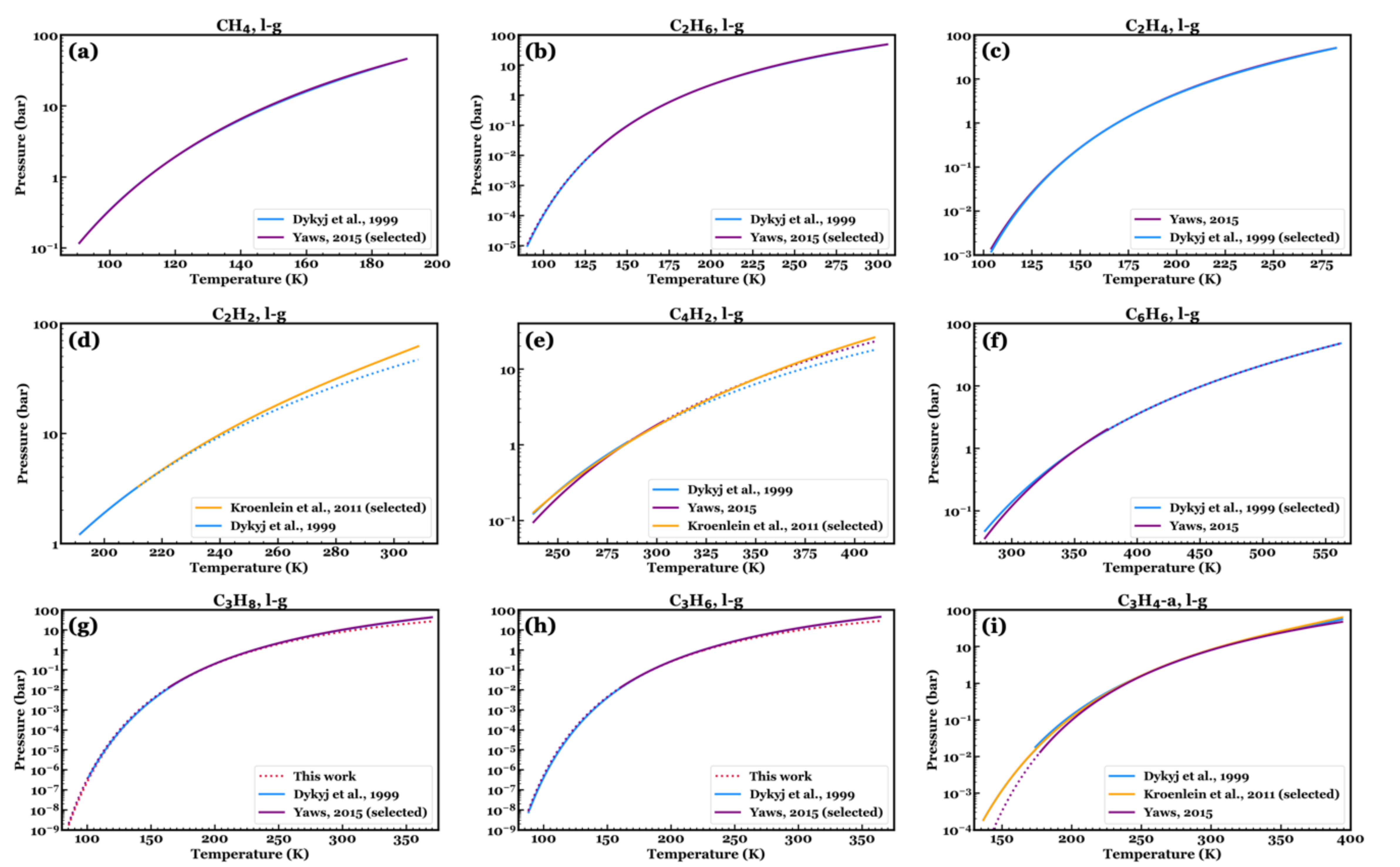}
    \caption{Comparison of the selected liquid-gas (l-g) saturation vapor pressure (SVP) functions in Table~\ref{table:SVP} to the other evaluated l-g SVP functions in Table~\ref{table:SVP_app}, for the first half of the species in Table~\ref{table:abundance}. Notations are the same as Figure~\ref{fig:s-g_comparsion1}. Similar to Figure~\ref{fig:s-g_comparsion1}, the different SVP functions agree relatively well for the applicable temperature range, and they start to deviate from each other for the extrapolated temperature range.} 
    \label{fig:l-g_comparsion1}
\end{figure}

\begin{figure}
    \centering
    \includegraphics[width=\textwidth]{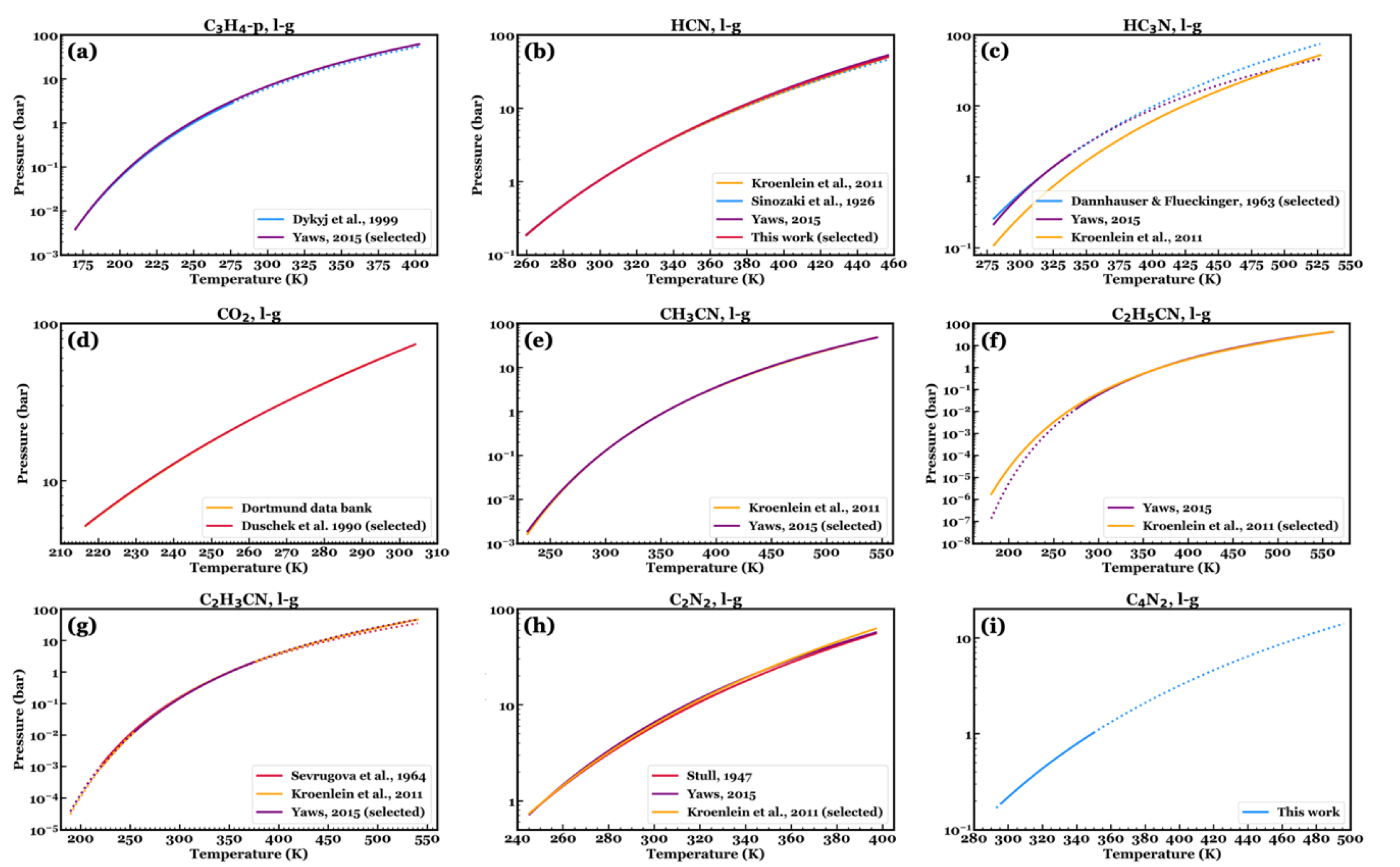}
    \caption{Comparison of the selected liquid-gas (l-g) saturation vapor pressure (SVP) functions in Table~\ref{table:SVP} to the other evaluated l-g SVP functions in Table~\ref{table:SVP_app}, for the second half of the species in Table~\ref{table:abundance}. Notations are the same as Figure~\ref{fig:s-g_comparsion1}. The different SVP functions agree relatively well for the applicable temperature range, except HC$_3$N, which has one outlier l-g SVP. Similar to Figure~\ref{fig:s-g_comparsion1}, the different SVP functions start to deviate from each other for the extrapolated temperature range.} 
    \label{fig:l-g_comparsion2}
\end{figure}

\clearpage

\setlength\LTleft{-40pt}    
\setlength\LTright{-40pt}           
\setlength\LTcapwidth{\linewidth}
\begin{longtable}{ccccc}
  \caption{Organic liquid density data, references, adopted density using expressions in Table~\ref{table:density_liquid}, and the deviation of the fitting to the experimental data in percentage, defined as the absolute difference in percentage between the experimental data and the model fit.}\\
    \endfirsthead
    \caption{Organic liquid density data (cont.)} \\
\hline
    \endhead
\multicolumn{5}{r}{Continued on the next page}
    \endfoot
    \endlastfoot
\toprule
\textbf{CH$_4$} & Temp & Density & Adopted & Deviation \\
& (K) & (kg$\cdot$m$^{-3}$) & density (kg$\cdot$m$^{-3}$) & (\%)\\
\hline
\cite{2020chcp.book.....L} & 90.6941 & 451.5 & 451.6 & 0.02\\
& 100.0 & 438.9 & 438.8 & 0.01\\
& 110.0 & 424.8 & 424.7 & 0.03\\
& 120.0 & 409.9 & 409.9 & 0.01\\
& 130.0 & 394.0 & 394.1 & 0.02\\
& 140.0 & 376.9 & 377.0 & 0.02\\
& 150.0 & 357.9 & 358.0 & 0.02\\
& 160.0 & 336.3 & 336.3 & 0.01\\
& 170.0 & 310.5 & 310.4 & 0.04\\
& 180.0 & 276.2 & 276.1 & 0.05\\
& 190.0 & 200.8 & 201.1 & 0.13\\
& 190.564 & 162.7 & 162.7 & 0.00\\
\toprule
\textbf{C$_2$H$_6$} & Temp & Density & Adopted & Deviation \\
& (K) & (kg$\cdot$m$^{-3}$) & density (kg$\cdot$m$^{-3}$) & (\%)\\
\hline
\cite{2020chcp.book.....L} & 90.368 & 651.5 & 651.5 & 0.01\\
& 100.0 & 640.9 & 640.9 & 0.00\\
& 120.0 & 618.9 & 618.8 & 0.01\\
& 140.0 & 596.6 & 596.5 & 0.02\\
& 160.0 & 573.6 & 573.4 & 0.01\\
& 180.0 & 549.5 & 549.6 & 0.01\\
& 200.0 & 524.0 & 524.1 & 0.02\\
& 220.0 & 496.3 & 496.4 & 0.02\\
& 240.0 & 465.3 & 465.3 & 0.01\\
& 260.0 & 429.1 & 429.0 & 0.03\\
& 280.0 & 382.7 & 382.5 & 0.04\\
& 300.0 & 303.5 & 303.7 & 0.07\\
& 305.322 & 206.2 & 206.2 & 0.00\\
\toprule
\textbf{C$_2$H$_4$} & Temp & Density & Adopted & Deviation \\
& (K) & (kg$\cdot$m$^{-3}$) & density (kg$\cdot$m$^{-3}$) & (\%)\\
\hline
\cite{2020chcp.book.....L} & 103.986 & 654.6 & 654.7 & 0.02\\
& 120.0 & 634.2 & 634.1 & 0.02\\
& 140.0 & 608.0 & 607.9 & 0.02\\
& 160.0 & 580.9 & 580.9 & 0.01\\
& 180.0 & 552.2 & 552.3 & 0.02\\
& 200.0 & 521.2 & 521.4 & 0.03\\
& 220.0 & 486.7 & 486.7 & 0.01\\
& 240.0 & 446.1 & 446.0 & 0.02\\
& 260.0 & 393.5 & 393.3 & 0.06\\
& 280.0 & 290.7 & 291.0 & 0.09\\
& 282.350 & 214.2 & 214.2 & 0.00\\
\toprule
\textbf{C$_4$H$_2$} & Temp & Density & Adopted & Deviation \\
& (K) & (kg$\cdot$m$^{-3}$) & density (kg$\cdot$m$^{-3}$) & (\%)\\
\hline
\cite{JPhchem...57.2053F} & 236.75 & 927.1 & 926.9 & 0.02\\
\hline
\cite{1926JCh..36..829W} & 273.15 & 736.4 & 736.5 & 0.02\\
\hline
\cite{1925Hebd...5..289L} & 278.15 & 710.7 & 710.4 & 0.05\\
\toprule
\textbf{C$_6$H$_6$} & Temp & Density & Adopted & Deviation \\
& (K) & (kg$\cdot$m$^{-3}$) & density (kg$\cdot$m$^{-3}$) & (\%)\\
\hline
\cite{2020chcp.book.....L} & 278.674 & 894.0 & 894.4 & 0.04\\
& 280.0 & 892.6 & 892.9 & 0.04\\
& 300.0 & 871.6 & 871.4 & 0.02\\
& 320.0 & 850.4 & 849.8 & 0.07\\
& 340.0 & 828.6 & 828.0 & 0.07\\
& 360.0 & 806.2 & 805.8 & 0.04\\
& 380.0 & 783.1 & 783.0 & 0.01\\
& 400.0 & 758.9 & 759.2 & 0.04\\
& 420.0 & 733.6 & 734.2 & 0.08\\
& 440.0 & 706.7 & 707.5 & 0.11\\
& 460.0 & 677.8 & 678.4 & 0.09\\
& 480.0 & 646.1 & 646.3 & 0.03\\
& 500.0 & 610.1 & 609.6 & 0.09\\
& 520.0 & 567.0 & 565.7 & 0.24\\
& 540.0 & 508.9 & 507.6 & 0.25\\
& 560.0 & 385.7 & 388.9 & 0.84\\
& 562.02 & 304.7 & 304.7 & 0.00\\
\toprule
\textbf{C$_3$H$_8$} & Temp & Density & Adopted & Deviation \\
& (K) & (kg$\cdot$m$^{-3}$) & density (kg$\cdot$m$^{-3}$) & (\%)\\
\hline
\cite{2020chcp.book.....L} & 85.5& 733.1 & 733.1 & 0.00\\
& 100.0 & 718.1 & 718.2 & 0.01\\
& 120.0 & 697.8 & 697.8 & 0.00\\
& 140.0 & 677.6 & 677.6 & 0.01\\
& 160.0 & 657.3 & 657.2 & 0.02\\
& 180.0 & 636.6 & 636.5 & 0.01\\
& 200.0 & 615.4 & 615.4 & 0.00\\
& 220.0 & 593.4 & 593.5 & 0.11\\
& 240.0 & 570.4 & 570.5 & 0.01\\
& 260.0 & 545.8 & 545.9 & 0.02\\
& 280.0 & 519.2 & 519.2 & 0.00\\
& 300.0 & 489.4 & 488.6 & 0.00\\
& 320.0 & 454.9 & 454.8 & 0.03\\
& 340.0 & 411.8 & 411.6 & 0.05\\
& 360.0 & 345.6 & 345.8 & 0.07\\
& 369.89 & 220.5 & 220.5 & 0.00\\
\toprule
\textbf{C$_3$H$_6$} & Temp & Density & Adopted & Deviation \\
& (K) & (kg$\cdot$m$^{-3}$) & density (kg$\cdot$m$^{-3}$) & (\%)\\
\hline
\cite{2020chcp.book.....L} & 87.9 & 768.1 & 768.0 & 0.02\\
& 100.0 & 754.0 & 754.1 & 0.01\\
& 120.0 & 731.2 & 731.3 & 0.02\\
& 140.0 & 708.8 & 708.9 & 0.01\\
& 160.0 & 686.4 & 686.4 & 0.00\\
& 180.0 & 663.9 & 663.8 & 0.02\\
& 200.0 & 640.8 & 640.7 & 0.01\\
& 220.0 & 616.9 & 616.8 & 0.01\\
& 240.0 & 591.8 & 591.8 & 0.00\\
& 260.0 & 565.0 & 565.1 & 0.01\\
& 280.0 & 535.8 & 535.9 & 0.02\\
& 300.0 & 503.1 & 503.1 & 0.00\\
& 320.0 & 464.5 & 464.5 & 0.01\\
& 340.0 & 414.7 & 414.7 & 0.01\\
& 360.0 & 325.7 & 325.7 & 0.00\\
& 364.21 & 229.6 & 229.6 & 0.00\\
\toprule
\textbf{C$_3$H$_4$-p} & Temp & Density & Adopted & Deviation \\
& (K) & (kg$\cdot$m$^{-3}$) & density (kg$\cdot$m$^{-3}$) & (\%)\\
\hline
\cite{2022nist.292..686K} & 273 & 650 & 644.5 & 0.85 \\
& 276 & 650 & 640.6 & 1.45 \\ 
& 282 & 637 & 632.7 & 0.67 \\ 
& 288 & 628 & 624.6 & 0.54 \\ 
& 294 & 618 & 616.2 & 0.28 \\ 
& 300 & 608 & 607.6 & 0.07 \\ 
& 306 & 599 & 598.6 & 0.06 \\ 
& 312 & 589 & 589.3 & 0.06 \\ 
& 318 & 578 & 579.6 & 0.28 \\ 
& 324 & 568 & 569.5 & 0.26\\ 
& 330 & 557 & 558.9 & 0.34\\ 
& 336 & 546 & 547.7 & 0.31\\ 
& 342 & 534 & 535.9 & 0.36\\ 
& 348 & 522 & 523.4 & 0.27\\ 
& 354 & 509 & 510.1 & 0.21\\ 
& 360 & 495 & 495.8 & 0.16\\ 
& 366 & 480 & 480.3 & 0.06\\ 
& 372 & 464 & 463.4 & 0.14\\ 
& 378 & 446 & 444.5 & 0.34\\ 
& 384 & 425 & 423.0 & 0.48\\ 
& 390 & 400 & 397.3 & 0.68\\ 
& 396 & 365 & 363.5 & 0.40\\ 
& 402 & 290 & 296.4 & 2.21\\ 
& 402.7 & 244.90 & 244.9 & 0.00\\ 
\cite{2020chcp.book.....L} & 298.15 & 607.0 & 610.3 & 0.54\\
\cite{1986Chem...351.8G} & 246.0 & 675.0 & 677.4 & 0.36\\
& 217.0 & 710.0 & 710.8 & 0.11\\
& 185.0 & 749.0 & 747.4 & 0.21\\
\cite{1921JAmC...32..116M} & 217.85 & 712.6 & 709.8 & 0.39\\
& 221.15 & 709.1 & 706.1 & 0.43\\
& 222.3 & 707.4 & 704.8 & 0.37\\
& 232.8 & 694.1 & 692.8 & 0.19\\
& 240.85 & 684.1 & 683.5 & 0.09\\
& 247.85 & 675.9 & 675.3 & 0.09\\
& 253.85 & 668.0 & 668.1 & 0.02\\
& 260.25 & 659.8 & 660.4 & 0.09\\
\cite{1931CJRes...5..306M} & 216.15 & 710.9 & 711.8 & 0.12\\
& 220.05 & 706.5 & 707.3 & 0.12\\
& 221.25 & 703.9 & 706.0 & 0.29\\
& 225.75 & 699.2 & 700.8 & 0.23\\
& 230.75 & 693.6 & 695.1 & 0.22\\
& 232.25 & 691.1 & 693.4 & 0.33\\
& 236.35 & 687.3 & 688.7 & 0.20\\
& 240.15 & 681.4 & 684.3 & 0.42\\
& 240.65 & 682.7 & 683.7 & 0.15\\
& 246.65 & 675.9 & 676.7 & 0.12\\
& 248.95 & 671.1 & 674.0 & 0.43\\
& 250.45 & 671.0 & 672.2 & 0.18\\
& 255.65 & 665.2 & 666.0 & 0.12\\
& 258.75 & 661.7 & 662.2 & 0.08\\
& 260.45 & 658.2 & 660.1 & 0.30\\
& 272.65 & 644.7 & 644.9 & 0.03\\
\toprule
\textbf{CO$_2$} & Temp & Density & Adopted & Deviation \\
& (K) & (kg$\cdot$m$^{-3}$) & density (kg$\cdot$m$^{-3}$) & (\%)\\
\hline
\cite{2020chcp.book.....L} & 216.6 ($T_{tri}$) & 1178.0 & 1178.2 & 0.01\\
& 220.0 & 1166.0 & 1165.9 & 0.01\\
& 230.0 & 1129.0 & 1128.7 & 0.03\\
& 240.0 & 1089.0 & 1089.0 & 0.00\\
& 250.0 & 1046.0 & 1046.2 & 0.02\\
& 260.0 & 998.9 & 999.0 & 0.01\\
& 270.0 & 945.8 & 945.8 & 0.00\\
& 280.0 & 883.6 & 883.5 & 0.01\\
& 290.0 & 804.7 & 804.5 & 0.02\\
& 300.0 & 679.2 & 679.4 & 0.03\\
& 304.128 & 467.8 & 467.6 & 0.00\\
\toprule
\label{table:liquid_density_data}
\end{longtable}

\clearpage

\setlength\LTleft{-40pt}    
\setlength\LTright{-40pt}           
\setlength\LTcapwidth{\linewidth}
\begin{longtable}{cccccc}
  \caption{Organic ice density data that are used for the intrinsic density expression fittings in Table~\ref{table:density_solid} (including diffraction and pycnometry data), references, adopted intrinsic density fits using expressions in Table~\ref{table:density_solid}, and the deviation of the fitting to the experimental data in percentage. All the ices measured in this table are in their crystalline form.}\\
    \endfirsthead
    \caption{Organic ice density data (cont.)} \\
\hline
    \endhead
\multicolumn{6}{r}{Continued on the next page}
    \endfoot
    \endlastfoot
\toprule
\textbf{CH$_4$} & Temp & Density & Technique & Adopted & Deviation\\
 & (K) & (kg$\cdot$m$^{-3}$) &  & density (kg$\cdot$m$^{-3}$) & (\%)\\
\hline
\cite{1930PhChem.147..466H} & 20.4 & 522.5 & Pycnometry & 528.1 & 1.08\\
\cite{1931Natur.127R.707M} & 20.5 & 521.41 & X-ray diffraction & 528.1 & 1.28\\
\cite{1939RSPSA.171..569S} & 13.9 & 534.9 & X-ray diffraction & 529.9 & 0.93\\
 & 14.25 & 540.4 &  & 529.9 & 1.95\\
 & 14.3 & 540.4 &  & 529.9 & 1.95\\
 & 15.0 & 557.5 &  & 529.8 & 4.97\\
 & 17.0 & 557.5 &  & 529.4 & 5.04\\
 & 18.5 & 533.5 &  & 529.0 & 0.86\\
 & 20.35 & 530.8 &  & 528.2 & 0.50\\
 & 20.35 & 551.7 &  & 528.2 & 4.27\\
 & 22.0 & 532.2 &  & 525.0 & 1.36\\
 & 27.5 & 533.5 &  & 523.8 & 1.82\\
 & 35.0 & 521.4 &  & 521.8 & 0.08\\
\cite{1964Sov...32..116M} & 77.4 & 494.2 & Pycnometry & 501.3 & 1.43\\
 & 85.4 & 489.2 &  & 495.8 & 1.35\\
\cite{1966sov..Tolka} & 20.4 & 522.5 & Pycnometry & 528.1 & 1.08\\
\cite{1969Zap..57.2053F} & 4.0 & 528.1 & X-ray diffraction & 530.0 & 0.36\\
 & 5.0 & 526.2 &  & 530.0 & 0.73\\
 & 17.0 & 521.9 &  & 529.4 & 1.43\\
 & 20.0 & 519.0 &  & 528.3 & 1.79\\
 & 21.0 & 517.4 &  & 525.1 & 1.49\\
 & 25.0 & 516.9 &  & 524.4 & 1.44\\
 & 31.0 & 515.6 &  & 523.0 & 1.42\\
 & 35.0 & 511.7 &  & 521.8 & 1.98\\
 & 40.0 & 510.7 &  & 520.2 & 1.86\\
 & 50.0 & 508.4 &  & 516.3 & 1.56\\
 & 55.0 & 502.7 &  & 514.0 & 2.23\\
 & 58.0 & 500.2 &  & 512.5 & 2.45\\
 & 61.0 & 497.7 &  & 511.0 & 2.66\\
 & 65.0 & 490.3 &  & 508.8 & 3.77\\
 & 70.0 & 492.8 &  & 505.9 & 2.66\\
 & 75.0 & 485.4 &  & 502.8 & 3.57\\
\cite{1971JChem..17904721B} & 11.0 & 524.5 & X-ray diffraction & 530.1 & 1.06\\
 & 12.0 & 524.4 &  & 530.1 & 1.08\\
 & 13.0 & 524.1 &  & 530.0 & 1.13\\
 & 14.0 & 523.8 &  & 529.9 & 1.17\\
 & 15.0 & 523.2 &  & 529.8 & 1.26\\
 & 16.0 & 522.3 &  & 529.6 & 1.40\\
 & 17.0 & 521.5 &  & 529.4 & 1.52\\
 & 18.0 & 520.6 &  & 529.1 & 1.64\\
 & 19.0 & 519.7 &  & 528.8 & 1.75\\
 & 20.0 & 518.5 &  & 528.3 & 1.90\\
 & 21.0 & 517.5 &  & 525.1 & 1.48\\
 & 22.0 & 517.4 &  & 525.0 & 1.46\\
 & 25.0 & 516.8 &  & 524.4 & 1.47\\
 & 30.0 & 515.2 &  & 523.2 & 1.56\\
 & 35.0 & 513.0 &  & 521.8 & 1.72\\
 & 40.0 & 511.3 &  & 520.2 & 1.74\\
 & 45.0 & 510.0 &  & 518.3 & 1.63\\
 & 50.0 & 507.0 &  & 516.3 & 1.83\\
 & 55.0 & 504.0 &  & 514.0 & 1.98\\
 & 60.0 & 500.0 &  & 511.5 & 2.29\\
 & 65.0 & 497.0 &  & 508.8 & 2.37\\
 & 70.0 & 494.0 &  & 505.9 & 2.40\\
\cite{1975PhDT........68A} & 2.5 & 527.9 & X-ray Diffraction & 530.0 & 0.39\\
 & 4.3 & 528.0 &  & 530.0 & 0.38\\
 & 5.7 & 528.0 &  & 530.0 & 0.39\\
 & 7.0 & 528.0 &  & 530.1 & 0.38\\
 & 7.9 & 528.0 &  & 530.1 & 0.39\\
 & 8.9 & 528.1 &  & 530.1 & 0.38\\
 & 9.8 & 528.0 &  & 530.1 & 0.39\\
 & 10.6 & 528.1 &  & 530.1 & 0.38\\
 & 11.9 & 528.0 &  & 530.1 & 0.40\\
 & 13.1 & 527.9 &  & 530.0 & 0.40\\
 & 13.9 & 527.8 &  & 529.9 & 0.40\\
 & 14.7 & 527.7 &  & 529.8 & 0.40\\
 & 15.2 & 527.6 &  & 529.8 & 0.40\\
 & 15.5 & 527.6 &  & 529.7 & 0.40\\
 & 16.1 & 527.4 &  & 529.6 & 0.41\\
 & 16.9 & 527.2 &  & 529.5 & 0.42\\
 & 17.6 & 527.1 &  & 529.2 & 0.41\\
 & 18.5 & 526.7 &  & 529.0 & 0.43\\
 & 19.0 & 526.4 &  & 528.8 & 0.45\\
 & 19.4 & 526.2 &  & 528.6 & 0.46\\
 & 19.7 & 525.9 &  & 528.5 & 0.50\\
 & 19.8 & 525.8 &  & 528.4 & 0.50\\
 & 19.9 & 525.7 &  & 528.4 & 0.51\\
 & 20.0 & 525.6 &  & 528.3 & 0.52\\
 & 20.1 & 525.5 &  & 528.3 & 0.53\\
 & 20.2 & 525.4 &  & 528.3 & 0.55\\
 & 20.3 & 525.1 &  & 528.2 & 0.58\\
 & 20.4 & 524.7 &  & 528.2 & 0.65\\
 & 20.5 & 523.8 &  & 528.1 & 0.82\\
 & 20.6 & 523.1 &  & 525.2 & 0.41\\
 & 20.7 & 523.0 &  & 525.2 & 0.41\\
 & 20.8 & 523.0 &  & 525.2 & 0.41\\
 & 20.9 & 523.0 &  & 525.2 & 0.41\\
 & 21.1 & 522.9 &  & 525.1 & 0.42\\
 & 21.2 & 522.9 &  & 525.1 & 0.42\\
 & 21.4 & 522.9 &  & 525.1 & 0.42\\
 & 21.5 & 522.8 &  & 525.1 & 0.42\\
 & 21.8 & 522.7 &  & 525.0 & 0.43\\
 & 21.9 & 522.7 &  & 525.0 & 0.43\\
 & 22.0 & 522.7 &  & 525.0 & 0.44\\
 & 22.3 & 522.6 &  & 524.9 & 0.44\\
 & 22.6 & 522.6 &  & 524.9 & 0.44\\
 & 23.1 & 522.5 &  & 524.8 & 0.44\\
 & 24.1 & 522.1 &  & 524.6 & 0.46\\
 & 25.1 & 521.9 &  & 524.4 & 0.46\\
 & 26.0 & 521.7 &  & 524.2 & 0.48\\
 & 27.1 & 521.4 &  & 523.9 & 0.49\\
 & 28.1 & 521.1 &  & 523.7 & 0.50\\
 & 30.1 & 520.5 &  & 523.2 & 0.52\\
 & 32.1 & 519.8 &  & 522.6 & 0.54\\
 & 34.1 & 519.2 &  & 522.1 & 0.56\\
 & 36.0 & 518.5 &  & 521.5 & 0.59\\
 & 38.1 & 517.7 &  & 520.8 & 0.60\\
 & 40.0 & 517.0 &  & 520.2 & 0.62\\
 & 42.0 & 516.2 &  & 519.5 & 0.63\\
 & 44.0 & 515.3 &  & 518.7 & 0.65\\
 & 46.1 & 514.4 &  & 517.9 & 0.67\\
 & 48.1 & 513.5 &  & 517.1 & 0.69\\
 & 50.2 & 512.5 &  & 516.2 & 0.72\\
 & 52.1 & 511.6 &  & 515.3 & 0.72\\
 & 54.2 & 510.7 &  & 514.4 & 0.72\\
 & 55.9 & 509.7 &  & 513.5 & 0.74\\
 & 58.1 & 508.6 &  & 512.4 & 0.75\\
 & 60.1 & 507.6 &  & 511.4 & 0.76\\
 & 62.2 & 506.4 &  & 510.3 & 0.77\\
 & 64.4 & 505.3 &  & 509.1 & 0.76\\
 & 66.3 & 504.2 &  & 508.0 & 0.76\\
 & 68.4 & 502.9 &  & 506.8 & 0.78\\
 & 70.7 & 501.6 &  & 505.5 & 0.77\\
 & 73.0 & 500.1 &  & 504.1 & 0.79\\
 & 74.9 & 498.9 &  & 502.8 & 0.79\\
 & 76.8 & 497.7 &  & 501.7 & 0.80\\
 & 78.9 & 496.4 &  & 500.2 & 0.78\\
 & 81.0 & 495.0 &  & 498.9 & 0.79\\
 & 82.9 & 493.6 &  & 497.6 & 0.81\\
 & 84.8 & 492.3 &  & 496.2 & 0.79\\
 & 86.2 & 491.3 &  & 495.2 & 0.80\\
\cite{1983xxx..12..419C} & 10.0 & 529.9 & X-ray diffraction & 530.1 & 0.04\\
 & 12.0 & 529.8 &  & 530.1 & 0.05\\
 & 14.0 & 529.6 &  & 529.9 & 0.07\\
 & 16.0 & 529.2 &  & 529.6 & 0.09\\
 & 18.0 & 528.5 &  & 529.1 & 0.12\\
 & 19.0 & 527.9 &  & 528.8 & 0.16\\
 & 20.0 & 527.1 &  & 528.3 & 0.23\\
 & 20.3 & 526.7 &  & 528.2 & 0.29\\
 & 20.6 & 524.9 &  & 525.2 & 0.06\\
 & 20.8 & 524.7 &  & 525.2 & 0.10\\
 & 21.0 & 524.5 &  & 525.1 & 0.13\\
 & 21.5 & 524.0 &  & 525.1 & 0.19\\
 & 22.0 & 523.8 &  & 525.0 & 0.23\\
 & 23.0 & 523.4 &  & 524.8 & 0.26\\
 & 24.0 & 523.1 &  & 524.6 & 0.28\\
 & 26.0 & 522.5 &  & 524.2 & 0.31\\
 & 28.0 & 521.9 &  & 523.7 & 0.34\\
 & 30.0 & 521.3 &  & 523.2 & 0.37\\
 & 35.0 & 519.7 &  & 521.8 & 0.40\\
 & 40.0 & 517.8 &  & 520.2 & 0.46\\
 & 45.0 & 515.8 &  & 518.3 & 0.49\\
 & 50.0 & 513.6 &  & 516.3 & 0.52\\
 & 55.0 & 511.2 &  & 514.0 & 0.54\\
 & 60.0 & 508.7 &  & 511.5 & 0.55\\
 & 65.0 & 506.0 &  & 508.8 & 0.55\\
 & 70.0 & 503.1 &  & 505.9 & 0.55\\
 & 75.0 & 500.1 &  & 502.8 & 0.54\\
 & 80.0 & 496.9 &  & 499.5 & 0.53\\
 & 85.0 & 493.7 &  & 496.1 & 0.49\\
 & 88.0 & 491.4 &  & 493.9 & 0.52\\
 & 89.0 & 490.6 &  & 493.2 & 0.53\\
 & 90.0 & 489.8 &  & 492.5 & 0.54\\
\cite{2020IUCrj292..683M} & 8.0 & 530.09 & Neutron diffraction & 530.1 & 0.00\\
 & 10.0 & 530.11 &  & 530.1 & 0.00\\
 & 12.0 & 530.05 &  & 530.1 & 0.00\\
 & 14.0 & 529.89 &  & 529.9 & 0.01\\
 & 16.0 & 529.62 &  & 529.6 & 0.00\\
 & 18.0 & 529.2 &  & 529.1 & 0.01\\
 & 20.0 & 528.31 &  & 528.3 & 0.01\\
 & 22.0 & 525.41 &  & 525.0 & 0.08\\
 & 24.0 & 524.76 &  & 524.6 & 0.03\\
 & 26.0 & 524.24 &  & 524.2 & 0.01\\
 & 28.0 & 523.64 &  & 523.7 & 0.01\\
 & 30.0 & 523.06 &  & 523.2 & 0.03\\
 & 32.0 & 522.59 &  & 522.7 & 0.02\\
 & 34.0 & 521.85 &  & 522.1 & 0.05\\
 & 36.0 & 521.35 &  & 521.5 & 0.03\\
 & 38.0 & 520.57 &  & 520.9 & 0.06\\
 & 40.0 & 519.93 &  & 520.2 & 0.05\\
 & 42.0 & 518.98 &  & 519.5 & 0.09\\
 & 44.0 & 518.53 &  & 518.7 & 0.04\\
 & 46.0 & 518.05 &  & 517.9 & 0.02\\
 & 48.0 & 517.6 &  & 517.1 & 0.09\\
 & 50.0 & 516.58 &  & 516.3 & 0.06\\
 & 52.0 & 515.58 &  & 515.4 & 0.04\\
 & 54.0 & 514.55 &  & 514.4 & 0.02\\
 & 56.0 & 513.33 &  & 513.5 & 0.03\\
 & 58.0 & 512.49 &  & 512.5 & 0.00\\
 & 60.0 & 511.4 &  & 511.5 & 0.01\\
 & 62.0 & 510.62 &  & 510.4 & 0.04\\
 & 64.0 & 509.66 &  & 509.3 & 0.06\\
 & 66.0 & 508.73 &  & 508.2 & 0.10\\
 & 68.0 & 506.26 &  & 507.1 & 0.16\\
 & 70.0 & 505.54 &  & 505.9 & 0.07\\
 & 72.0 & 504.67 &  & 504.7 & 0.00\\
 & 74.0 & 503.09 &  & 503.4 & 0.07\\
 & 76.0 & 502.91 &  & 502.2 & 0.15\\
 & 78.0 & 500.64 &  & 500.9 & 0.04\\
 & 80.0 & 499.7 &  & 499.5 & 0.03\\
 & 82.0 & 497.98 &  & 498.2 & 0.04\\
 & 84.0 & 497.49 &  & 496.8 & 0.14\\
 & 86.0 & 495.31 &  & 495.4 & 0.02\\
 & 88.0 & 493.37 &  & 493.9 & 0.12\\
\cite{2017LTP....43..724D} & 16.0 & 510.0 & Pycnometry & 529.6 & 3.85\\
 & 30.0 & 520.0 &  & 523.2 & 0.62\\
\toprule
\textbf{C$_2$H$_6$} & Temp & Density & Technique & Adopted & Deviation\\
 & (K)  & (kg$\cdot$m$^{-3}$) &  & density (kg$\cdot$m$^{-3}$) & (\%)\\
\hline
\cite{1925Zeiov...32..116M} & 88.0 & 708.0 & X-ray diffraction & 709.6 & 0.22\\
\cite{1930PhChem.147..466H} & 20.4 & 751.9 & Pycnometry & 739.8 & 1.61\\
\cite{1958JChPh..28..425S} & 77.0 & 713.0 & Pycnometry & 718.3 & 0.74\\
\cite{1978Chry..36..829W} & 90.0 & 669.0 & X-ray diffraction & 669.0 & 0.00\\
 & 85.0 & 719.0 &  & 712.1 & 0.96\\
\cite{2008LTP....34.1038K} & 5.0 & 740.2 & X-ray diffraction & 738.8 & 0.18\\
 & 10.0 & 740.1 &  & 739.5 & 0.09\\
 & 15.0 & 739.7 &  & 739.8 & 0.01\\
 & 20.0 & 739.2 &  & 739.8 & 0.08\\
 & 25.0 & 738.6 &  & 739.5 & 0.12\\
 & 30.0 & 737.8 &  & 738.9 & 0.15\\
 & 35.0 & 737.1 &  & 738.0 & 0.12\\
 & 40.0 & 735.9 &  & 736.8 & 0.12\\
 & 45.0 & 734.8 &  & 735.2 & 0.06\\
 & 50.0 & 733.4 &  & 733.4 & 0.00\\
 & 55.0 & 731.7 &  & 731.3 & 0.06\\
 & 60.0 & 729.8 &  & 728.9 & 0.13\\
 & 65.0 & 727.5 &  & 726.1 & 0.19\\
 & 70.0 & 724.6 &  & 723.1 & 0.21\\
 & 75.0 & 721.5 &  & 719.7 & 0.25\\
 & 80.0 & 717.4 &  & 716.1 & 0.19\\
 & 82.0 & 715.5 &  & 714.5 & 0.14\\
 & 84.0 & 713.3 &  & 712.9 & 0.05\\
 & 86.0 & 710.8 &  & 711.3 & 0.07\\
 & 88.0 & 707.9 &  & 709.6 & 0.24\\
 & 89.5 & 704.4 &  & 708.3 & 0.55\\
 & 90.0 & 656.2 &  & 669.0 & 1.95\\
\toprule
\textbf{C$_2$H$_4$} & Temp & Density & Technique & Adopted & Deviation\\
 & (K) & (kg$\cdot$m$^{-3}$) &  & density (kg$\cdot$m$^{-3}$) & (\%)\\
\hline
\cite{1930PhChem.147..466H} & 20.4 & 780.7 & Pycnometry & 780.9 & 0.03\\
\cite{1940Zei..12..315C} & 103.97 & 717.8 & Pycnometry & 722.6 & 0.68\\
\cite{1958JChPh..28..425S} & 77.0 & 732.0 & Pycnometry & 741.5 & 1.29\\
\cite{1979Chry..36..829W} & 85.0 & 750.4 & X-ray diffraction & 735.9 & 1.94\\
\toprule
 \textbf{C$_2$H$_2$} & Temp & Density & Technique & Adopted & Deviation\\
 & (K) & (kg$\cdot$m$^{-3}$) &  & density (kg$\cdot$m$^{-3}$) & (\%)\\
\hline
\cite{1907phychem..11...306M} & 188.15 & 730.0 & Pynometry & 731.7 & 0.23\\
\cite{1952SciRep..4..607W} & 156.0 & 747.2 & X-ray diffraction & 752.5 & 0.70\\
\cite{1965Tr..10..150A} & 77.0 & 810.0 & Pynometry & 803.7 & 0.78\\
 & 90.0 & 790.0 &  & 795.3 & 0.67\\
\cite{1979Acta..35..2850W} & 160.0 & 752.0 & X-ray diffraction & 749.9 & 0.28\\
 & 170.0 & 750.5 &  & 743.4 & 0.95\\
 & 141.0 & 765.4 &  & 762.2 & 0.42\\
\cite{1992Acta..176..316M} & 131.0 & 764.3 & Neutron diffraction & 768.7 & 0.58\\
 & 141.0 & 760.2 &  & 762.2 & 0.27\\
\toprule
\textbf{C$_6$H$_6$} & Temp & Density & Technique & Adopted & Deviation\\
 & (K) & (kg$\cdot$m$^{-3}$) &  & density (kg$\cdot$m$^{-3}$) & (\%)\\
\hline
\cite{1930PhChem.147..466H} & 20.4 & 1125.5 & Pycnometry & 1125.7 & 0.01\\
\cite{1958RSPSA.247....1C} & 270.15 & 1023.0 & X-ray diffraction & 1022.0 & 0.09\\
\cite{1964RSPSA.279...98B} & 218.15 & 1055.0 & Neutron diffraction & 1055.4 & 0.02\\
 & 138.15 & 1094.0 &  & 1094.7 & 0.02\\
\cite{Kohzin1954} & 77.0 & 1115.0 & X-ray diffraction & 1114.7 & 0.02\\
\cite{1921JAm..43...1538R} & 278.7 & 1015.1 & Pycnometry & 1015.9 & 0.08\\
\toprule
\textbf{C$_3$H$_8$} & Temp & Density & Technique & Adopted & Deviation\\
 & (K) & (kg$\cdot$m$^{-3}$) &  & density (kg$\cdot$m$^{-3}$) & (\%)\\
\hline
\cite{1930PhChem.147..466H} & 20.4 & 813.7 & Pycnometry & 813.5 & 0.00\\
\cite{1958JChPh..28..425S} & 77.0 & 763.0 & Pycnometry & 759.5 & 0.50\\
\cite{1999Ang..38.988B} & 30.0 & 803.0 & X-ray diffraction & 804.4 & 0.20\\
 & 90.0 & 743.0 &  & 747.1 & 0.60\\
\toprule
\textbf{C$_3$H$_6$} & Temp & Density & Technique & Adopted & Deviation\\
 & (K) & (kg$\cdot$m$^{-3}$) &  & density (kg$\cdot$m$^{-3}$) & (\%)\\
\hline
\cite{1958JChPh..28..425S} & 77.0 & 806.0 & Pycnometry & 806.0 & 0.00\\
 \toprule
\textbf{C$_3$H$_4$-a} & Temp & Density & Technique & Adopted & Deviation\\
 & (K) & (kg$\cdot$m$^{-3}$) &  & density (kg$\cdot$m$^{-3}$) & (\%)\\
\hline
\cite{1962JChPh..36.2654B} & 95.0 & 860.0 & Lattice structure model & 860.0 & 0.00\\
\toprule
\textbf{C$_3$H$_4$-p} & Temp & Density & Technique & Adopted & Deviation\\
 & (K) & (kg$\cdot$m$^{-3}$) &  & density (kg$\cdot$m$^{-3}$) & (\%)\\
\hline
\cite{2022LPICo2678.2321M} & 90.0 & 882.0 & X-ray diffraction & 882.0 & 0.00\\
\toprule
\textbf{HCN} & Temp & Density & Technique & Adopted & Deviation\\
 & (K) & (kg$\cdot$m$^{-3}$) &  & density (kg$\cdot$m$^{-3}$) & (\%)\\
\hline
\cite{1951Acta..4..330D} & 153.15 & 1032.8 & X-ray diffraction & 1028.2 & 0.45\\
 & 193.15 & 965.1 &  & 974.3 & 0.96\\
\cite{1929Werner} & 233.15 & 925.0 & Pycnometry & 920.4 & 0.50\\
\toprule
\textbf{HC$_3$N} & Temp & Density & Technique & Adopted & Deviation\\
 & (K) & (kg$\cdot$m$^{-3}$) &  & density (kg$\cdot$m$^{-3}$) & (\%)\\
\hline
\cite{1958ActaCry..28..425S} & 248.15 & 1075.3 & X-ray diffraction & 1075.3 & 0.00\\
\toprule
\textbf{CO$_2$} & Temp & Density & Technique & Adopted & Deviation\\
 & (K) & (kg$\cdot$m$^{-3}$) &  & density (kg$\cdot$m$^{-3}$) & (\%)\\
\hline
\cite{1930PhChem.147..466H} & 20.4 & 1702.1 & Pycnometry & 1708.3 & 0.36\\
\cite{1900AnP...308..733B} & 194.15 & 1560.0 & Pycnometry & 1566.8 & 0.43\\
\cite{1924SKons..27...839D} & 78.8 & 1638.7 & X-ray diffraction & 1683.6 & 2.74\\
\cite{1926RSPSA.111..224M} & 191.65 & 1569.0 & Pycnometry & 1570.3 & 0.08\\
 & 189.55 & 1573.0 &  & 1573.2 & 0.01\\
 & 187.65 & 1575.0 &  & 1575.8 & 0.05\\
 & 177.85 & 1589.0 &  & 1588.8 & 0.02\\
 & 167.95 & 1600.0 &  & 1601.2 & 0.08\\
 & 157.95 & 1610.0 &  & 1613.1 & 0.19\\
 & 144.35 & 1621.0 &  & 1628.3 & 0.45\\
 & 90.15 & 1663.0 &  & 1676.1 & 0.79\\
 & 193.55 & 1565.0 &  & 1567.6 & 0.17\\
 & 185.25 & 1582.0 &  & 1579.0 & 0.19\\
 & 178.05 & 1590.0 &  & 1588.5 & 0.09\\
 & 168.05 & 1602.0 &  & 1601.1 & 0.06\\
 & 158.25 & 1611.0 &  & 1612.8 & 0.11\\
 & 147.95 & 1623.0 &  & 1624.4 & 0.08\\
 & 137.95 & 1630.0 &  & 1634.9 & 0.30\\
 & 90.15 & 1674.0 &  & 1676.1 & 0.12\\
\cite{1934Phy.....1..655K} & 83.15 & 1687.6 & X-ray diffraction & 1680.8 & 0.41\\
\cite{1934Phy.....1..167K} & 20.0 & 1718.0 & X-ray diffraction & 1708.4 & 0.56\\
 & 35.0 & 1714.3 &  & 1704.3 & 0.58\\
 & 65.0 & 1701.3 &  & 1691.5 & 0.58\\
 & 86.0 & 1688.6 &  & 1678.9 & 0.57\\
 & 94.0 & 1682.2 &  & 1673.3 & 0.53\\
 & 111.0 & 1667.8 &  & 1660.0 & 0.47\\
 & 114.0 & 1663.4 &  & 1657.5 & 0.35\\
\cite{1969PhDT........76M} & 77.0 & 1670.0 & Pycnometry & 1684.7 & 0.88\\
\cite{1971ApOpt..10.2086S} & 82.0 & 1670.0 & Pycnometry & 1681.5 & 0.69\\
\cite{1990ActaCry..36..2750S} & 150.0 & 1643.9 & X-ray diffraction & 1622.1 & 1.33\\
\cite{2017Icar..294..201M} & 80.0 & 1684.6 & X-ray diffraction & 1682.8 & 0.10\\
 & 85.0 & 1681.4 &  & 1679.6 & 0.11\\
 & 90.0 & 1678.3 &  & 1676.2 & 0.13\\
 & 95.0 & 1674.0 &  & 1672.6 & 0.08\\
 & 100.0 & 1670.4 &  & 1668.9 & 0.09\\
 & 105.0 & 1667.4 &  & 1664.9 & 0.15\\
 & 110.0 & 1662.4 &  & 1660.9 & 0.09\\
 & 115.0 & 1658.7 &  & 1656.6 & 0.13\\
 & 120.0 & 1653.4 &  & 1652.2 & 0.07\\
 & 110.0 & 1664.1 &  & 1660.9 & 0.19\\
 & 115.0 & 1658.4 &  & 1656.6 & 0.11\\
 & 120.0 & 1654.0 &  & 1652.2 & 0.11\\
 & 125.0 & 1649.9 &  & 1647.6 & 0.14\\
 & 130.0 & 1646.3 &  & 1642.9 & 0.21\\
\toprule
\textbf{CH$_3$CN-$\alpha$} & Temp & Density & Technique & Adopted & Deviation\\
 & (K) & (kg$\cdot$m$^{-3}$) &  & density (kg$\cdot$m$^{-3}$) & (\%)\\
\hline
\cite{1981Act...37..2239B} & 215.0 & 1022.4 & X-ray diffraction & 1022.4 & 0.00\\
\cite{1997Chem..130...899} & 208.0 & 1025.3 & X-ray diffraction & 1025.5 & 0.01\\
\cite{ActaCry...57.2053F} & 201.0 & 1027.9 & X-ray diffraction & 1027.9 & 0.01\\
\cite{1989JPCM....1.5827T} & 229.0 & 1016.5 & Neutron diffraction & 1014.3 & 0.22\\
\cite{2020ECS.....4.1324C} & 225.0 & 1016.5 & Neutron diffraction & 1016.8 & 0.03\\
 & 230.0 & 1011.7 &  & 1013.6 & 0.19\\
 & 235.0 & 1010.0 &  & 1010.0 & 0.01\\
\toprule
\textbf{CH$_3$CN-$\beta$} & Temp & Density & Technique & Adopted & Deviation\\
 & (K) & (kg$\cdot$m$^{-3}$) &  & density (kg$\cdot$m$^{-3}$) & (\%)\\
\hline
\cite{1991darn.reptR....A} & 100.0 & 1104.5 & Neutron diffraction & 1107.8 & 0.30\\
\cite{ActaCry...57.2053F} & 206.0 & 1058.2 & X-ray diffraction & 1053.1 & 0.48\\
\cite{1989JPCM....1.5827T} & 12.0 & 1131.0 & Neutron diffraction & 1126.8 & 0.37\\
 & 60.0 & 1125.6 &  & 1119.4 & 0.56\\
 & 110.0 & 1107.3 &  & 1104.1 & 0.29\\
 & 170.0 & 1075.8 &  & 1075.6 & 0.02\\
 & 220.0 & 1044.5 &  & 1043.3 & 0.11\\
\cite{2020ECS.....4.1324C} & 5.0 & 1125.1 & Neutron diffraction & 1127.2 & 0.19\\
 & 10.0 & 1125.2 &  & 1126.9 & 0.15\\
 & 15.0 & 1124.9 &  & 1126.5 & 0.14\\
 & 20.0 & 1125.0 &  & 1126.0 & 0.09\\
 & 25.0 & 1124.7 &  & 1125.5 & 0.07\\
 & 30.0 & 1124.2 &  & 1124.8 & 0.05\\
 & 35.0 & 1123.9 &  & 1124.1 & 0.02\\
 & 40.0 & 1123.2 &  & 1123.3 & 0.01\\
 & 45.0 & 1122.6 &  & 1122.4 & 0.01\\
 & 50.0 & 1121.7 &  & 1121.5 & 0.02\\
 & 55.0 & 1120.7 &  & 1120.5 & 0.02\\
 & 60.0 & 1119.5 &  & 1119.4 & 0.01\\
 & 65.0 & 1118.4 &  & 1118.2 & 0.02\\
 & 70.0 & 1117.2 &  & 1116.9 & 0.02\\
 & 75.0 & 1115.6 &  & 1115.6 & 0.01\\
 & 80.0 & 1114.2 &  & 1114.2 & 0.00\\
 & 85.0 & 1112.7 &  & 1112.7 & 0.00\\
 & 90.0 & 1111.0 &  & 1111.1 & 0.01\\
 & 95.0 & 1109.2 &  & 1109.5 & 0.02\\
 & 100.0 & 1107.4 &  & 1107.8 & 0.03\\
 & 105.0 & 1105.6 &  & 1106.0 & 0.03\\
 & 110.0 & 1103.7 &  & 1104.1 & 0.04\\
 & 115.0 & 1101.7 &  & 1102.1 & 0.04\\
 & 120.0 & 1099.6 &  & 1100.1 & 0.05\\
 & 125.0 & 1097.6 &  & 1098.0 & 0.04\\
 & 130.0 & 1095.5 &  & 1095.8 & 0.03\\
 & 135.0 & 1093.0 &  & 1093.6 & 0.05\\
 & 140.0 & 1090.8 &  & 1091.2 & 0.04\\
 & 145.0 & 1088.7 &  & 1088.8 & 0.01\\
 & 150.0 & 1085.9 &  & 1086.3 & 0.04\\
 & 155.0 & 1083.3 &  & 1083.7 & 0.04\\
 & 160.0 & 1080.8 &  & 1081.1 & 0.02\\
 & 165.0 & 1078.0 &  & 1078.4 & 0.03\\
 & 170.0 & 1075.4 &  & 1075.6 & 0.02\\
 & 175.0 & 1072.5 &  & 1072.7 & 0.02\\
 & 180.0 & 1069.3 &  & 1069.7 & 0.04\\
 & 185.0 & 1066.0 &  & 1066.7 & 0.06\\
 & 190.0 & 1063.1 &  & 1063.6 & 0.05\\
 & 195.0 & 1059.7 &  & 1060.4 & 0.07\\
 & 200.0 & 1056.4 &  & 1057.1 & 0.07\\
 & 205.0 & 1053.0 &  & 1053.8 & 0.08\\
 & 210.0 & 1049.4 &  & 1050.4 & 0.09\\
 & 215.0 & 1045.8 &  & 1046.9 & 0.10\\
\toprule
\textbf{C$_2$H$_5$CN} & Temp & Density & Technique & Adopted & Deviation\\
 & (K) & (kg$\cdot$m$^{-3}$) &  & density (kg$\cdot$m$^{-3}$) & (\%)\\
\hline
\cite{2020Jsyn..27...212} & 100.0 & 1014.0 & X-ray diffraction & 1014.5 & 0.01\\
 & 110.0 & 1010.0 &  & 1009.9 & 0.01\\
 & 120.0 & 1005.0 &  & 1005.3 & 0.02\\
 & 130.0 & 1001.0 &  & 1000.7 & 0.03\\
 & 140.0 & 996.0 &  & 996.0 & 0.03\\
 & 150.0 & 991.0 &  & 991.4 & 0.05\\
\toprule
\textbf{C$_2$H$_3$CN} & Temp & Density & Technique & Adopted & Deviation\\
 & (K) & (kg$\cdot$m$^{-3}$) &  & density (kg$\cdot$ m$^{-3}$) & (\%)\\
\hline
\cite{1998xxx..48..299c} & 153.0 & 1027.4 & X-ray diffraction & 1027.4 & 0.00\\
\toprule
\textbf{C$_2$N$_2$} & Temp & Density & Technique & Adopted & Deviation\\
 & (K) & (kg$\cdot$m$^{-3}$) &  & density (kg$\cdot$m$^{-3}$) & (\%)\\
\hline
\cite{1963Acta...16..734} & 178.15 & 1250.2 & X-ray diffraction & 1250.2 & 0.00\\
\cite{1989JPCM....1.5827T} & 15.0 & 1282.9 & Neutron diffraction & 1282.9 & 0.00\\
\toprule
\textbf{C$_4$N$_2$} & Temp & Density & Technique & Adopted & Deviation\\
 & (K) & (kg$\cdot$m$^{-3}$) &  & density (kg$\cdot$m$^{-3}$) & (\%)\\
\hline
\cite{1953ActCry...6..350H} & 278.15 & 1229.5 & X-ray diffraction & N/A & N/A\\
\toprule
\label{table:ice_density_data_1}
\end{longtable}

\clearpage

\setlength\LTleft{-40pt}    
\setlength\LTright{-40pt}           
\setlength\LTcapwidth{\linewidth}
\begin{longtable}{cccccc}
  \caption{Organic ice average density data that are not used for the fittings in Table~\ref{table:density_solid}, the associated references, adopted intrinsic density fits using expressions in Table~\ref{table:density_solid}, and the phase of ice. The quartz crystal microbalance (QCM)/double laser interferometry combination technique is abbreviated as QCM in the table.}\\
    \endfirsthead
    \caption{Organic ice density data (cont.)} \\
\hline
    \endhead
\multicolumn{6}{r}{Continued on the next page}
    \endfoot
    \endlastfoot
\toprule
\textbf{CH$_4$} & Temp & Density & Technique & Adopted & Phase\\
 & (K) & (kg$\cdot$m$^{-3}$) & & density (kg$\cdot$m$^{-3}$) & \\
\hline
\cite{1980aro..rept.....R} & 20.0 & 426.0 & QCM & 528.3 & Undetermined\\
\cite{2008PSS...56.1748S} & 10.0 & 471.5 & QCM & 530.1 & Undetermined\\
\cite{2017rus...865..1989} & 16.0 & 480.0 & Estimated using $n_{vis}$ & 529.6 & Crystalline\\
 & 30.0 & 490.0 &  & 523.2 & Crystalline\\
\cite{2022Icar..37314799Y} & 30.0 & 508.0 & QCM & 523.2 & Crystalline\\
\toprule
\textbf{C$_2$H$_6$} & Temp & Density & Technique & Adopted & Phase\\
 & (K)  & (kg$\cdot$m$^{-3}$) &  & density (kg$\cdot$m$^{-3}$) & (\%)\\
\hline
\cite{2016ApJ...825..156M} & 30.0 & 539.0 & QCM & 738.9 & Amorphous\\
 & 18.0 & 414.0 &  & 739.8 & Amorphous\\
\cite{2017Icar..296..179S} & 13.0 & 411.0 & QCM & 739.7 & Amorphous\\
 & 18.0 & 413.0 &  & 739.8 & Amorphous\\
 & 20.0 & 441.0 &  & 739.8 & Amorphous\\
 & 25.0 & 474.0 &  & 739.5 & Amorphous\\
 & 30.0 & 530.0 &  & 738.9 & Amorphous\\
 & 35.0 & 541.0 &  & 738.0 & Amorphous\\
 & 40.0 & 600.0 &  & 736.8 & Amorphous\\
 & 45.0 & 602.0 &  & 735.2 & Crystalline\\
 & 65.0 & 604.0 &  & 726.1 & Crystalline\\
\cite{2022Icar..37314799Y} & 60.0 & 778.0 & QCM & 728.9 & Crystalline\\
\toprule
\textbf{C$_2$H$_4$} & Temp & Density & Technique & Adopted & Phase\\
 & (K) & (kg$\cdot$m$^{-3}$) &  & density (kg$\cdot$m$^{-3}$) & \\
\hline
 \cite{2017MNRAS.466.1894M} & 30.0 & 580.0 & QCM & 774.2 & Metastable crystalline\\
\cite{2017Icar..296..179S} & 13.0 & 463.0 & QCM & 786.1 & Amorphous\\
 & 16.0 & 507.0 &  & 784.0 & Amorphous\\
 & 18.0 & 530.0 &  & 782.6 & Amorphous\\
 & 20.0 & 564.0 &  & 781.2 & Amorphous\\
 & 22.0 & 570.0 &  & 779.8 & Metastable crystalline\\
 & 26.0 & 584.0 &  & 777.0 & Metastable crystalline\\
 & 28.0 & 588.0 &  & 775.6 & Metastable crystalline\\
 & 30.0 & 580.0 &  & 774.2 & Metastable crystalline\\
 & 33.0 & 620.0 &  & 772.1 & Metastable crystalline\\
 & 35.0 & 643.0 &  & 770.7 & Metastable crystalline\\
 & 40.0 & 634.0 &  & 767.3 & Crystalline\\
 & 45.0 & 650.0 &  & 763.8 & Crystalline\\
 & 50.0 & 629.0 &  & 760.3 & Crystalline\\
 & 55.0 & 636.0 &  & 756.8 & Crystalline\\
 & 60.0 & 629.0 &  & 753.3 & Crystalline\\
 & 65.0 & 630.0 &  & 749.8 & Crystalline\\
\cite{2022Icar..37314799Y} & 60.0 & 796.0 & QCM & 753.3 & Crystalline\\
\toprule
\textbf{C$_6$H$_6$} & Temp & Density & Technique & Adopted & Phase\\
 & (K) & (kg$\cdot$m$^{-3}$) &  & density (kg$\cdot$m$^{-3}$) & \\
\hline
\cite{2010Icar..205..695R} & 100.0 & 952.0 & Estimated using $n_{vis}$ & 1108.2 & Crystalline\\
\cite{2022Icar..37314799Y} & 100.0 & 1085.0 & QCM & 1108.2 & Crystalline\\
\toprule
\textbf{C$_3$H$_8$} & Temp & Density & Technique & Adopted & Phase\\
 & (K) & (kg$\cdot$m$^{-3}$) &  & density (kg$\cdot$m$^{-3}$) & \\
\hline
\cite{2021Icar..35414033H} & 15.0 & 653.0 & QCM & 818.7 & Amorphous\\
 & 65.0 & 797.0 &  & 771.0 & Crystalline\\
\cite{2022Icar..37314799Y} & 65.0 & 797.0 & QCM & 771.0 & Crystalline\\
\toprule
\textbf{C$_3$H$_6$} & Temp & Density & Technique & Adopted & Phase\\
 & (K) & (kg$\cdot$m$^{-3}$) &  & density (kg$\cdot$m$^{-3}$) & \\
\hline
\cite{2021Icar..35414033H} & 15.0 & 663.0 & QCM & N/A & Amorphous\\
 & 65.0 & 782.0 &  &  N/A & Crystalline\\
\toprule
\textbf{C$_3$H$_4$-p} & Temp & Density & Technique & Adopted & Phase\\
 & (K) & (kg$\cdot$m$^{-3}$) &  & density (kg$\cdot$m$^{-3}$) & \\
\hline
\cite{2021Icar..35414033H} & 15.0 & 705.0 & QCM & N/A & Amorphous\\
 & 80.0 & 866.0 &  & N/A & Crystalline\\
\toprule
\textbf{HCN} & Temp & Density & Technique & Adopted & Phase\\
 & (K) & (kg$\cdot$m$^{-3}$) &  & density (kg$\cdot$m$^{-3}$) & \\
\hline
\cite{2020Icar..33813548H} & 15.0 & 808.0 & QCM & 1214.4 & Amorphous\\
\cite{2022Icar..37314799Y} & 120.0 & 1037.0 & QCM & 1072.9 & Crystalline\\
\toprule
\textbf{CO$_2$} & Temp & Density & Technique & Adopted & Phase\\
 & (K) & (kg$\cdot$m$^{-3}$) &  & density (kg$\cdot$m$^{-3}$) & \\
\hline
\cite{1980CP.....52..381S} & 4.1 & 1088.0 & Estimated using $n_{vis}$ & 1712.5 & Undetermined\\
 & 6.0 & 1066.0 &  & 1712.3 & Undetermined\\
 & 7.2 & 1035.0 &  & 1712.1 & Undetermined\\
 & 8.1 & 1028.0 &  & 1712.0 & Undetermined\\
 & 9.7 & 1013.0 &  & 1711.7 & Undetermined\\
 & 12.0 & 1013.0 &  & 1711.4 & Undetermined\\
 & 14.9 & 1016.0 &  & 1710.9 & Undetermined\\
 & 17.1 & 1030.0 &  & 1710.4 & Undetermined\\
 & 18.2 & 1042.0 &  & 1710.2 & Undetermined\\
 & 20.2 & 1069.0 &  & 1709.8 & Undetermined\\
 & 23.0 & 1101.0 &  & 1709.1 & Undetermined\\
 & 25.2 & 1135.0 &  & 1708.6 & Undetermined\\
 & 28.1 & 1197.0 &  & 1707.8 & Undetermined\\
 & 29.7 & 1224.0 &  & 1707.3 & Undetermined\\
 & 31.3 & 1245.0 &  & 1706.9 & Undetermined\\
 & 33.1 & 1268.0 &  & 1706.3 & Undetermined\\
 & 35.2 & 1307.0 &  & 1705.6 & Undetermined\\
 & 36.5 & 1324.0 &  & 1705.2 & Undetermined\\
 & 37.8 & 1362.0 &  & 1704.7 & Undetermined\\
 & 40.8 & 1399.0 &  & 1703.7 & Undetermined\\
 & 44.9 & 1465.0 &  & 1702.1 & Undetermined\\
 & 48.7 & 1499.0 &  & 1700.6 & Undetermined\\
 & 56.8 & 1588.0 &  & 1696.9 & Undetermined\\
 & 60.4 & 1652.0 &  & 1695.2 & Undetermined\\
 & 65.1 & 1664.0 &  & 1692.7 & Undetermined\\
 & 69.7 & 1711.0 &  & 1690.2 & Undetermined\\
 & 73.4 & 1747.0 &  & 1688.0 & Undetermined\\
 & 76.9 & 1763.0 &  & 1685.9 & Undetermined\\
 & 79.5 & 1762.0 &  & 1684.3 & Undetermined\\
 & 81.9 & 1769.0 &  & 1682.8 & Undetermined\\
 & 84.9 & 1769.0 &  & 1680.8 & Undetermined\\
 & 88.0 & 1776.0 &  & 1678.7 & Undetermined\\
\cite{1982JOSA...72..720W} & 20.0 & 1080.0 & QCM & 1709.8 & Undetermined\\
 & 80.0 & 1670.0 &  & 1684.0 & Undetermined\\
\cite{2016ApJ...827...98L} & 14.0 & 1190-1370 & QCM & 1711.0 & Predominantly crystalline\\
 & 20.0 & 1270-1450 &  & 1709.8 & Predominantly crystalline\\
 & 30.0 & 1330-1560 &  & 1707.2 & Predominantly crystalline\\
 & 40.0 & 1520-1640 &  & 1704.0 & Predominantly crystalline\\
 & 50.0 & 1630-1700 &  & 1700.0 & Predominantly crystalline\\
 & 60.0 & 1670-1710 &  & 1695.4 & Predominantly crystalline\\
 & 70.0 & 1670-1700 &  & 1690.0 & Predominantly crystalline\\
\cite{2008PSS...56.1748S} & 10.0 & 990.0 & QCM & 1711.7 & Amorphous\\
 & 12.0 & 985.0 &  & 1711.4 & Amorphous\\
 & 15.0 & 985.0 &  & 1710.8 & Amorphous\\
 & 20.0 & 1030.0 &  & 1709.8 & Partially amorphous\\
 & 30.0 & 1160.0 &  & 1707.2 & Crystalline\\
 & 31.0 & 1160.0 &  & 1706.9 & Crystalline\\
 & 34.0 & 1190.0 &  & 1706.0 & Crystalline\\
 & 42.0 & 1250.0 &  & 1703.2 & Crystalline\\
 & 46.0 & 1350.0 &  & 1701.7 & Crystalline\\
 & 50.0 & 1455.0 &  & 1700.0 & Crystalline\\
 & 58.0 & 1485.0 &  & 1696.3 & Crystalline\\
 & 66.0 & 1485.0 &  & 1692.2 & Crystalline\\
 & 73.0 & 1480.0 &  & 1688.3 & Crystalline\\
 & 77.0 & 1560.0 &  & 1685.9 & Crystalline\\
 & 80.0 & 1540.0 &  & 1684.0 & Crystalline\\
 & 84.0 & 1510.0 &  & 1681.4 & Crystalline\\
 & 90.0 & 1470.0 &  & 1677.3 & Crystalline\\
\cite{2022Icar..37314799Y} & 70.0 & 1666.0 & QCM & 1690.0 & Crystalline\\
\cite{2015JLTP..181....1D} & 10.0 & 986.0 & QCM & 1711.7 & Undetermined\\
 & 12.0 & 983.0 &  & 1711.4 & Undetermined\\
 & 14.0 & 983.0 &  & 1711.0 & Undetermined\\
 & 18.0 & 1032.0 &  & 1710.2 & Undetermined\\
 & 29.0 & 1161.0 &  & 1707.5 & Undetermined\\
 & 33.0 & 1183.0 &  & 1706.3 & Undetermined\\
 & 40.0 & 1255.0 &  & 1704.0 & Undetermined\\
 & 44.0 & 1345.0 &  & 1702.5 & Undetermined\\
 & 48.0 & 1455.0 &  & 1700.9 & Undetermined\\
 & 55.0 & 1485.0 &  & 1697.8 & Undetermined\\
 & 63.0 & 1484.0 &  & 1693.8 & Undetermined\\
 & 70.0 & 1480.0 &  & 1690.0 & Undetermined\\
 & 74.0 & 1559.0 &  & 1687.7 & Undetermined\\
 & 77.0 & 1532.0 &  & 1685.9 & Undetermined\\
 & 80.0 & 1513.0 &  & 1684.0 & Undetermined\\
 & 86.0 & 1468.0 &  & 1680.0 & Undetermined\\
\toprule
\textbf{CH$_3$CN-$\beta$} & Temp & Density & Technique & Adopted & Phase\\
 & (K) & (kg$\cdot$m$^{-3}$) &  & density (kg$\cdot$m$^{-3}$) & \\
\hline
\cite{2020Icar..33813548H} & 15.0 & 778.0 & QCM & 1126.5 & Amorphous\\
\cite{2022Icar..37314799Y} & 130.0 & 1073.0 & QCM & 1095.8 & Crystalline\\
\toprule
\textbf{C$_2$H$_5$CN} & Temp & Density & Technique & Adopted & Phase\\
 & (K) & (kg$\cdot$m$^{-3}$) &  & density (kg$\cdot$m$^{-3}$) & \\
\hline
\cite{2020Icar..33813548H} & 15.0 & 703.0 & QCM & 1053.8 & Amorphous\\
\cite{2022Icar..37314799Y} & 110.0 & 992.0 & QCM & 1009.9 & Crystalline\\
\toprule
\label{table:ice_density_data_2}
\end{longtable}

\clearpage
\begin{table}
\caption{Additional experimental parameters for each tholin production facility included in this work. The unit for the gas flow rate is standard cubic centimeters per minute (sccm). *: Reaction time is 72 h for the samples made with cold plasma discharge and 144 h for the samples made with UV lamp. References: [a] \cite{2006PNAS..10318035T}; [b] \cite{2013ApJ...770L..10H}; [c] \cite{2008Icar..194..186S}; [d] \cite{2004Icar..168..344I}; [e] \cite{2012Icar..218..247I}; [f] \cite{2017ApJ...841L..31H}; [g] \cite{2022PSJ.....3....2L}; [h] \cite{2006P&SS...54..394S}; [i] \cite{2017Icar..289..214S}.}
\addtolength{\leftskip} {-1.8cm}
\resizebox{7.9in}{!}{
\begin{tabular}{crcccccc}
\toprule
Laboratory location && Pressure & Temperature & Reaction time &  Irradiation  & Gas flow rate & Ref.\\
(Setup) && (mbar) & (K) & (h) & time &  (sccm) & \\
\hline
UC Boulder && 800-850 & 300 & N/A & N/A  & 100 & [a, b]\\
(Tolbert setup) &&&&&&\\
\hline
NASA ARC 1 && 0.26-1.6 & 300 & 31-65 & N/A  & 24-41 & [c, d, e] \\
 (tholin-RF system) &&&&&&\\
\hline
JHU && 2.67 & 100-300 & 72-144$^*$
& 3 s & 10 & [f, g]\\
(PHAZER) &&&& &&\\
\hline
LATMOS && 0.2-10 & 300 & 8-32 & N/A  & N/A  & [h]\\
(PAMPRE) &&&&&&\\ 
\hline
UNI && 670 & 300 & 72 & 3 min & 20 & [g] \\
(PAC) &&&&&&\\ 
\hline
NASA ARC 2 && 30 & 150 & 5-13 & 3.5 $\mu$s & 2000 & [g, i] \\
(COSmIC) &&&&&&\\ 
\toprule
\end{tabular}}
\label{table:additional_parameters}
\end{table}

\clearpage

\begin{table}[h]
\addtolength{\leftskip} {-2cm} 
\renewcommand{\arraystretch}{0.8}
\caption{Calculated total surface tension ($\gamma_{l}$) of Titan-relevant organic liquids using the three empirical approaches (Equations~\ref{eq:parachor}, \ref{eq:bbmethod}, and \ref{eq:pitzer}) at the triple point of each species and the accuracy (the ``Acc" column) of the calculated values compared to \cite{Yaws2014}.}
\begin{tabular}{crcccrccrcc}
\toprule
    \centering
     Species && \multicolumn{3}{c}{MacLeod-Sugden method} && \multicolumn{2}{c}{Brock-Bird method} && \multicolumn{2}{c}{Pitzer method}\\
      && \multicolumn{3}{c}{(Equation~\ref{eq:parachor})} && \multicolumn{2}{c}{(Equation~\ref{eq:bbmethod})} && \multicolumn{2}{c}{(Equation~\ref{eq:pitzer})}\\
    \cline{3-5} \cline{7-8} \cline{10-11}
     && Parachor  & $\gamma_l$ ($T_{tri}$) & Acc. && $\gamma_l$($T_{tri}$) & Acc. && $\gamma_l$ ($T_{tri}$) & Acc.\\
     && (mN$\cdot$m$^{-1}$)$^{1/4}\times$(cm$^3\cdot$mol$^{-1}$) & (mN$\cdot$m$^{-1}$) & (\%)  && (mN$\cdot$m$^{-1}$)& (\%) && (mN$\cdot$m$^{-1}$)& (\%)\\
    \hline 
    CH$_4$ && 71.0 & 15.97 & 13.63 && 16.24 & 12.17 && 18.05 & 2.37\\
    \hline 
    C$_2$H$_6$ && 111.0 & 33.46 & 2.96 && 31.94 & 1.73 && 34.27 & 5.45\\
    \hline 
    C$_2$H$_4$ && 99.1 & 28.63 & 3.19 && 27.55 & 6.85 && 29.80 & 0.75\\
    \hline 
    C$_2$H$_2$ && 89.6 & 20.46 & 9.03 && 20.11 & 7.13 && 20.91 & 11.40\\
    \hline 
    C$_4$H$_2$ && 148.2 & 55.53 & N/A && 16.10 & N/A && 16.91 & N/A\\
    \hline 
    C$_6$H$_6$ && 205.7 & 30.70 & 0.35 && 29.83 & 3.15 && 31.34 & 1.73\\
    \hline 
    C$_3$H$_8$ && 151.0 & 39.72 & 11.33 && 36.88 & 3.35 && 39.16 & 9.76\\
    \hline 
    C$_3$H$_6$ && 139.1 & 41.56 & 8.26 && 37.62 & 1.99 && 39.97 & 4.11\\
    \hline 
    C$_3$H$_4$-a && 129.6 & 40.40 & 34.95 && 41.25 & 37.81 && 44.03 & 47.08\\
    \hline 
    C$_3$H$_4$-p && 129.6 & 37.58 & 25.91 && 33.95 & 13.76 && 35.04 & 17.41\\
    \hline 
    HCN && 82.6 & 25.22 & 13.78 && 26.76 & 20.72 && 28.04 & 26.50 \\
    \hline 
    HC$_3$N && 141.2 & 27.75 & N/A && 24.11 & N/A && 24.76 & N/A\\
    \hline 
    CO$_2$ && 68.1 & 11.05 & 33.52 && 19.41 & 16.80 && 17.11 & 2.96\\
    \hline 
    CH$_3$CN && 122.6 & 41.44 & 10.34 && 39.12 & 4.15 && 41.01 & 9.19\\
    \hline 
    C$_2$H$_5$CN && 162.6 & 48.93 & 21.46 && 44.20 & 9.72 && 46.34 & 15.02\\
    \hline 
    C$_2$H$_3$CN && 150.7 & 43.58 & 9.71 && 42.68 & 7.45 && 44.60 & 12.28\\
    \hline 
    C$_2$N$_2$ && 134.2 & 38.98 & 69.02 && 24.80 & 7.54 && 25.86 & 12.13\\
    \hline 
    C$_4$N$_2$ && 192.8 & 36.67 & N/A && 44.48 & N/A && 48.01 & N/A \\
\toprule
    \end{tabular}
    \label{table:calc_SE_liq}
\end{table}

\end{document}